\documentclass[twocolumn,
showpacs,
showkeys,
preprintnumbers,
nofootinbib,
superscriptaddress,
amsmath,
amssymb,
floatfix,
secnumarabic,
aps,
pra,
a4paper,
notitlepage,
final,
]{revtex4-1}%

\usepackage[colorlinks=true,urlcolor=blue]{hyperref}
\usepackage{graphicx}
\usepackage{epsfig}
\usepackage{float}

\newcommand{\beq}{\begin{equation}}
\newcommand{\eeq}{\end{equation}}
\newcommand{\beqar}{\begin{eqnarray}}
\newcommand{\eeqar}{\end{eqnarray}}
\newcommand{\bal}{\begin{aligned}}
\newcommand{\eal}{\end{aligned}}

\def\dalam{\hbox
{\vrule\vbox{\hrule\hbox to 1ex{ \hfill}\kern 1 ex\hrule}\vrule}}

\def\1/2{\hbox{$ {1 \over 2}$ }}

\def\h{\hbar}
\def\i/h{{i \over \h}}

\def\arctg{\hbox{arctg}}

\def\inf{\infty}
\def\pd{\partial} 
\def\v{\vec}

\def\a{\alpha}  
\def\b{\beta}  
 
\def\g{\gamma} \def\G{\Gamma} 
\def\d{\delta} \def\D{\Delta} \def\td{\tilde{\delta}}
\def\l{\lambda}  \def\L{\Lambda}
\def\e{\epsilon} \def\E{\hbox{$\cal E $}}
 
 \def\tN{\tilde {N}}

\def\s{\sigma}
 \def\vr{\varrho}
\def\x{\xi}\def\tx{\tilde{\xi}}
\def\c{\chi} 
\def\vf{\varphi} 
 \def\F{\Phi} \def\tF{\tilde{\Phi}}
 
\def\p{\psi} \def\P{\Psi}

\def\m{\mu}
\def\n{\nu}

  \def\vk{\varkappa}

\def\z{\zeta}

\def\W{\Omega}
\def\tt{\theta}

\def\<{\langle}
\def\>{\rangle}

\def\({\left(}
\def\[{\left[}
\def\){\right)}
\def\]{\right]}

\usepackage{subfigure}

\usepackage[notcite,notref,color]{showkeys}  %Р”Р»СЏ РѕС‚РѕР±СЂР°Р¶РµРЅРёСЏ СЃРѕРґРµСЂР¶РёРјРѕРіРѕ label{} Рє С„РѕСЂРјСѓР»Р°Рј, СЂРёСЃСѓРЅРєР°Рј, С‚Р°Р±Р»РёС†Р°Рј, Р»РёС‚РµСЂР°С‚СѓСЂРµ  %(С‡С‚РѕР±С‹ РЅРµ РїРµСЂРµРёРјРµРЅРѕРІС‹РІР°С‚СЊ С„РѕСЂРјСѓР»С‹, РїРµСЂРµРґ РѕС‚РїСЂР°РІРєРѕР№ РІ Р¶СѓСЂРЅР°Р» РїР°РєРµС‚ РѕС‚РєР»СЋС‡РёС‚СЊ)

\usepackage{tikz}
\usetikzlibrary{decorations.pathreplacing}
\usetikzlibrary{decorations.pathmorphing}    %Р”Р»СЏ СЂРёСЃРѕРІР°РЅРёСЏ РєРѕРЅС‚СѓСЂРѕРІ РёРЅС‚РµРіСЂРёСЂРѕРІР°РЅРёСЏ 2
\usepackage{multirow}
\usepackage{dcolumn}		%РІС‹СЂР°РІРЅРёРІР°РЅРёРµ РїРѕ С‚РѕС‡РєРµ РІ С‚Р°Р±Р»РёС†Р°С…, РёСЃРїРѕР»СЊР·РѕРІР°С‚СЊ d РІРјРµСЃС‚Рѕ c РІ СЃРїРµС†РёС„РёРєР°С†РёРё РєРѕР»РѕРЅРєРё
\newcolumntype{.}{D{.}{.}{-1}}
\newcolumntype{i}[1]{D{.}{.}{#1}}

\newcommand{\myfrac}[2]{{\ifmmode{}^{#1}\!/_{\!#2}\else${}^{#1}\!/_{\!#2}$\fi}}

\begin{document}
\sloppy

\title{Super-critical QED-effects via VP-energy}

\author{K.Sveshnikov and Yu.Voronina}
\email{k.sveshnikov@physics.msu.ru}
\email{voroninayu@physics.msu.ru}
\affiliation{Department of Physics and
Institute of Theoretical Problems of MicroWorld, Moscow State
University, 119991, Leninsky Gory, Moscow, Russia}

%%remember to change the email and affiliation!!!!!

%Remember to change the date
\date{\today}

%%%%%%%%%%%%%%%%%%%

\begin{abstract}
The properties of the QED-vacuum energy $\E_{VP}$ under Coulomb super-criticality condition $Z>Z_{cr,1}$ are explored in essentially non-perturbative approach. It is shown that  in the supercritical region  $\E_{VP}$ is the decreasing function of the Coulomb source parameters, resulting in decay into the negative range as $\sim - Z^4/R$. The conditions, under which the emission of vacuum positrons  can be unambiguously detected on the nuclear conversion pairs background, are also discussed. In particular, for a super-critical dummy nucleus with charge $Z$ and $R(Z) \simeq 1.2 \ (2.5\, Z)^{1/3}$ fm the lowest $1s$-level dives into  the lower continuum at $Z_{cr,1} \simeq 170$ (the model of uniformly charged ball) or $Z_{cr,1} \simeq 173$ (the spherical shell model), whereupon there appears the vacuum shell with the induced charge (-$2|e|$) and the QED-vacuum becomes charged, but in both cases the actual threshold for reliable spontaneous positrons detection is not less than  $Z^\ast \simeq 210$.
\end{abstract}


\maketitle

\section{Introduction}

Nowadays the behavior of QED-vacuum under influence of a supercritical Coulomb source is subject to an active research~\cite{Rafelski2016,Kuleshov2015a,*Kuleshov2015b,*Godunov2017,Davydov2017,*Sveshnikov2017,
*Voronina2017,Popov2018,*Novak2018,*Maltsev2018,Roenko2018,Maltsev2019,*Maltsev2020}. Of the main interest is that  in such external fields there should take place  a deep vacuum state reconstruction, caused by discrete levels diving into the lower continuum and accompanied by such nontrivial effects as spontaneous positron emission combined with vacuum  shells formation (see e.g., Refs.~\cite{Greiner1985a,Plunien1986,Greiner2012,Ruffini2010,Rafelski2016} and citations therein). In 3+1 QED  such effects are expected for extended Coulomb sources of nucleus size with charges $Z>Z_{cr,1} \simeq 170$, which are large enough for direct  observation and probably could be created in low energy heavy ions collisions  at new heavy ion facilities like  FAIR (Darmstadt), NICA  (Dubna), HIAF (Lanzhou)~\cite{FAIR2009,Ter2015,MA2017169}.

In the present paper the non-perturbative VP-effects, caused by quasi-static supercritical Coulomb sources with $Z>Z_{cr,1}$, are explored in terms of VP-energy $\E_{VP}$. Actually, $ \E_{VP} $ is nothing else but the Casimir vacuum energy for the electron-positron system in the external EM-field~\cite{Plunien1986}. As a component of the total electrostatic energy of the whole system, it plays an essential role in attaining the region of super-criticality and its properties. Moreover,  the most important consequences of discrete levels diving into the lower continuum beyond the threshold of super-criticality including vacuum  shells formation and spontaneous  positron emission, show up in the behavior of $\E_{VP}$ even more clear compared to VP-density $\vr_{VP}(\v r)$. In particular, $\E_{VP}$ being considered as a function of $Z$ reveals with growing $Z$ a pronounced decline into the negative range, accompanied with negative jumps, exactly equal to the electron rest mass, which occur each time when the subsequent discrete level attains the threshold of the lower continuum. Furthermore, it is indeed the decline of $\E_{VP}$, which supplies the spontaneous positrons with corresponding energy for emission. At the same time,  from its properties there follows a significant shift $\D Z \simeq 30-40$ with respect to corresponding diving point $Z_{cr}$ of the threshold for  reliable vacuum positron detection on the nuclear conversion pairs background. In view of intimate relation between spontaneous positron emission and lepton number conservation~\cite{Krasnov2022} the last circumstance requires additional attention.

These questions are explored within the  Dirac-Coulomb problem (DC) with external static or adiabatically slowly varying spherically-symmetric Coulomb potential, created by uniformly charged sphere
\beq
\label{1.5a}
V(r)=-Z \a\,\( {1 \over R(Z)}\, \tt(R(Z)-r)+ {1 \over r}\, \tt(r-R(Z)) \)  \ ,
\eeq
or  charged ball
\begin{multline}\label{1.6}
V(r)=- Z \a\,\( {3\, R^2(Z) - r^2 \over 2\,R^3(Z) }\, \tt(R(Z)-r) \ +  \right. \\ \left.  + \ {1 \over r}\, \tt(r-R(Z)) \)  \ .
\end{multline}
Here and henceforth
\beq
\label{1.5b}
Q=Z \a \ ,
\eeq
while the relation between the radius  of the Coulomb source and its charge is given by
\beq
\label{1.8}
R(Z) \simeq 1.2\, (2.5\, |Z|)^{1/3} \ \hbox{fm} \ ,
\eeq
which roughly imitates  the size of super-heavy nucleus with charge $Z$. In what follows $R(Z)$ will be quite frequently denoted simply as $R$.

It  should be specially noted that the parameter $Q$ plays actually the role of effective coupling constant for VP-effects under question. At the same time, the main input parameter is the source charge $Z$, which determines both $Q$ and the shape of the Coulomb potential. The size of the source $R(Z)$ and its shape are also the additional input parameters, but their role in VP-effects is quite different from $Q$ and in some important questions, in particular, in the renormalization procedure this difference must be clearly tracked. Furthermore, the difference between the charged sphere and ball, which seems more preferable as a model of super-heavy nucleus or heavy-ions cluster, in VP-effects of super-criticality under consideration is very small. At the same time,   the spherical shell model allows for almost completely analytical study of the problem, which has clear advantages in many positions. The ball model doesn't share such options, since explicit solution of DC problem in this case is absent and so one has to use from beginning the numerical methods or special approximations. We will briefly consider one of these approximations in Sect. VII.

As in other works on this topic ~\cite{Wichmann1956,Gyulassy1975, McLerran1975a,*McLerran1975b,*McLerran1975c,Greiner1985a,Plunien1986,Greiner2012,Ruffini2010, Rafelski2016},  radiative corrections from virtual photons are neglected. Henceforth, if it is not stipulated separately, relativistic units  $\hbar=m_e=c=1$ and the standard representation of  Dirac matrices are used. Concrete calculations, illustrating the general picture, are performed for $\a=1/137.036$ by means of Computer Algebra Systems (such as Maple 21) to facilitate  the analytic calculations  and GNU Octave code for boosting the numerical work.

\section{Perturbative approach to $\E_{VP}$}

It this section it would be pertinent to show explicitly the dependence on $m$. To the leading order, the perturbative VP-energy $\E^{(1)}_{VP}$ is obtained from the general first-order relation
\beq
\label{2.1}
\E^{(1)}_{VP}=\frac{1}{2} \int d \v r\, \vr^{(1)}_{VP}(\vec{r})\,A_{0}^{ext}(\vec{r}) \ ,
\eeq
where $\vr_{VP}^{(1)}(\v r)$ is the first-order perturbative VP-density, which is obtained from the one-loop (Uehling) potential $A^{(1)}_{VP,0}(\vec{r})$ in the next way ~\cite{ Greiner2012,Itzykson1980}
\beq
\label{2.2}
\vr^{(1)}_{VP}(\vec{r})=-\frac{1}{4 \pi} \D A^{(1)}_{VP,0}(\vec{r}) \ ,
\eeq
where
\beq\begin{gathered}
\label{2.3}
A^{(1)}_{VP,0}(\vec{r})=\frac{1}{(2 \pi)^3} \int d \v q\,\, \mathrm{e}^{i \vec{q} \vec{r}}\, \Pi_{R}(-{\v q}^2)\,\widetilde{A}_{0}(\vec{q}) \ , \\
\widetilde{A}_{0}(\vec{q})=\int d \v r'\,\mathrm{e}^{-i \vec{q} \vec{r\,}' }\,A^{ext}_{0}(\vec{r}\,' ) \ .
\end{gathered}\eeq
The polarization function  $\Pi_R(q^2)$, which enters eq. (\ref{2.3}), is defined via general relation $\Pi_R^{\m\n}(q)=\(q^\m q^\n - g^{\m\n}q^2\)\Pi_R(q^2)$ and so is dimensionless. In the adiabatic case under consideration  $q^0=0$ and $\Pi_R(-{\v q}^2)$ takes the  form
\begin{multline}
\label{2.4}
\Pi_R(-{\v q}^2) = \\ = {2 \a \over \pi}\, \int \limits_0^1 \! d\b\,\b(1-\b)\,\ln\[1+\b(1-\b)\,{{\v q}^2 \over m^2-i\e}\] = \\ = {\a \over \pi}\, S\(|\v q|/m \) \ ,
\end{multline}
where
\begin{multline}\label{2.4a}
S(x)= -5/9 + 4/3 x^2 + (x^2- 2)\, \sqrt{x^2+4} \ \times \\ \times \ \ln \[ \(\sqrt{x^2+4}+x\) \Big/ \(\sqrt{x^2+4}-x\) \]/3x^3  \ .
\end{multline}

Proceeding further, from (\ref{2.1}-\ref{2.3}) one finds
\begin{multline}
\label{2.11}
\E^{(1)}_{VP}=\frac{1}{64 \pi^4}\, \int \! d \v q \ {\v q}^2\, \Pi_R(-{\v q}^2) \ \times \\ \times \  \Big| \int \! d \v r\, \mathrm{e}^{i \vec{q} \vec{r}}\,A_{0}^{ext}(\vec{r}) \Big|^2  \ .
\end{multline}
Note that since the function  $S(x)$ is strictly positive, the perturbative VP-energy is positive too.

In the spherically-symmetric case  with $A_{0}^{ext}(\vec{r})=A_0(r)$  the perturbative VP-term belongs to the $s$-channel and equals to
\begin{multline}
\label{2.12}
\E^{(1)}_{VP}=\frac{1}{\pi}\, \int\limits_0^{\inf} \! d q \  q^4\, \Pi_R(-q^2) \ \times \\ \times \  \( \int\limits_0^{\inf} \! r^2\, d r\, j_0(q r)\, A_{0}(r) \)^2  \ ,
\end{multline}
 whence for the sphere there follows
\beq
\label{2.15}
\E^{(1)}_{VP,\,sphere}={ Q^2  \over 2 \pi R}\, \int\limits_0^{\inf} \! {d q \over q} \ S(q/m) \ J_{1/2}^2(q R) \ ,
\eeq
while for the ball one obtains
\beq
\label{2.17}
\E^{(1)}_{VP,\,ball}={ 9\,  Q^2  \over 2 \pi R^3}\, \int\limits_0^{\inf} \! {d q \over q^3} \ S(q/m) \ J_{3/2}^2(q R) \ .
\eeq
By means of the condition
\beq
\label{2.18}
m R(Z) \ll 1 \ ,
\eeq
which is satisfied by the Coulomb source with relation (\ref{1.8}) between its charge and radius up to $Z \sim 1000$, the integrals (\ref{2.15},\ref{2.17}) can be  calculated analytically (see Ref.~\cite{Plunien1986} for details). In particular,
\beq\label{2.20}
\E^{(1)}_{VP,\,sphere}={ Q^2  \over 3 \pi R}\,  \[ \ln\( {1 \over 2 m R}\) - \g_E + {1 \over 6}\]
\eeq
for the sphere and
\beq
\label{2.21}
\E^{(1)}_{VP,\,ball}={2\, Q^2 \over 5 \pi R}\,  \[ \ln\( {1 \over 2 m R}\) - \g_E +  {1 \over 5} \]
\eeq
for the ball.

It would be worth to note that the expressions  (\ref{2.20},\ref{2.21}) take place only under condition (\ref{2.18}), when $ \ln\( 1 / 2 m R \) \gg 1$. If the last condition is satisfied, then the relation for $\E^{(1)}_{VP}$ between ball and sphere
\beq
\label{2.22}
\E^{(1)}_{VP,\,ball}/\E^{(1)}_{VP,\,sphere} \simeq 6/5 \
 \eeq
is almost  the same as for their classical electrostatic self-energies $ 3 Z^2 \a/5 R$ and $ Z^2 \a/2 R$.

\section{VP-energy in the non-perturbative approach}

The starting expression for $ \E_{VP} $ is
 \beq
\label{3.29a}
\E_{VP}=\frac{1}{2}\(\sum\limits_{\e_{n}<\e_{F}} \e_n  -   \sum\limits_{\e_{n}\geqslant \e_{F}} \e_n \) \ ,
\eeq
where $\e_F=-1$ is the Fermi level, which in such problems with strong Coulomb fields is chosen at the lower threshold, while $\e_{n}$  the eigenvalues of  corresponding DC.

The expression  (\ref{3.29a}) is obtained from the Dirac hamiltonian, written in the form which is  consistent with Schwinger prescription for the current  (for details see, e.g., Ref.~\cite{Plunien1986}) and is defined up to a constant, depending on the choice of the energy reference point. In (\ref{3.29a}) VP-energy is negative and divergent even in absence of external fields $ A_{ext} = 0 $. But since the VP-charge density is defined so that  it vanishes for $A_{ext} = 0 $, the natural choice of the reference point for $ \E_{VP} $ should be the same. Furthermore, in presence of the external Coulomb potential of the type (\ref{1.5a},\ref{1.6}) there appears in the sum (\ref{3.29a}) also an (infinite) set of discrete levels. To pick out  exclusively the interaction effects it is therefore necessary to subtract from each discrete level the mass of the free electron at rest.

Thus, in the physically motivated form and in agreement with $\vr_{VP}$, the initial expression for the VP-energy should be written as
\begin{multline}
\label{3.29}
\E_{VP}=\1/2 \(\sum\limits_{\e_n<\e_F} \e_n-\sum\limits_{\e_n \geqslant \e_F} \e_n + \sum\limits_{-1\leqslant \e_n<1} \! 1
\)_A \ - \\ - \ \1/2 \(\sum\limits_{\e_n \leqslant -1} \e_n-\sum\limits_{\e_n
	\geqslant 1} \e_n \)_0 \ ,
\end{multline}
where the label A denotes the  non-vanishing external field $A_{ext}$, while the label 0 corresponds to   $A_{ext}=0$.  Defined in such a way,  VP-energy vanishes by turning off the external field, while by turning on it contains only the interaction effects, and so  the expansion of $ \E_{VP} $ in (even) powers of the external field starts from $ O\(Q^2\) $.

For what follows it would be pertinent to introduce a number of additional definitions and notations. The reason is that the purely Coulomb problem with spherical symmetry is just a start-up for more complicated problems, where only the axial symmetry is preserved. These are, in particular, two-center Coulomb one, which imitates the slow collision of two heavy ions, and one-center Coulomb in presence of an axial magnetic field, when the total angular moment $j$ is not conserved, there remains only its projection $m_j$. Thence the angular quantum number $k=\pm (j+1/2)$, which is very suitable for enumerating the Coulomb states of the Dirac fermion both in momentum and parity, lies aside. In this situation it is useful to represent the Dirac bispinor with fixed $m_j$ in the form
\beq\label{3.5}
\p_{m_j}(\vec{r})= \begin{pmatrix}
 \vf_{m_j}(\v r)\\
-i \c_{m_j}(\v r)
\end{pmatrix} \ ,
\eeq
 where the spinors $\p\, , \c$  are defined as the partial series over integer orbital momentum $l$
\beq\begin{gathered}
\label{3.6}
\vf_{m_j}(\v r)= \sum\limits_{l=0}^{\inf} \  \( u_l(r)\, \W_{l m_j}^{(+)} (\v n) +  v_l(r)\, \W_{l+1, m_j}^{(-)} (\v n)\) \ , \\
\c_{m_j}(\v r)= \sum\limits_{l=0}^{\inf} \ \( p_l(r)\, \W_{l m_j}^{(+)} (\v n) + q_l(r)\, \W_{l+1, m_j}^{(-)} (\v n)\) \ ,
\end{gathered}\eeq
with $\v n=\v r /r$ and   $\W_{l m_j}^{(\pm)}(\v n)$ being the spherical spinors with the total momentum $j=l\pm 1/2$ and fixed $j_z=m_j$. Each term in  parentheses in series (\ref{3.6})
correspond to $j=l+1/2$, while from the structure of DC problem there follows that the radial functions $u_l(r), v_l(r), p_l(r), q_l(r)$  can be always chosen real.

The phase of spherical spinors is rigorously fixed by their explicit form
\beq\begin{gathered}
\label{3.7}
\W_{l m_j}^{(+)} (\v n)=  \begin{pmatrix} \sqrt{{l+m_j+1/2 \over 2l+1}}\, Y_{l, m_j-1/2} \\ \sqrt{{l-m_j+1/2 \over 2l+1}}\, Y_{l, m_j+1/2} \ \end{pmatrix} \ , \\
\W_{l m_j}^{(-)} (\v n)=  \begin{pmatrix} \sqrt{{l-m_j+1/2 \over 2l+1}}\, Y_{l, m_j-1/2} \\ -\sqrt{{l+m_j+1/2 \over 2l+1}}\, Y_{l, m_j+1/2} \ \end{pmatrix}  \ ,
\end{gathered}\eeq
thence $(\v \s \v n )\W_{l m_j}^{(+)} (\v n)=\W_{l+1, m_j}^{(-)} (\v n)$, whereas the phase of spherical functions is chosen in a standard way, providing  $l_{\pm}\,Y_{lm}=\sqrt{(l\mp m)(l\pm m +1)}\,Y_{l, m \pm 1}$ and $Y_{l, -|m|}=(-1)^{|m|}\,Y_{l, |m|}^{\ast}$.

For the spherically-symmetric Coulomb potential $V(r)$  of the type (\ref{1.5a},\ref{1.6}) the spectral DC problem for the energy level $\e$  divides into two radial subsystems, containing either  $(u_l\,, q_l)$- or $(p_l\,, v_l)$-pairs, of the following form
\beq
\label{3.6a}
\left\lbrace\bal
&\(\pd_r -{l \over r}\)\, u_l=(\e-V(r)+1)\, q_l \ ,\\
&\(\pd_r +{l+2 \over r}\)\, q_l=-(\e-V(r)-1)\, u_l \ ,
\eal\right.
\eeq
\beq
\label{3.6b}
\left\lbrace\bal
&\(\pd_r -{l \over r}\)\, p_l=-(\e-V(r)-1)\, v_l \ ,\\
&\(\pd_r +{l+2 \over r}\)\, v_l=(\e-V(r)+1)\, p_l \ .
\eal\right.
\eeq
Eqs.(\ref{3.6a},\ref{3.6b}) are subject of crossing symmetry: under simultaneous change of the sign of external potential and energy $Q \to -Q\,, \e \to - \e$ the pairs $(u_l\,, q_l)$ and $(p_l\,, v_l)$ interchange. This symmetry will be used further by calculation the VP-energy via phase integral method.

Now let us extract  from (\ref{3.29}) separately the contributions from the discrete and continuous spectra for each value of orbital momentum $l$, and afterwards use for the difference of integrals over the continuous spectrum $ (\int d {\v k}\, \sqrt{k^2+1})_A-(\int d {\v k}\,\sqrt{k^2+1})_0 $ the well-known technique, which represents this difference in the form of an integral of the elastic scattering phase $ \d_l(k)$~\cite{Sveshnikov2017,Rajaraman1982,Sveshnikov1991,Jaffe2004}.  Omitting a number of almost obvious intermediate steps, which have been considered in detail in Ref.~\cite{Sveshnikov2017}, let us  write the final answer for $ \E_{VP}(Z) $ as a partial series
\beq
\label{3.30}
\E_{VP}(Z)=\sum\limits_{l=0} \E_{VP,l}(Z) \ ,
\eeq
where
\begin{multline}
\label{3.31}
\E_{VP,l}(Z) = (l+1)\, \(\frac{1}{\pi} \int\limits_0^{\inf} \!   \  \frac{k \, dk }{\sqrt{k^2+1}} \ \d_{tot}(l,k) \ + \right. \\ \left. + \ \sum\limits_{\pm} \sum\limits_{-1 \leqslant \e_{n,l}^{\pm}<1} \(1-\e_{n,l}^{\pm}\)\) \ .
\end{multline}
In (\ref{3.31}) $ \d_{tot}(l,k) $ is the total phase shift for the given values of the wavenumber $k$ and orbital momentum $l$, including the  contributions of the scattering states of the problem (\ref{3.6a},\ref{3.6b}) from both  continua and  both parities. In the discrete spectrum contribution to $\E_{VP,l}(Z)$ the additional sum $\sum_{\pm}$ takes  also account of both parities. Note also that the  multiplier $l+1$ in (\ref{3.31}) appears as a product  of the degeneracy factor $2(l+1)=2j+1$ and $1/2$ in (\ref{3.29}).

Such approach to evaluation of $ \E_{VP} $ turns out to be quite effective, since for the external potentials of the type (\ref{1.5a},\ref{1.6}) each partial VP-energy turns out to be finite without any special regularization. First, $ \d_{tot}(l,k) $ behaves both in IR and UV-limits in the $k$-variable much better, than each of the scattering phase shifts separately.  Namely,  $ \d_{tot}(l,k) $ is finite for $ k \to 0 $ and behaves like  $ O(1/k^3) $ for $ k \to \inf $, hence, the phase integral in (\ref{3.31}) is always convergent. Moreover, $ \d_{tot}(l,k) $ is by construction an even function of the external field, more precisely, of the effective coupling constant $Q$. Thereby the complete dependence on $Z$ is more diverse, since the latter defines also the shape of the external source and potential in a quite different way via (\ref{1.8}). Second, in the bound states contribution to $ \E_{VP,l}(Z) $  the condensation point $ \e_{n,l}^{\pm} \to 1 $ turns out to be regular for each  $l$ and parity, because  $ 1-\e_{n,l}^{\pm} \sim O(1/n^2) $ for $ n \to \inf $. The latter circumstance permits to avoid intermediate cutoff of the Coulomb asymptotics of the external potential for $ r \to \inf $, what significantly simplifies all the subsequent calculations.

The principal problem of convergence of the partial series (\ref{3.30}) can be solved along the lines of Ref.~\cite{ Davydov2018b}, demonstrating the convergence of the similar  expansion for $ \E_{VP}(Z) $ in 2+1 D . Let us explore first the behavior of  partial VP-energy  (\ref{3.31})  for large $l \gg Q$. More precisely, the last condition implies
\beq\label{3.33a}
(l+1)^2 \gg Q^2 + 2\, Q R \ .
\eeq
For such $l$ the main component  $ \sim Q^2$ of the total scattering phase $\d_{tot}(l,k)$ per each parity (or equivalently, for pairs $(u_l\,, q_l)$ or $(p_l\,, v_l)$)   can be reliably estimated via quasiclassical (WKB) approximation:
\beq\label{3.33}
\d_\mathrm{WKB}(l,k) = \d_{+}(l,k)+\d_{-}(l,k)-2\,\d_{0}(l,k) \ ,
\eeq
where
\beq\begin{gathered}
\label{3.34}
\d_{\pm}(l,k)=\int \!   dr \ \sqrt{\(\e(k) \mp V(r)\)^2-1-{(l+1)^2 \over r^2}} \ , \\
\d_{0}(l,k)=\int \!   dr \ \sqrt{k^2-{(l+1)^2 \over r^2}} \ ,
\end{gathered}\eeq
and
\beq\label{3.35}
\e(k)=\sqrt{k^2+1} \ ,
\eeq
while the integration is performed over regions, where the expressions under  square root are non-negative. In turn, the total WKB-phase shift is twice the contribution of each parity
\beq\label{3.35a}
\d^\mathrm{WKB}_{tot}(l,k)=2\,\d_\mathrm{WKB}(l,k) \ .
\eeq
For a rigorous justification of such formulae for WKB-phase shifts in DE with spherically-symmetric Coulomb-like potentials see Refs.~\cite{Lazur2005, *Zon2012} (and refs. therein).

\begin{figure}
\center
\includegraphics[scale=1.0]{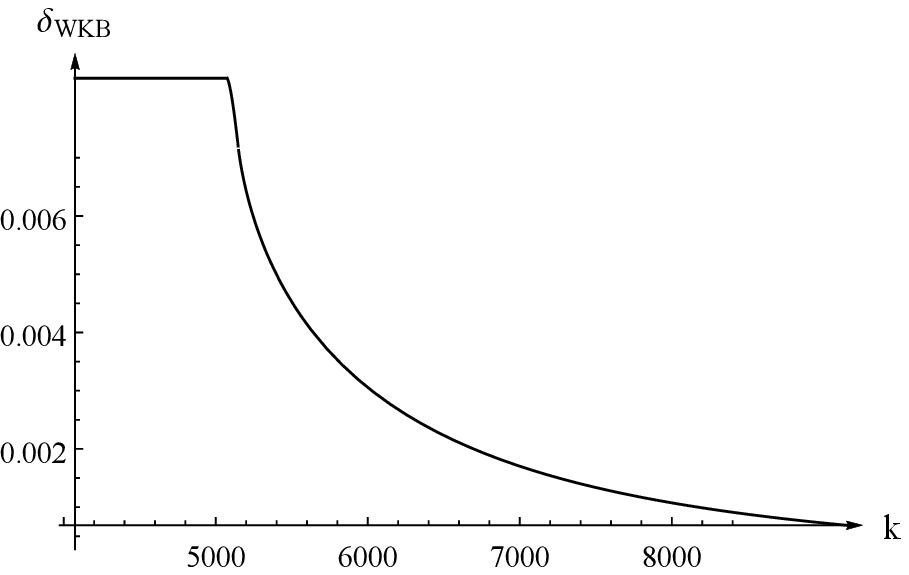} \\
\caption{\small $\d_\mathrm{WKB}(l,k)$ for $Z=100$ and $l=100$.}
\label{WKB}
\end{figure}

For the case of a charged sphere (\ref{1.5a})  all the  calculations can be performed analytically and lead to the following result
\begin{widetext}
\begin{multline}
\label{3.36}
\d_\mathrm{WKB}(l,k) =  \tt(0 \leqslant k  \leqslant k_-)\, \pi\(l+1-\vk_l\)  +  \tt(k_- \leqslant k  \leqslant k_+)\, \[\pi\,(l+1) \ -  \ {\pi \vk_l \over 2} \ - \ (l+1)\, \arctg\({A_+(k) \over l+1}\) \ + \right. \\ \left. + \ \vk_l\,\arctg\( {\e(k) Q R - \vk_l^2 \over \vk_l\,A_+(k)}\) \ +  \ {\e(k) Q \over k}\,\ln \({\sqrt{Q^2+k^2\,(l+1)^2} \over k\,A_+(k)+ k^2 R +\e(k)Q }\)\] \ + \\ + \ \tt(k_+ \leqslant k  \leqslant \inf)\,\[(l+1)\(\pi - \arctg\({A_+(k) \over l+1}\)-\arctg\({A_-(k) \over l+1}\)\) \ + \right. \\ \left. + \  \vk_l\, \(\arctg\( {\e(k) Q R - \vk_l^2\over \vk_l\,A_+(k)}\) \ - \ \arctg\( {\e(k) Q R + \vk_l^2\over \vk_l\,A_-(k)}\)\)  + \ {\e(k) Q \over k}\,\ln \({k\,A_-(k)+ k^2 R -\e(k)Q \over k\,A_+(k)+ k^2 R +\e(k)Q }\)\] \ ,
\end{multline}
\end{widetext}
where
\beq
\label{3.37}
\vk_l=\sqrt{(l+1)^2-Q^2} \ ,
\eeq
\beq
\label{3.38}
k_{\pm}={1 \over R}\,\sqrt{(l+1)^2+Q^2 \pm 2 Q\,\sqrt{(l+1)^2+R^2}} \ ,
\eeq
and
\beq
\label{3.39}
A_{\pm}(k)=\sqrt{(k R)^2 \pm 2 \e(k) Q R + Q^2 -(l+1)^2} \ .
\eeq
The typical behavior of $\d_\mathrm{WKB}(l,k)$ is shown for $Z=100$ and $l=100$ in Fig.\ref{WKB}.

 The main properties of the total WKB-phase (\ref{3.35a}) are the following. For $k < k_- < (l+1)/R$ it is a constant
\beq\begin{gathered}
\label{3.39a}
\d^\mathrm{WKB}_{tot}(l,k)=2 \pi\(l+1-\vk_l\) \ , \\  k < k_- < (l+1)/R \ ,
\end{gathered}\eeq
while for large $k \gg k_+ > (l+1)/R$ it vanishes $\sim 1/k^3$, namely
\beq\begin{gathered}
\label{3.39b}
\d^\mathrm{WKB}_{tot}(l,k) \to {4\, Q^2\, ((l+1)^2-3 R^2) \over 3\, (kR)^3} \ , \\ k \gg k_+ > (l+1)/R \ .
\end{gathered}\eeq
In the region between $k_-$ and $k \gg k_+$  it behaves as a smooth interpolation function. Moreover, the smaller $R$, the greater the value of $k$ is needed (the correct condition reads $k R \gg l$) to alter the behavior of $\d^\mathrm{WKB}_{tot}(l,k)$ from the constant value (\ref{3.39a}) into the decreasing one (\ref{3.39b}).

 These results reproduce quite well the behavior of the exact total phase   $\d_{tot}(l,k)$ with the following remarks.
 First, the decreasing  either $\sim 1/k^3$ or  $\sim 1/|y|^3$ asymptotics is a common feature of the integrands in VP-integral expressions, regardless  which VP-observable is under consideration. This applies equally to calculating  the VP-energy by means of (\ref{3.31}) within the phase integral method or to elaborated recently in Refs.~\cite{Voronina2019c, *Voronina2019d} even more sophisticated approach to evaluation the VP-energy, based on the $\ln [\mathrm{Wronskian}]$ techniques, which turns out to be effective beyond the partial expansions. Second, the quasiclassical approximation does not reproduce  oscillations of the exact phase for large  $ k $, which are caused by diffraction on a sphere of the radius  $R$. In more details this topic is discussed within 2+1 D case in Ref.~\cite{ Davydov2018b}.  At the same time, the behavior of both $ \d^\mathrm{WKB}_{tot}(l,k) $ and $ \d_{tot}(l,k) $ for $kR \ll l$ can be easily understood by comparing them to the total phase for a point-like Coulomb source with the potential $V(r) = -Q/r$ (for $l+1 > Q$). The analytic solution of  corresponding DE is well-known (see, e.g., Ref.~\cite{landau2012qed}) and gives the following exact answer for each of the  partial phase shifts. Namely, for $(u_l\,, q_l)$-pair
\begin{multline}\label{3.40}
	\d^{\pm}_{uq}(l,k)=\frac{\pi} {2}(l+1) \pm \frac{\e(k) Q}{ k}\ln{2 k r} - \frac{\pi \vk_l} {2} \mp \\  \mp   \mathrm{Arg}\[ \G(1+\vk_l + i \e(k) Q/k)\]  + \1/2 \mathrm{Arg} \[ \frac{ l+1+i Q/k} { \vk_l \mp i \e(k) Q/k} \] \ , \end{multline}
where  the signs $\pm$ correspond to the phase shifts for the upper and lower continua, while  for $(p_l\,, v_l)$-pair the phase shifts  are obtained from (\ref{3.40}) via simple replacement  $l+1 +i Q/k\to  l+1 -i Q/k$ in the last term. From these results one obtains that for all $0 \leq k \leq \inf$ the exact total phase  for a point-like source equals to a constant
\beq
\label{3.41}
 \d_{tot}(l,k)|_{R \to 0}=2 \pi\(l+1-\vk_l\) \ ,
\eeq
which exactly coincides with the answer  coming from WKB-approximation (\ref{3.39a}) for $k < k_- < (l+1)/R$. This result should be quite clear, since for large $l$ the condition $k R \ll l$ implies  scattering with large sighting distances $d$ to target $d \gg  R$. In the last case, the difference between the  sphere of the size $R$ and a point-like source is negligibly small. It should be remarked, however, that for such behavior of the exact phase  the WKB-condition $l \gg Q$ is crucial, otherwise $ \d_{tot}(l,k) $ for $k \to 0$ remains finite, but its limiting value in this case can be sufficiently different from (\ref{3.41}), especially in the case $l+1< Q$, when $\vk_l$ becomes imaginary (see  below).

In the next step let us consider the  behavior of the partial phase integrals in (\ref{3.31})
\beq
\label{3.42}
I(l)= {1 \over \pi}\,\int\limits_0^{\inf} \!   \  \frac{k \, dk }{\sqrt{k^2+1}} \ \d_{tot}(l,k)
 \eeq
for $l \gg Q$, or more precisely, subject to  condition (\ref{3.33a}). The details of calculation are given in Appendix. The result is
\beq\label{3.53}
 I(l) \to {2\over \pi}\,\int\limits_0^{\infty} \! dr \ V^2(r)-{Q^2 \over l+1} + O\(Q^4 \over (l+1)^3\) \ , \quad l \to \inf \ .
\eeq

At the same time, the discrete spectrum with the same conditions on  $l$ (that means $l\to \inf$ or at least $l$ subject to condition (\ref{3.33a})) corresponds with a high precision to well-known solution of the Coulomb-Schroedinger problem for a point-like source with the same $Z$.  In this limit the main contribution to the total sum of discrete levels comes from the vicinity of condensation point $\e_{n,l} \to 1$, where both parities reveal the same properties and so can be freely treated at the same footing. The leading $ \sim Q^2$ terms in bound energies of discrete levels per each parity are given by the Bohr formula
\beq\label{3.54}
1-\e_{n_r,l}^{\pm} ={Q^2 \over 2\,(n_r+l+1)^2} \ , \quad l \to \inf \ .
\eeq
Upon summing the bound energies (\ref{3.54}) over $n_r$ one obtains
\begin{multline}\label{3.55}
\sum\limits_{n_r=0}^{\inf} (1-\e_{n_r,l}^{\pm}) ={Q^2 \over 2}\,\hbox{Polygamma}\[1,l+1\] \ = \\ = \ {Q^2 \over 2}\,\({1 \over (l+1)} + {1 \over 2\,(l+1)^2} +{1 \over 6\,(l+1)^3} \ + \right. \\ \left. + \ O\({1 \over (l+1)^4}\)\) \ .
\end{multline}
The next-to-leading  $ \sim Q^4$ terms in bound energies in this limit are given by the first relativistic (fine-structure) corrections to  (\ref{3.54})
\beq\label{3.56}
Q^4\, \({1 \over 2 (l+1) (n_r+l+1)^3} - {3 \over 8 (n_r+l+1)^4}\) \
\eeq
and upon summing over $n_r$ yield the terms  $ \sim Q^4/(l+1)^3$ in the  sum over discrete levels in $\E_{VP,l}(Z)$. At the same time, the correction to Bohr levels (\ref{3.54}),  caused by the non-vanishing  size of the Coulomb source, equals to (for the external potential (\ref{1.5a}))
\begin{multline}\label{3.56a}
- {Q^2 \over (n_r+l+1)^2}\, \( {2\, Q R \over n_r+l+1 }\)^{2l+2}\, \ \times \\ \times \ {(n_r + 2l+1)! \over n_r!} { (2l+2)\,(2l+3) \over (2l+3)!^2} \ ,
\end{multline}
and for growing $l$ turns out to be negligibly (in fact exponentially) small, since in this limit the dominating factor in (\ref{3.56a}) is $1/(2l+3)!^2$. It should be noted however, that for small or even moderate $l$ and especially for the case $l+1 < Q$ the nonzero size of the Coulomb source plays an essential role in all VP-effects and so by no means should be treated at the same footing with the other most important input parameters.

As a result, the partial VP-energy (\ref{3.31}) in the large $l$ limit can be represented as
\begin{multline}
\label{3.57}
\E_{VP,l}(Z)  = (l+1)\,\[\frac{2}{\pi} \int\limits_0^{\inf} \! dr \ V^2(r) -{Q^2 \over l+1} +  \right. \\ \left. + Q^2\,\hbox{Polygamma}\[1,l+1\]  + O\({Q^4 \over (l+1)^3}\) \] \ .
\end{multline}
Thus,  in complete agreement with similar results in 1+1 and 2+1 D cases ~\cite{Davydov2017,*Sveshnikov2017,
*Voronina2017,Davydov2018a,Davydov2018b,Sveshnikov2019a,Sveshnikov2019b},  the partial series (\ref{3.30})  for $\E_{VP}$ diverges quadratically in the leading $O(Q^2)$-order and so requires  regularization and subsequent renormalization, although each partial $\E_{VP,l}(Z)$ in itself is finite without any additional manipulations. The degree of divergence of the partial series (\ref{3.30}) is formally the same as in 3+1 QED  for the   fermionic loop with two external lines. The latter circumstance shows that by calculation of $ \E_{VP} $ via  principally different non-perturbative approach, that does not reveal any connection with perturbation theory (PT) and Feynman graphs, we nevertheless meet actually the same  divergence of the theory, as in PT \footnote{In  3+1 D the loop with 4 legs is also (logarithmically) divergent. But it is apart of interest in our approach since the calculation of VP-density and VP-energy implies averaging over the vortex indices, which removes the divergency of the 4-legs diagram (see, e.g., Ref.~\cite{landau2012qed}). Second, our approach deals only with gauge-invariant quantities, and so does not require for any additional regularization.}. Actually,  it should be indeed so, since both approaches deal with the same physical phenomenon (VP-effects caused by the strong Coulomb field) with the main difference in  the methods of calculation. Therefore in the present approach the cancelation of divergent terms should follow the same rules as in PT, based on  regularization of the fermionic loop with two external lines, that preserves the physical content of the whole renormalization procedure and simultaneously provides the mutual agreement between perturbative and non-perturbative approaches to the calculation of $ \E_{VP} $.  This conclusion is in  complete agreement with results obtained earlier in Refs.~\cite{Gyulassy1975, Mohr1998}.

One more but quite general reason is that for $Q \to 0$, but with fixed $R(Z)$, both the total renormalized VP-density and VP-energy should coincide with results obtained within PT by means of (\ref{2.1}-\ref{2.3}). Due to spherical symmetry of the external field they both  belong to the partial s-channel with $l=0$.  However,  in the general case the non-renormalized  (but already finite) partial VP-density $ \vr_{VP,0}(r)$ and VP-energy $ \E_{VP,0}$ do not reproduce the corresponding perturbative answers for $Q \to 0$.  For $ \E_{VP}^{(1)} $ and $ \E_{VP.0}$ the difference is quite transparent, since the perturbative VP-energy is built  only  from the distorted continuum and so has nothing to do with the discrete levels. To  the contrary,  $ \E_{VP,0}$  contains by construction a non-vanishing $O(Q^2)$-contribution from the latters, since for Coulomb-like potentials the discrete spectrum exists for any infinitesimally small $Q$.

 Thus, in the complete analogy with the renormalization of VP-density~\cite{Davydov2017,*Sveshnikov2017,
*Voronina2017,Krasnov2022,Gyulassy1975, Mohr1998,Davydov2018a,Davydov2018b,Sveshnikov2019a,*Sveshnikov2019b} we should pass to the renormalized VP-energy by means of  relations
\beq
\label{3.58}
\E^{ren}_{VP}(Z) =\sum\limits_{l=0} \E^{ren}_{VP,l}(Z) \ ,
\eeq
where
\beq
\label{3.59}
\E^{ren}_{VP,l}(Z)=\E_{VP,l}(Z)+ \z_l Z^2 \ ,
\eeq
with the renormalization coefficients $\z_l$ defined in the next way
\beq
\label{3.60}
\z_l = \lim\limits_{Z_0 \to 0}  \[{\E_{VP}^{(1)}(Z_0)\,\d_{l,0}-\E_{VP,l}(Z_0) \over Z_0^2}\]_{R=R(Z)}
\ . \eeq
The  essence of relations  (\ref{3.58}-\ref{3.60}) is to remove (for fixed $Z$ and  $R(Z)$!) the divergent $O(Q^2)$-components from the non-renormalized partial terms  $ \E_{VP,l}(Z) $ in the series (\ref{3.30}) and  replace them further by renormalized via fermionic loop  perturbative contribution to VP-energy $\E^{(1)}_{VP}\,\d_{l,0}$. Such procedure provides simultaneously the convergence of the regulated this way partial series (\ref{3.58}) and the correct limit of $\E^{ren}_{VP}(Z)$  for $Q \to 0$ with fixed $R(Z)$.

So  renormalization via fermionic loop turns out to be a universal method, which removes the divergencies of the theory simultaneously  in purely perturbative and essentially non-perturbative approaches to VP. However, the concrete implementation of this general method depends on the VP-quantity under consideration.   The considered approach to evaluation of $ \E_{VP}$ treats separately the contributions to $ \E_{VP,l}(Z)$ from discrete spectrum and continua. The subtraction in this case proceeds directly on the level of $ \E_{VP,l}(Z)$ by means of (\ref{3.59},\ref{3.60}), where the renormalization coefficients  $ \z_l $ are determined through a special limit, in which the effective coupling constant $Q_0=Z_0\a$ tends to zero, but the shape of the external field (in the present case it is the radius $R(Z)$) is preserved. So $ \z_l $  contain a non-trivial dependence on  $R(Z)$, and hence, on the current charge  $Z$ of the Coulomb source. This dependence, however, has nothing to do with the renormalization procedure presented above, since in fact the latter deals with  the dependence on  $Q=Z\a$, but not on the shape of the external potential.

Moreover, the complete analogy between  renormalizations of VP-density and VP-energy implies the validity of Schwinger relation~\cite{Sveshnikov2017,Plunien1986, Greiner2012} for renormalized quantities
\beq\label{3.61}
\d \E^{ren}_{VP}=\int \! \mathrm{d} \v r \ \vr^{ren}_{VP}(\v r)\, \d A_0^{ext}(\v r) +\d \E_N \ ,
\eeq
since $\E_N$ is responsible only for jumps in the VP-energy caused by discrete levels crossing through the border of the lower continuum and so is an essentially non-perturbative quantity, which  doesn't need any  renormalization. The relation (\ref{3.61}) can be represented also in the partial form
\beq\label{3.62}
\d \E^{ren}_{VP,l}=\int\limits_0^{\inf} \! r^2\, dr \ \vr^{ren}_{VP,l}(r)\, \d A_0^{ext}(r) +\d \E_{N,l} \ ,
\eeq
from which there follows that the  convergence of partial series for  VP-density implies the convergence of  partial series for  VP-energy and vice versa.  $\E_{N,l}$ is always finite and, moreover, for any finite $Z$ vanishes for $l \geq l_{max}(Z)$, therefore doesn't influence the convergence of the partial series.

\section{Evaluation of the total elastic phase $\d_{tot}(l,k)$ for the  potential (\ref{1.5a})}

Now --- having dealt with the first principles of essentially nonperturbative evaluation of VP-energy by means of the phase integral method this way --- let us turn to the explicit evaluation of $\E^{ren}_{VP}(Z)$  for the external potential (\ref{1.5a}). It would be pertinent to present the details of this procedure in terms of separate pairs  $(u_l\,, q_l)$ and $(p_l\,, v_l)$, introduced via general expansion (\ref{3.6}). Although each pair contains a set of states with different parity, more detailed description of solutions of DC  per each parity is here of no use,  since   the calculation of $\E^{ren}_{VP}(Z)$ itself implies the summation over both parities.

The evaluation of total elastic phase $\d_{tot}(l,k)$ proceeds as follows. It suffices to consider only $(u_l\,, q_l)$-pair, since  the  contribution of $(p_l\,, v_l)$-pair to $\d_{tot}(l,k)$  can be achieved via crossing symmetry of the initial DC  (\ref{3.6a},\ref{3.6b}). First we consider the region $l+1 > Q$, that means for real $\vk_l$ defined in (\ref{3.37}). In this case in the upper continuum with
\beq\label{4.0}
\e(k)=+\sqrt{k^2+1}\geqslant 1
\eeq
the solutions of  DC  up to a common normalization factor can be represented in the next form.

For $r\leq R(Z)$
\beq
\label{4.1}
\left\lbrace\bal
& u_l(k,r)=\sqrt{\e(k)+V_0+1}\, J_{l+1/2} \(\x(k) r\) /\sqrt{r} \ ,\\
& q_l(k,r)=-\sqrt{\e(k)+V_0-1}\, J_{l+3/2} \(\x(k) r\) /\sqrt{r} \ ,
\eal\right.
\eeq
with $J_{\nu}(z)$ being the Bessel functions,
\beq\label{4.2}
\x(k)=\sqrt{(\e(k)+V_0)^2-1} \ , \quad V_0=Q/R \ .
\eeq
For $r\geq R(Z)$ the scattering states of DC  should be written in terms of the Kummer $\F(b,c,z)$ and Tricomi $\P(b,c,z)$ or modified Kummer $\tF(b,c,z)=z^{1-c}\P(b-c+1,2-c,z)$ functions~\cite{Bateman1953}. The reason is that due to the Kummer relation $\F(b,c,x)=e^x\F(c-b,c,-x)$ for real $\vk_l$  such solutions cannot be represented in terms of $\F(b,c,z)$ and $\F^{\ast}(b,c,z)$. For our purposes  the modified Kummer function $\tF(b,c,z)$ is more preferable than the Tricomi one due to the reasons of numerical calculations. The Tricomi function contains combination of two Kummer's ones and so is calculated twice longer.

The parameters $b,c$ of the Kummer's functions are defined as
\beq\begin{gathered}
\label{4.4}
b_l=\vk_l-i \e(k) Q/k \ , \quad c_l=1+2\vk_l \  .
\end{gathered}\eeq
Upon introducing the subsidiary phases $\x_1\, , \x_2$
\beq\begin{gathered}
\label{4.4a}
\mathrm{e}^{-2\, i\, \x_1}={b_l \over l+1 +i Q/k} \ , \quad \mathrm{e}^{2\, i\, \x_2}={b_l \over l+1 -i Q/k} \ ,
\end{gathered}\eeq
and corresponding functions
\beq\begin{gathered}
\label{4.5}
\F_1(l,k,r)=\mathrm{e}^{i\,(kr+\x_1)}\,\F(b_l\, , c_l\, , -2\, i\, k\, r) \ , \\ \F_2(l,k,r)=\mathrm{e}^{i\,(kr+\x_2-\pi \vk_l)}\,\tF(b_l\, , c_l\, ,-2\, i\, k\, r) \ ,
\end{gathered}\eeq
the real-valued scattering solutions  in the upper continuum are defined as follows
\begin{widetext}
\beq\begin{gathered}
\label{4.6}
\left\lbrace\bal
& u_l(k,r)=\sqrt{\e(k)+1}\,r^{\vk_l-1}\,\(\mathrm{Re}\[\F_1(l,k,r)\]+\l_{uq}^{+}(l,k)\,\mathrm{Im}\[\F_2(l,k,r)\]\) \ , \\
& q_l(k,r)=\sqrt{\e(k)-1}\,r^{\vk_l-1}\,\(-\mathrm{Im}\[\F_1(l,k,r)\]+\l_{uq}^{+}(l,k)\,\mathrm{Re}\[\F_2(l,k,r)\]\) \ ,
\eal\right.
\end{gathered}\eeq
with $\l_{uq}^{+}(l,k)$ being  the matching coefficient  between inner $r\leq R(Z)$ and outer $r\geq R(Z)$ solutions
 \beq
\label{4.7}
\l_{uq}^{+}(l,k)={\sqrt{\(\e-1\)\,\(\e+V_0+1\)}\, J_{l+1/2}\(\x R\)\mathrm{Im}\[\F_1\] - \sqrt{\(\e+1\)\,\(\e+V_0-1\)}\, J_{l+3/2}\(\x R\)\mathrm{Re}\[\F_2\] \over
\sqrt{\(\e-1\)\,\(\e+V_0+1\)}\, J_{l+1/2}\(\x R\)\mathrm{Re}\[\F_2\] + \sqrt{\(\e+1\)\,\(\e+V_0-1\)}\, J_{l+3/2}\(\x R\)\mathrm{Im}\[\F_2\]} \ .
\eeq
\end{widetext}
In the r.h.s. of (\ref{4.7}) $\e \equiv \e(k)=+\sqrt{k^2+1}\geqslant 1\, , \ \x=\x(k)\, , \ R=R(Z)$, while the arguments of functions $\F_1$ and $\F_2$  are $\(l\, , k\, , R(Z)\)$.

The phase shift $\d^+_{uq}(l,k)$  in the upper continuum is defined in the standard way via asymptotics for $r \to \inf$
and equals to
\beq\begin{gathered}
\label{4.8}
\d^+_{uq}(l,k)={\pi l \over 2} + {\e(k)\,Q \over k}\, \ln \(2\,k r\) + \td^+_{uq}(l,k) \  ,
\end{gathered}\eeq
where
\beq\begin{gathered}
\label{4.9}
\td^+_{uq}(l,k)=-{\mathrm{Re}\[A(l,k)\] - \l_{uq}^{+}(l,k)\,\mathrm{Im}\[B(l,k)\] \over  \mathrm{Im}\[A(l,k)\] + \l_{uq}^{+}(l,k)\,\mathrm{Re}\[B(l,k)\] }\  ,
\end{gathered}\eeq
while the functions $ A(l,k)\, , B(l,k)$ are defined as follows
\beq\begin{gathered}
\label{4.10}
A(l,k)=\mathrm{e}^{-i\pi \vk_l/2}\, { \G\(1+2\,\vk_l\) \over \G \(\vk_l + i \e(k)\,Q/k\)} \, \sqrt{{l+1 +i Q/k \over \vk_l + i \e(k) Q/k}} \  ,\\
B(l,k)=\mathrm{e}^{i\pi \vk_l/ 2}\, { \G\(1-2\,\vk_l\) \over \G \(-\vk_l + i \e(k)\,Q/k\)} \, \sqrt{{l+1 +i Q/k \over \vk_l - i \e(k) Q/k}} \ .
\end{gathered}\eeq

In the lower continuum, where
\beq\label{4.18}
\e(k)=-\sqrt{k^2+1} < -1 \ ,
\eeq
the solutions are more diverse, since now the whole half-axis $0\leq k \leq \inf$  should be divided in 3 intervals $0\leq k \leq k_1\, , \ k_1 \leq k \leq k_2\, , \ k_2 \leq k \leq \inf$, where
\beq\label{4.19}
k_1=\sqrt{(V_0-1)^2-1} \ , \quad  k_2=\sqrt{(V_0+1)^2-1} \ .
\eeq
Note that in the case under consideration $V_0$ turns out to be  about several dozens or even hundreds, so $k_1$ is always well-defined.

According to this division the inner solutions of DC problem in the lower continuum are defined as follows. For $0\leq k \leq k_1$
\beq
\label{4.20}
\left\lbrace\bal
& u_l(k,r)=\sqrt{\e(k)+V_0+1}\, J_{l+1/2}\(\x(k) r\)/\sqrt{r} \ ,\\
& q_l(k,r)=-\sqrt{\e(k)+V_0-1}\, J_{l+3/2}\(\x(k) r\)/\sqrt{r} \ ,
\eal\right.
\eeq
where $\x(k)$ is defined as before in (\ref{4.2}), while for the second interval  $k_1 \leq k \leq k_2$
\beq
\label{4.21}
\left\lbrace\bal
& u_l(k,r)=\sqrt{\e(k)+V_0+1}\, I_{l+1/2}\(\tx(k) r\)/\sqrt{r} \ ,\\
& q_l(k,r)=\sqrt{|\e(k)|-V_0+1}\, I_{l+3/2}\(\tx(k) r\)/\sqrt{r} \ ,
\eal\right.
\eeq
with $I_{\nu}(z)$ being the Infeld functions and
\beq\label{4.22}
\tx(k)=\sqrt{1-(\e(k)+V_0)^2} \ .
\eeq
In the third interval $k_2 \leq k \leq \inf$
\beq
\label{4.23}
\left\lbrace\bal
& u_l(k,r)=\sqrt{|\e(k)|-V_0-1}\, J_{l+1/2}\(\x(k) r\)/\sqrt{r} \ ,\\
& q_l(k,r)=\sqrt{|\e(k)|-V_0+1}\, J_{l+3/2}\(\x(k) r\)/\sqrt{r} \ ,
\eal\right.
\eeq
The real-valued scattering solutions  in the lower  continuum are
\begin{widetext}
\beq\begin{gathered}
\label{4.24}
\left\lbrace\bal
& u_l(k,r)=\sqrt{|\e(k)|-1}\,r^{\vk_l-1}\,
\(\mathrm{Re}\[\F_1(l,k,r)\]+\l_{uq}^{-}(l,k)\,\mathrm{Im}\[\F_2(l,k,r)\]\) \ , \\
& q_l(k,r)=\sqrt{|\e(k)|+1}\,r^{\vk_l-1}\,
\(\mathrm{Im}\[\F_1(l,k,r)\]-\l_{uq}^{-}(l,k)\,\mathrm{Re}\[\F_2(l,k,r)\]\) \ ,
\eal\right.
\end{gathered}\eeq
while the corresponding matching coefficients  contain now  triplets $\l_{i,uq}^{-}(l,k)$, $i=1\,,2\,,3$, which belong to 3 intervals $0\leq k \leq k_1\, , \ k_1 \leq k \leq k_2\, , \ k_2 \leq k \leq \inf$ and take the following form
 \beq
\label{4.25}
\left\lbrace\bal
&\l_{1,uq}^{-}(l,k)={\sqrt{\Big|\(1-\e \)\,\(\e+V_0+1\)\Big|}\, J_{l+1/2}\(\x R\)\mathrm{Im}\[\F_1\] + \sqrt{\Big|\(1+\e\)\,\(\e+V_0-1\)\Big|}\, J_{l+3/2}\(\x R\)\mathrm{Re}\[\F_1\] \over
 \sqrt{\Big|\(1-\e \)\,\(\e+V_0+1\)\Big|}\,J_{l+1/2}\(\x R\)\mathrm{Re}\[\F_2\] - \sqrt{\Big|\(1+\e\)\,\(\e+V_0-1\)\Big|}\,J_{l+3/2}\(\x R\)\mathrm{Im}\[\F_2\]} \ , \\
&\l_{2,uq}^{-}(l,k)={\sqrt{\Big|\(1-\e\)\,\(\e+V_0+1\)\Big|}\, I_{l+1/2}\(\tx R\)\mathrm{Im}\[\F_1\] - \sqrt{\Big|\(1+\e\)\,\(\e+V_0-1\)\Big|}\,I_{l+3/2}\(\tx R\)\mathrm{Re}\[\F_1\] \over
 \sqrt{\Big|\(1-\e\)\,\(\e+V_0+1\)\Big|}\,I_{l+1/2}\(\tx R\)\mathrm{Re}\[\F_2\] + \sqrt{\Big|\(1+\e\)\,\(\e+V_0-1\)\Big|}\,I_{l+3/2}\(\tx R\)\mathrm{Im}\[\F_2\]} \ , \\
&\l_{3,uq}^{-}(l,k)={\sqrt{\Big|\(1-\e\)\,\(\e+V_0+1\)\Big|}\, J_{l+1/2}\(\x R\)\mathrm{Im}\[\F_1\] - \sqrt{\Big|\(1+\e\)\,\(\e+V_0-1\)\Big|}\,J_{l+3/2}\(\x R\)\mathrm{Re}\[\F_1\] \over
 \sqrt{\Big|\(1-\e\)\,\(\e+V_0+1\)\Big|}\,J_{l+1/2}\(\x R\)\mathrm{Re}\[\F_2\] + \sqrt{\Big|\(1+\e\)\,\(\e+V_0-1\)\Big|}\,J_{l+3/2}\(\x R\)\mathrm{Im}\[\F_2\]} \ ,
\eal\right.
\eeq
\end{widetext}
In the r.h.s. of (\ref{4.25}) $\e \equiv \e(k) < -1\, , \ \x=\x(k)\, , \ \tx=\tx(k)\, , \ R=R(Z)$, the parameters  $b,c$ of the Kummer's functions remain the same as in (\ref{4.4}), but with  negative $\e(k)<-1$, while the arguments $\(l\, , k\, , R(Z)\)$ in the functions $\F_1\, , \F_2$ remain the same as above.

The phase shifts  $\d_{uq}^-(l,k)$   are given by the same expressions (\ref{4.8}-\ref{4.10}) as for the upper one with two main differences. First, $\e(k)$ is negative and, second, the matching coefficients are defined via (\ref{4.25}) in  accordance with 3 intervals $0\leq k \leq k_1\, , \ k_1 \leq k \leq k_2\, , \ k_2 \leq k \leq \inf$.

The calculation of phase shifts  for the region $l+1 < Q$, that means for imaginary  $\vk_l$, proceeds as follows. The inner solutions for $r \leq R(Z)$ remain unchanged, while for  $r \geq R(Z)$ the scattering states of  DC  can be written now in terms of the Kummer $\F(b,c,z)$ and conjugated  Kummer $\F^{\ast}(b,c,z)$ ones, since for imaginary  $\vk_l$ the Kummer relation doesn't destroy the independence of solutions built in such a way.

Upon introducing the notation
 \beq\label{4.28}
\vk_l=i\, \eta_l \ ,
\eeq
 where
 \beq\label{4.29}
\eta_l=\sqrt{Q^2-(l+1)^2} \ ,
\eeq
and the parameters $b\,,c$ of Kummer's functions in the form
\beq\begin{gathered}
\label{4.30}
b_l=i\, \(\eta_l- \e(k) Q/k\) \ , \quad c_l=1+2\,i\,\eta_l \ ,
\end{gathered}\eeq
the real-valued scattering solution for $(u_l\,, q_l)$-pair in the upper continuum is represented as
\begin{widetext}
\beq\begin{gathered}
\label{4.31}\left\lbrace\bal
& u_l(k,r)={1 \over r}\,\sqrt{\e(k)+1}\,\mathrm{Re}\[\mathrm{e}^{i k(r-R)}\, \({r \over R}\)^{i \eta_l}\, N_R^{\ast}(l,k)\, F_1 (l,k,r)\] \ , \\
& q_l(k,r)={1 \over r}\,\sqrt{\e(k)-1}\,\mathrm{Im}\[\mathrm{e}^{i k(r-R)}\, \({r \over R}\)^{i \eta_l} N_R^{\ast}(l,k)\, F_2(l,k,r)\] \ ,
\eal\right.\end{gathered}\eeq
where
\beq
\begin{gathered}
\label{4.32}
F_1(l,k,r)=b_l\,\F(b_l+1\, , c_l\, , -2\, i\, k\, r) + (l+1 + i Q/k)\,\F(b_l\, , c_l\, , -2\, i\, k\, r)   \ , \\
F_2(l,k,r)=b_l\,\F(b_l+1\, , c_l\, , -2\, i\, k\, r) - (l+1 + i Q/k)\,\F(b_l\, , c_l\, , -2\, i\, k\, r)   \ ,
\end{gathered}\eeq
and
 \beq
\label{4.33}
N_R(l,k)=\sqrt{\(\e-1\)\,\(\e+V_0+1\)}\, J_{l+1/2}\(\x R\)\, F_{2R} + i\, \sqrt{\(\e+1\)\,\(\e+V_0-1\)}\, J_{l+3/2}\(\x R\)\, F_{1R} \ .
\eeq
In the r.h.s. of (\ref{4.33}) $\e \equiv \e(k)=+\sqrt{k^2+1}\geqslant 1\, , \ \x=\x(k)\, , \ R=R(Z)$ with the same arguments $\(l\, , k\, , R(Z)\)$ in functions $F_{1R}$ and $F_{2R}$.

From (\ref{4.31}-\ref{4.33}) for the phase shift $\d^+_{uq}(l,k)$ one finds
\beq\begin{gathered}
\label{4.34}
\d^+_{uq}(l,k)={\pi l \over 2} + {\e(k)\,Q \over k}\, \ln \(2\,k r\) + \mathrm{Arg}\[C(l,k)\] \  ,
\end{gathered}\eeq
where
\beq\begin{gathered}
\label{4.35}
C(l,k)=i\,\[ \mathrm{e}^{-\pi \eta_l/2}\,\tN_R(l,k)\, \({\G(c_l) \over \G(b_l)}\)^{\ast} +  \mathrm{e}^{\pi \eta_l/2}\,\tN^{\ast}_R(l,k)\,(l+1 + i Q/k)\,{ \G(c_l) \over \G\(1+i\( \eta_l + \e(k) Q /k\) \)}\]  ,
\end{gathered}\eeq
and
\beq\begin{gathered}
\label{4.36}
\tN_R(l,k)=\mathrm{e}^{i \( k R + \eta_l \ln \( 2 k R\)\)}\,N_R(l,k)  \  ,
\end{gathered}\eeq

In the lower continuum with $\e(k) < -1$ we again deal with 3 separate intervals $0\leq k \leq k_1\, , \ k_1 \leq k \leq k_2\, , \ k_2 \leq k \leq \inf$. The general answer for scattering solutions in this case reads
\beq\begin{gathered}
\label{4.43}\left\lbrace\bal
& u_l(k,r)={1 \over r}\,\sqrt{|\e(k)|-1}\,\mathrm{Im}\[\mathrm{e}^{i k(r-R)}\, \({r \over R}\)^{i \eta_l}\, N_R^{\ast}(l,k)\, F_1 (l,k,r)\] \ , \\
& q_l(k,r)={1 \over r}\,\sqrt{|\e(k)|+1}\,\mathrm{Re}\[\mathrm{e}^{i k(r-R)}\, \({r \over R}\)^{i \eta_l} N_R^{\ast}(l,k)\, F_2(l,k,r)\] \ ,
\eal\right.
\end{gathered}\eeq
where
 \beq\begin{gathered}
\label{4.44}
N_R(l,k)=\left\lbrace\bal
& \sqrt{\Big|\(1+\e\)\,\(\e+V_0-1\)\Big|}\, J_{l+3/2}\(\x R\)\, F_{1R} + i\, \sqrt{\Big|\(1-\e\)\,\(\e+V_0+1\)\Big|}\, J_{l+1/2}\(\x R\)\, F_{2R} \ , \quad  0\leq k \leq k_1 \ ,  \\
& \sqrt{\Big|\(1+\e\)\,\(\e+V_0-1\)\Big|}\, I_{l+3/2}\(\tx R\)\, F_{1R} - i\, \sqrt{\Big|\(1-\e\)\,\(\e+V_0+1\)\Big|}\,I_{l+1/2}\(\tx R\)\, F_{2R} \ , \quad  k_1 \leq k \leq k_2  \ , \\
& \sqrt{\Big|\(1+\e\)\,\(\e+V_0-1\)\Big|}\, J_{l+3/2}\(\x R\)\, F_{1R} - i\, \sqrt{\Big|\(1-\e\)\,\(\e+V_0+1\)\Big|}\,J_{l+1/2}\(\x R\)\, F_{2R} \ , \quad  k_2 \leq k \leq \inf \ .
\eal\right.\\
\end{gathered}\eeq
From (\ref{4.43},\ref{4.44}) for the phase shift one finds
\beq\begin{gathered}
\label{4.46}
\d^-_{uq}(l,k)={\pi l \over 2} + {\e(k)\,Q \over k}\, \ln \(2\,k r\) + \mathrm{Arg}\[D(l,k)\] \  ,
\end{gathered}\eeq
where
\beq\begin{gathered}
\label{4.47}
D(l,k)=  \mathrm{e}^{\pi \eta_l/2}\,\tN^{\ast}_R(l,k)\,(l+1 + i Q/k)\,{ \G(c_l) \over \G\(1+i\( \eta_l + \e(k) Q /k\) \)} - \mathrm{e}^{-\pi \eta_l/2}\,\tN_R(l,k)\, \({\G(c_l) \over \G(b_l)}\)^{\ast} \  ,
\end{gathered}\eeq
\end{widetext}

Finally, the total elastic phase $\d_{tot}(l,k)$ is obtained by summing up all the four separate phases for $\(u_l\,, q_l\)$- and $\(p_l\,, v_l\)$-pairs
\beq\label{4.50}
\d_{tot}(l,k)=\d_{uq}^+(l,k)+\d_{uq}^-(l,k)+\d_{pv}^+(l,k)+\d_{pv}^-(l,k)\ ,
\eeq
where $\d_{pv}^{\pm}(l,k)$ are obtained from $\d_{uq}^{\pm}(l,k)$ via crossing-symmetry prescription.
$\d_{tot}(l,k)$ automatically includes  contributions from both continua and both parities, and reveals the following general properties. First,  each separate phase in (\ref{4.50}) is determined from  the asymptotics of scattering solutions only up to additional term $\pi s$. Therefore the term $2 \pi l$ in $\d_{tot}(l,k)$ can be freely replaced by any other $2 \pi s$ and so should be chosen by means of additional grounds, the main of which is  $\d_{tot}(l,k \to \inf) \to 0$. Second, the Coulomb logarithms $\pm Q\, (|\e|/k)\, \ln (2 k r)$, which enter each separate phase, in the total phase cancel each other. However, after canceling Coulomb logarithms the separate phase shifts contain still the singularities for both $k \to 0$ and $k \to \inf$.  In particular, for $k\to\infty$ in the asymptotics of separate phases there remain the singular terms $\mp Q|\e|\ln(2 k R)/k$, but they also cancel mutually in the total phase just as in 2+1 D case ~\cite{Davydov2018b,Sveshnikov2019b}. As  a result, $\d_{tot}(l,k \to \inf)$ shows up for any $\vk_l$ the following vanishing asymptotics \footnote{Here there are shown only 3 first orders of asymptotical expansion of $\d_{tot}(l,k \to \inf)$ in inverse powers of $k R$. At the same time, by means of computer algebra tools it is possible to derive in explicit form as many orders of this expansion as needed. It is quite important, since for  concrete calculation of phase integral such explicit terms of asymptotics are very useful, because it allows to perform the integration over large $k$ in the phase integral with given accuracy purely analytically.  }
\begin{widetext}
\beq\begin{gathered}
\label{4.56}
\d_{tot}(l,k \to \inf)={1 \over (k R)^3}\, \[ {4 \over 3}\, Q^2 \( (l+1)^2-3\,R^2\) -(l+1)\, Q \sin( 2\,Q)\, \cos(2\,k R - \pi l) \]  \  + \\ + \
{\sin(2\,k R - \pi l) \over (k R)^4}\,(l+1)\, \[ 3\,Q^2 \cos( 2\,Q) + Q\,\sin( 2\,Q)\, \( (l+1)^2-5/2\)\]
\ + \\ + \ {1 \over (k R)^5}\, \Big\{  {8\,Q^4 \over 15}\,  \( 3(l+1)^2-5\,R^2\) +
{Q^2 \over 2}\,\[{(l+1)^2 \over 5}\, \(4\,(l+1)^2-5\)+{R^2 \over 3}\,\(12(l+1)^2 -1 \) \] \ + \\ +
{l+1 \over 2}\, \[ \(l^2(l+1)^2 - 6\, l(l+1) +6 -R^2 - 12\,Q^2\)\, Q\,\sin( 2\,Q) + \(8\,(l+1)^2-23 - 2\,R^2\)\,Q^2\cos(2\,Q)\]\cos(2\,k R - \pi l) \Big\} + \\ + O\({1 \over (k R)^6}\) \ .
\end{gathered}\eeq
As it was already stated above in the Section III, the leading $O\(1/(kR)^3\)$ non-oscillating term of  $\d_{tot}(l,k \to \inf)$  coincides exactly with the leading-order-WKB asymptotics (\ref{3.39b}), while the oscillating ones in $ \d_{tot}(l,k \to \inf) $ are responsible for diffraction on a sphere of the radius  $R$. It would be also worth-while noticing that the explicit asymptotics (\ref{4.56}) confirms that the total phase is an even function of $Q$, but not of $Z$, since the dependence on $Z$ includes the function $R=R(Z)$, which is quite different.

The IR-asymptotics of separate phases contain also the singularities of the form $\pm Q\,(1-\ln(Q/k))/k$. However, these singularities again cancel each other in  $\d_{tot}(l,k \to 0)$, and so the total phase for  $k \to 0$ possesses a finite limit, which for large $l \gg Q$ coincides with the WKB-approximation and reproduces the answer for a  point-like source, but in general case turns out to be quite different, especially for imagine $\vk_l$. The explicit expressions for $\d_{tot}(l,k \to 0)$ depend strongly on either $\vk_l$ is real or imagine. In the case of real $\vk_l$ by means of the subsidiary functions
\beq\begin{gathered}
\label{4.51}
f_l={ J_{2\vk_l}\(\sqrt{8\,Q R}\)\, \[ z_0\,j_1 -  (l+1 +\vk_l)\,j_2 \]  +
j_2\,\sqrt{2\, Q R }\, J_{1+2\,\vk_l}\(\sqrt{8\,Q R}\) \over
J_{-2\vk_l}\(\sqrt{8\,Q R}\)\, \[ z_0\,j_1 - ( l+1 -\vk_l)\,j_2 \]  +
j_2\,\sqrt{2\, Q R }\, J_{1-2\,\vk_l}\(\sqrt{8\,Q R}\) } \ , \\
g_l={J_{2\vk_l}\(\sqrt{8\,Q R}\)\, \[ (l+1 -\vk_l)\,j_1 - z_0\,j_2 \]  +
j_1\,\sqrt{2\, Q R }\, J_{1+2\,\vk_l}\(\sqrt{8\,Q R}\) \over
J_{-2\vk_l}\(\sqrt{8\,Q R}\)\, \[(l+1 +\vk_l)\,j_1 - z_0\,j_2 \]  +
j_1\,\sqrt{2\, Q R }\, J_{1-2\,\vk_l}\(\sqrt{8\,Q R}\) } \ ,
\end{gathered}\eeq
where
\beq\label{4.52}
z_0=\sqrt{Q^2+2\,Q R} \ , \quad j_1=J_{l+1/2}(z_0) \ , \quad j_2=J_{l+3/2}(z_0) \ ,
\eeq
the answer for $\d_{tot}(l,k \to 0)$  reads
\beq\label{4.53}
\tan \(\d_{tot}(l,k \to 0)\)=\tan \Big\{ \mathrm{Arg}\[\(1-\mathrm{e}^{2 \pi i \vk_l}\,f_l\) \(1-\mathrm{e}^{2 \pi i \vk_l}\,g_l\)\]-2 \pi  \vk_l \Big\} \ ,
\eeq
At the same time, for imagine $\vk_l=i \eta_l$ by means of two subsidiary phases
\beq\begin{gathered}
\label{4.54}
\vf_l=-\mathrm{Arg}\Big\{ \sqrt{1+V_0/2}\,j_1\, 2 Q J_{2\,i\eta_l}\(\sqrt{8\,Q R}\) + \sqrt{2\,V_0}\,j_2\, \[\sqrt{2\, Q R }\, J_{1+2\,i\eta_l}\(\sqrt{8\,Q R}\) -  (l+1 +i \eta_l)\,J_{2\,i \eta_l}\(\sqrt{8\,Q R}\) \] \Big\} \ , \\
\c_l=-\mathrm{Arg}\Big\{\sqrt{2\,V_0}\,j_1\, \[\sqrt{2\, Q R }\, J_{1+2\,i\eta_l}\(\sqrt{8\,Q R}\) +  (l+1 -i \eta_l)\,J_{2\,i \eta_l}\(\sqrt{8\,Q R}\) \] - \sqrt{1+V_0/2}\,j_2\, 2 Q J_{2\,i\eta_l}\(\sqrt{8\,Q R}\) \Big\} \ ,
\end{gathered}\eeq
instead  of (\ref{4.53}) one finds
\beq\label{4.55}
\tan \(\d_{tot}(l,k \to 0)\)=\tan \Big\{ \mathrm{Arg}\[\(\mathrm{e}^{ \pi \eta_l+i\,\vf_l}-\mathrm{e}^{- \pi  \eta_l -i\,\vf_l}\)\(\mathrm{e}^{ \pi \eta_l+i\,\c_l}-\mathrm{e}^{- \pi  \eta_l -i\,\c_l}\)\]  \Big\} \ .
\eeq
\end{widetext}

\begin{figure*}[t!]
\subfigure[]{
		\includegraphics[width=\columnwidth]{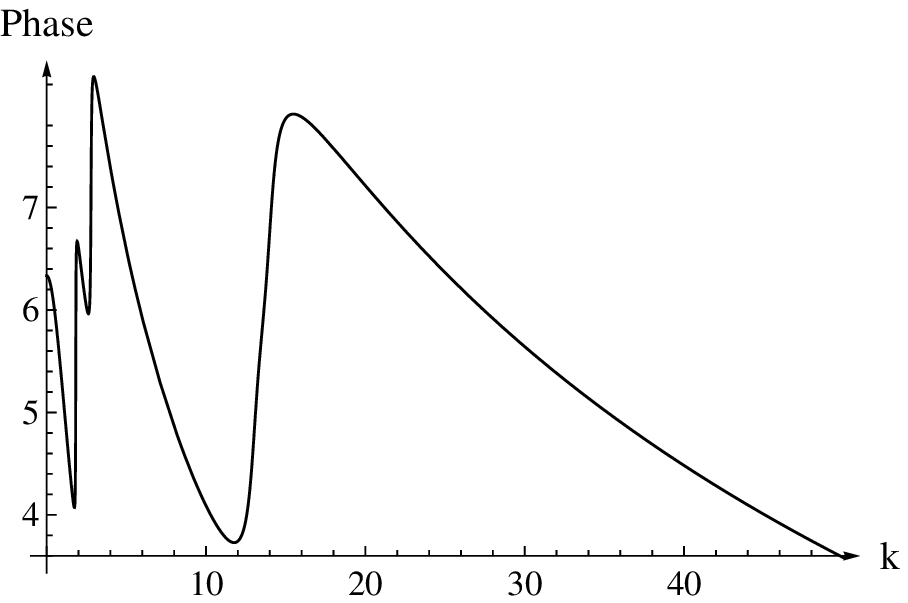}
}
\hfill
\subfigure[]{
		\includegraphics[width=\columnwidth]{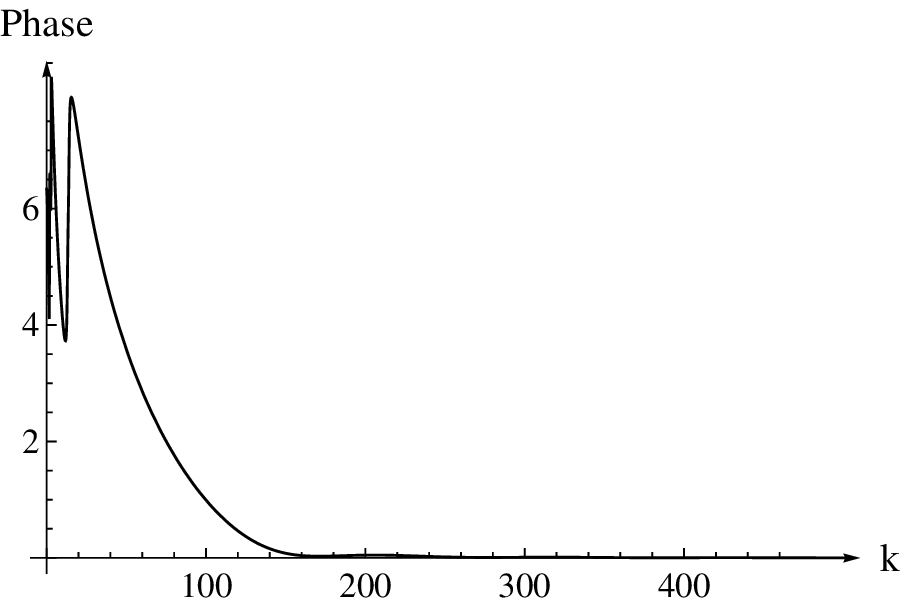}
}
\vfill
\subfigure[]{
		\includegraphics[width=\columnwidth]{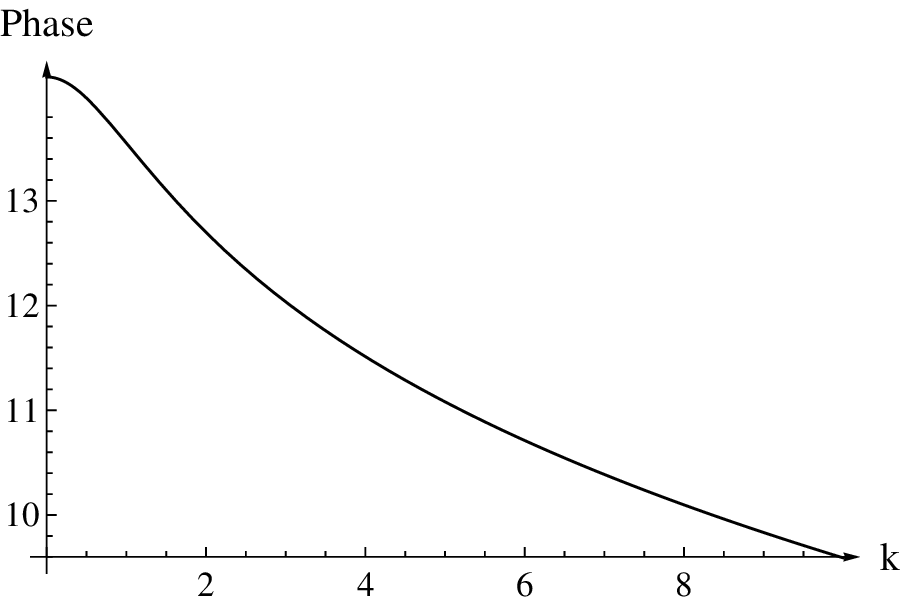}
}
\hfill
\subfigure[]{
		\includegraphics[width=\columnwidth]{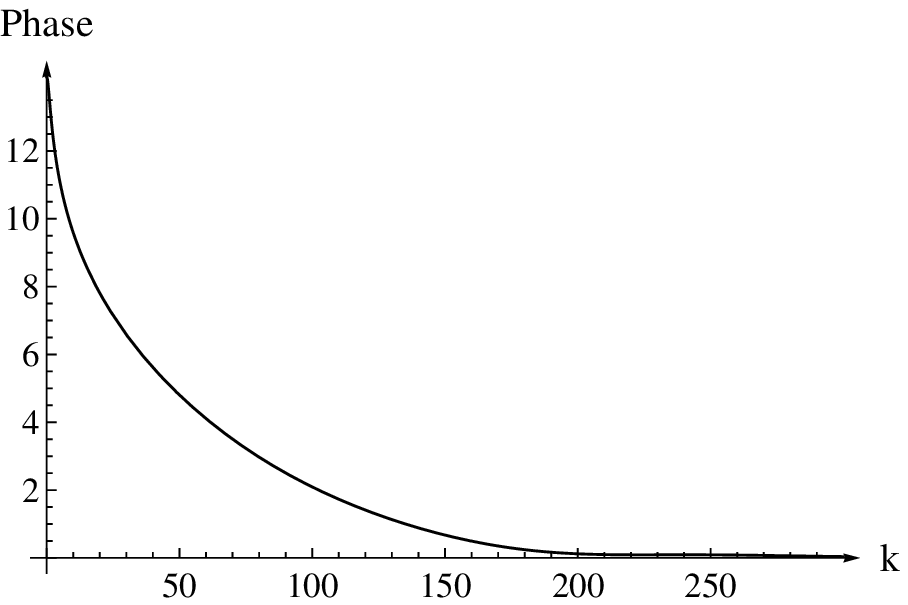}
}
\vfill
\subfigure[]{
		\includegraphics[width=\columnwidth]{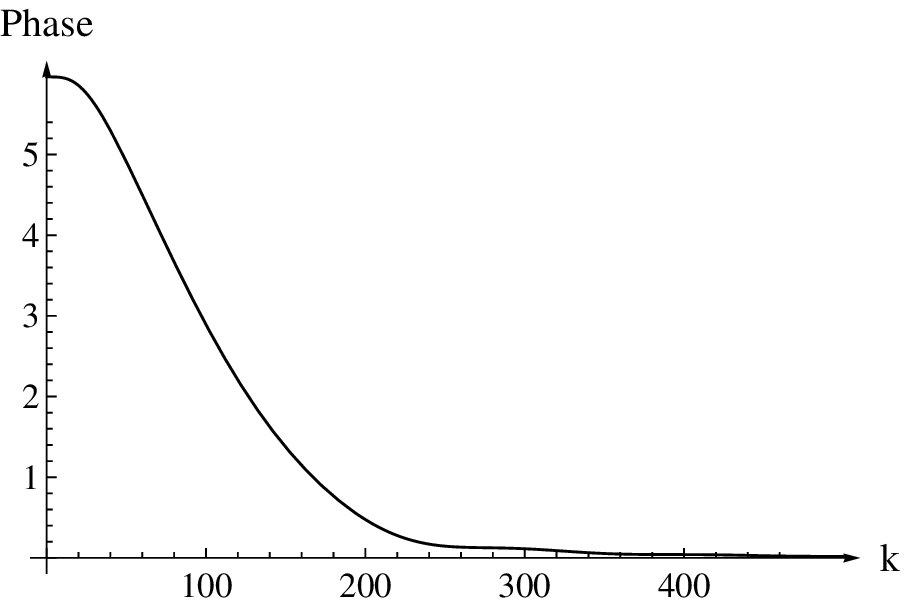}
}
\hfill
\subfigure[]{
		\includegraphics[width=\columnwidth]{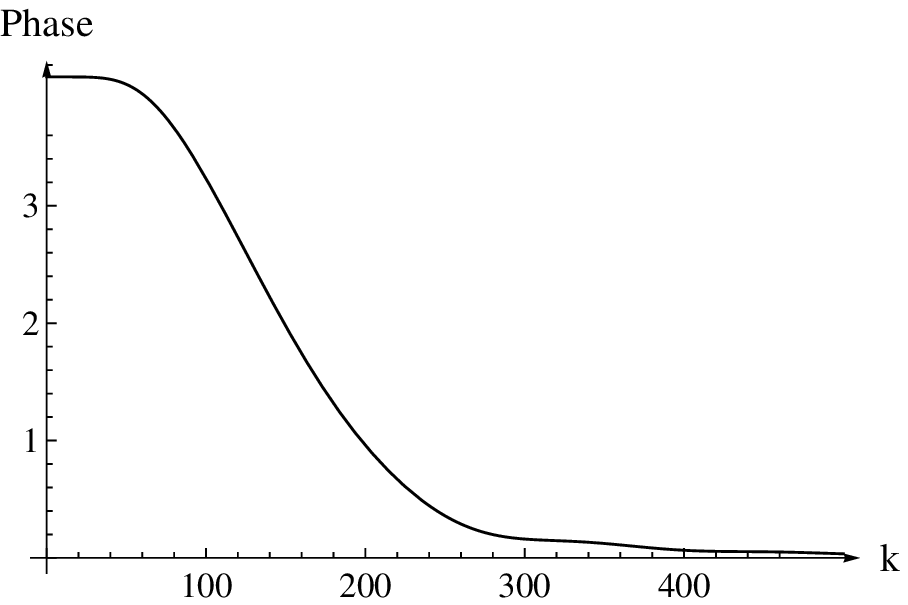}
}
\caption{$\d_{tot}(l,k)$ for $Z=300$ and (a,b): $l=0$; (c,d): $l=1$; (e): $l=2$; (f): $l=3$. }
	\label{phase300}	
\end{figure*}
\begin{figure*}[t!]
\subfigure[]{
		\includegraphics[width=0.87\columnwidth]{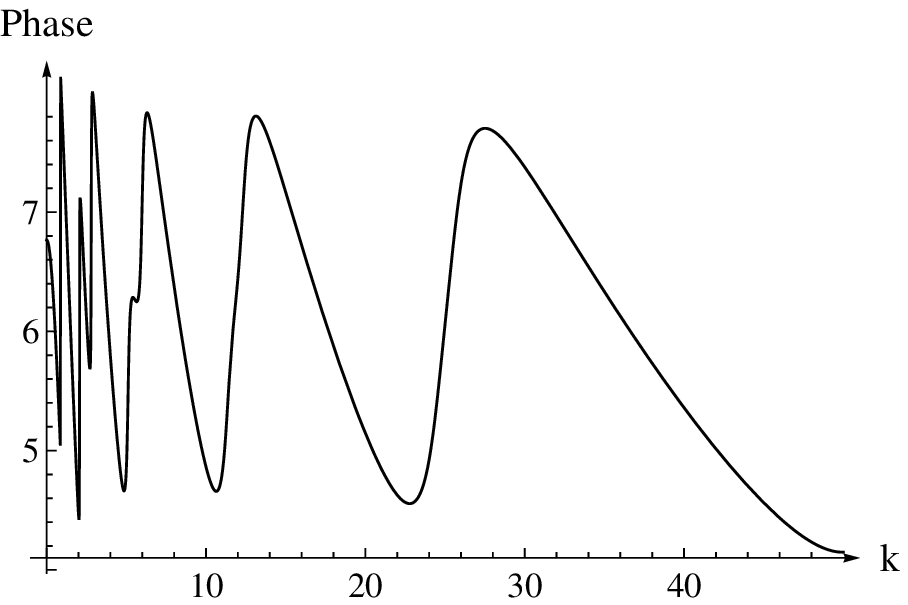}
}
\hfill
\subfigure[]{
		\includegraphics[width=0.87\columnwidth]{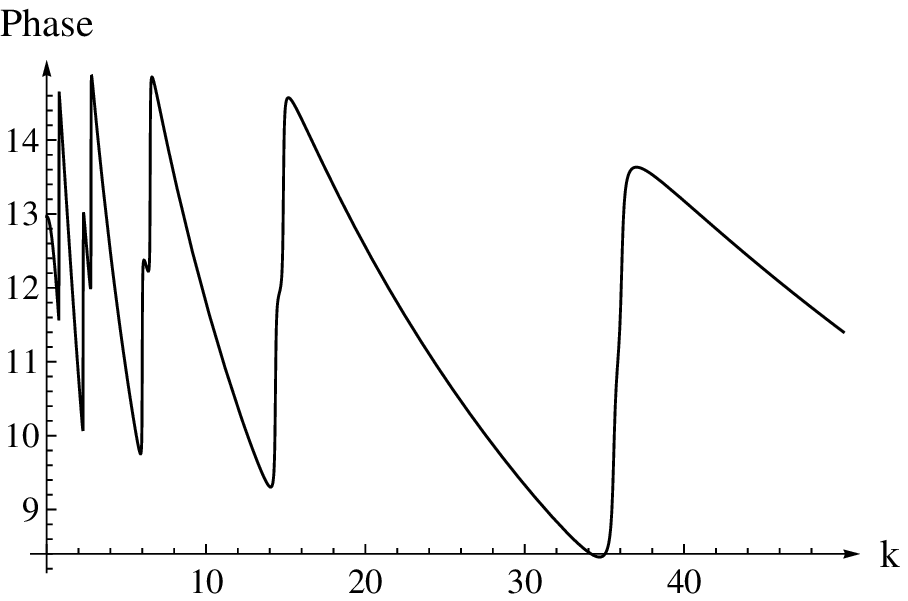}
}
\vfill
\subfigure[]{
		\includegraphics[width=0.86\columnwidth]{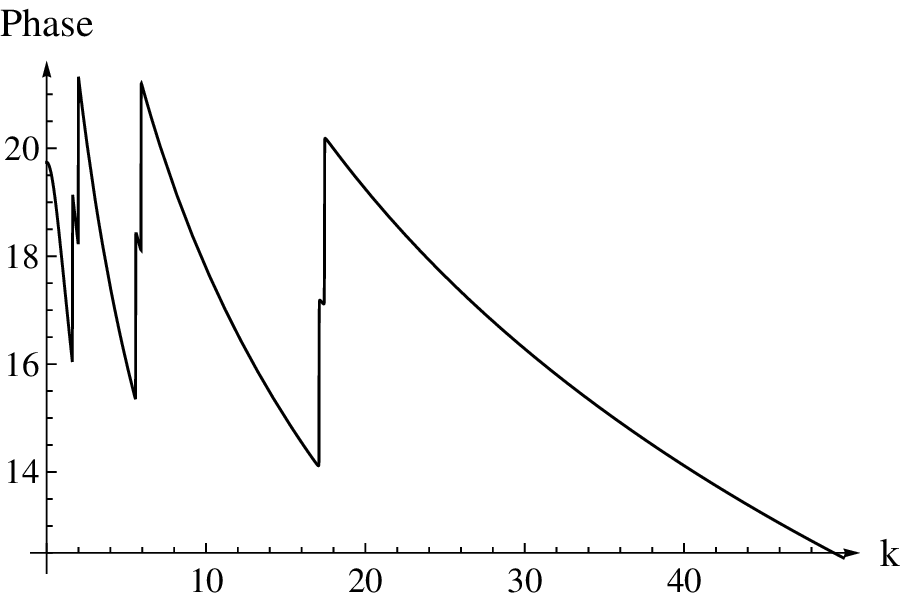}
}
\hfill
\subfigure[]{
		\includegraphics[width=0.86\columnwidth]{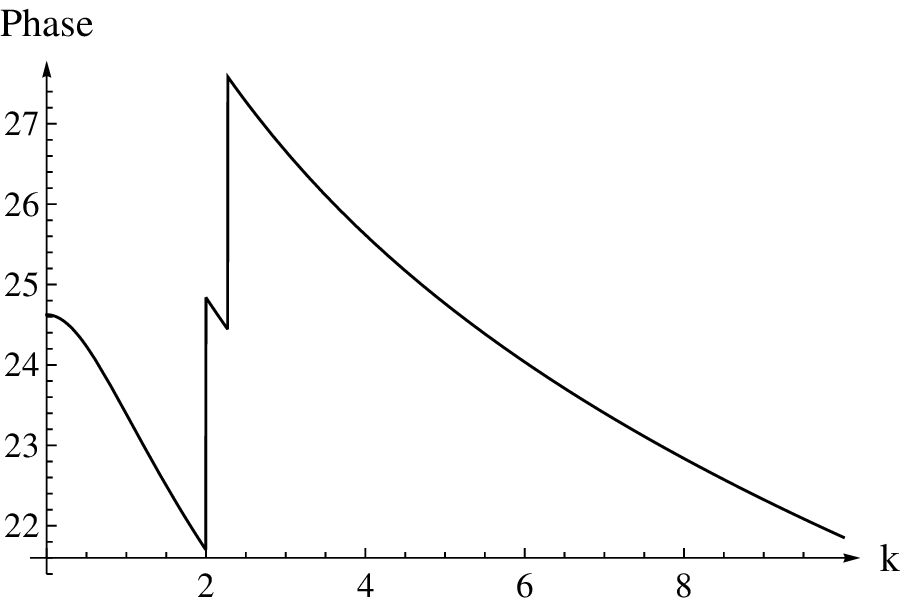}
}
\vfill
\subfigure[]{
		\includegraphics[width=0.86\columnwidth]{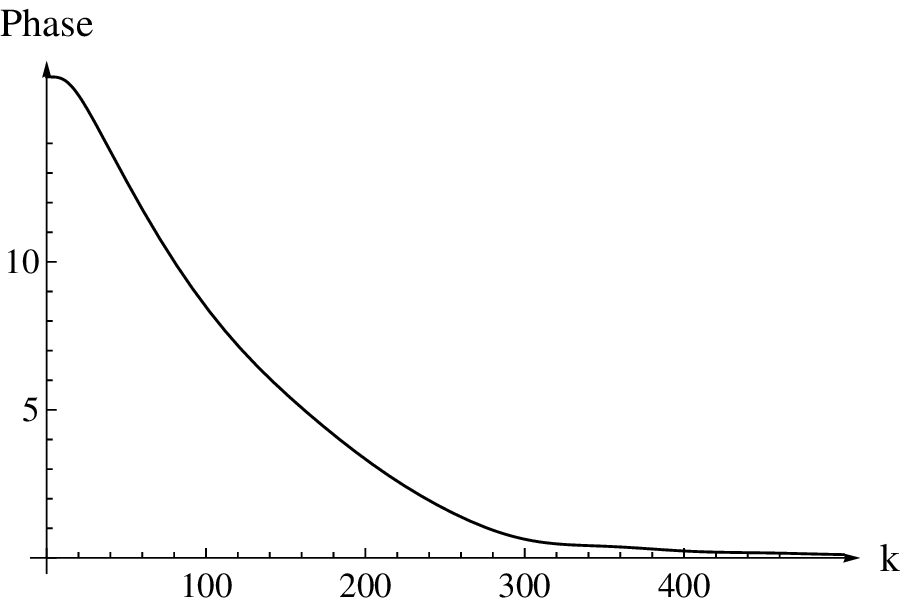}
}
\hfill
\subfigure[]{
		\includegraphics[width=0.86\columnwidth]{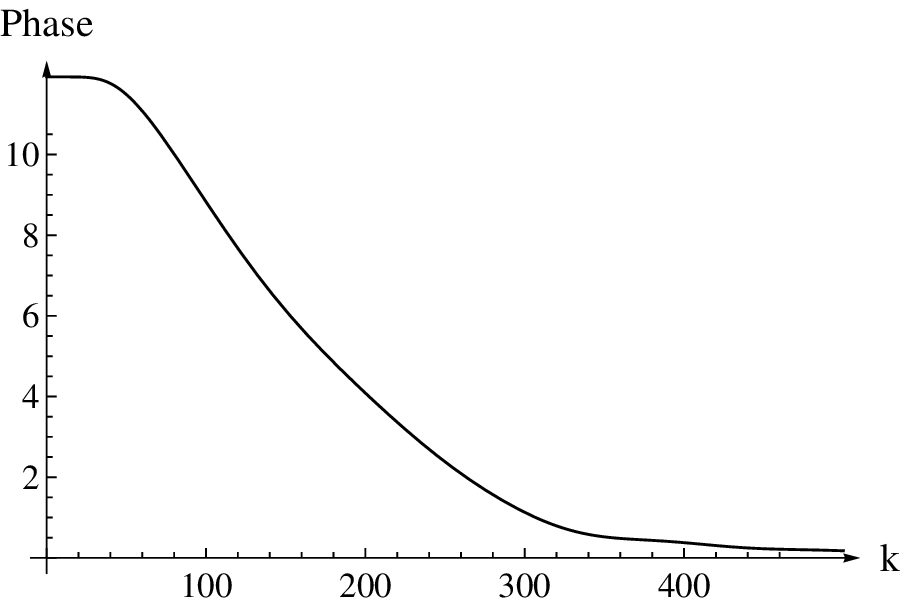}
}
\vfill
\subfigure[]{
		\includegraphics[width=0.86\columnwidth]{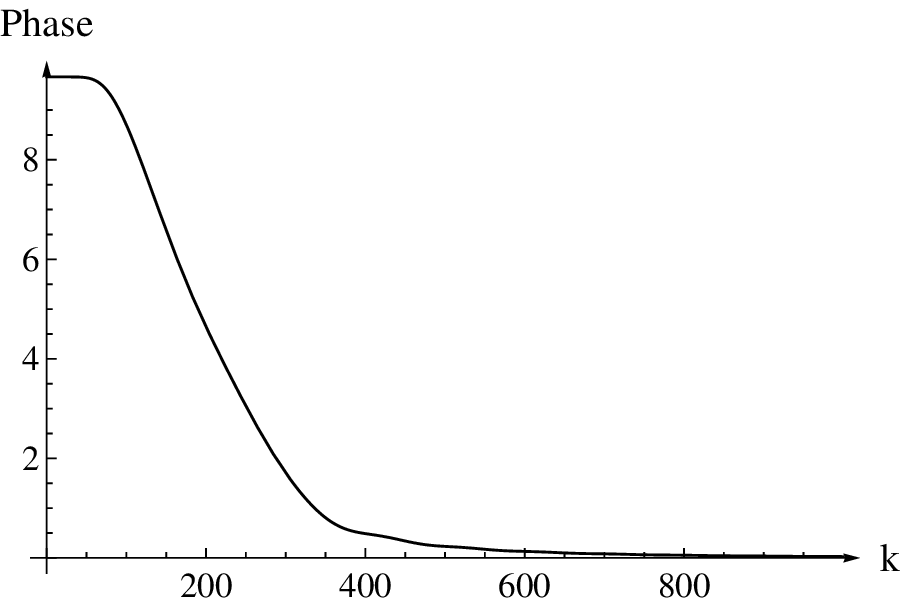}
}
\caption{$\d_{tot}(l,k)$ for $Z=600$ and (a): $l=0$; (b): $l=1$; (c): $l=2$; (d): $l=3$; (e): $l=4$; (f): $l=5$; (g): $l=6$. }
	\label{phase600}	
\end{figure*}

The peculiar feature in the behavior of $ \d_{tot}(l,k) $ is the appearance of (positronic) elastic resonances upon diving of each subsequent  discrete level into the lower continuum. Here it should be  noted that the additional $\pi s$ in the separate phase shifts, mentioned above, are in principle unavoidable, since this arbitrariness comes from the possibility to alter the  common factor in the corresponding  wavefunctions. So one needs to apply a special procedure, which provides to distinguish between such artificial jumps in the phases by  $\pi$ of purely mathematical origin in the inverse $\tan$-function, and the physical ones, which are caused by resonances and for extremely narrow low-energy resonances look just like the same jumps by  $\pi$.  Removing the first ones, coming from the inverse $\tan$-function, we provide   the continuity of the phases, while the latter contain an important physical information. Hence, no matter how narrow they might be, they must necessarily be preserved in the phase function, and so this procedure should be performed with a very high level of accuracy.

In  Fig.\ref{phase300} the pertinent set of total phase curves for $Z=300$ is present.  For such $Z$ only 4 first levels from the $s$-channel have already dived into the lower continuum, namely $1s_{1/2}\, \(Z_{cr,1}\simeq 173.6\)\,, 2p_{1/2}\, \(Z_{cr,2}\simeq 188.5\)$\,, $2s_{1/2}\, \(Z_{cr,3}\simeq 244.3\)$ and $3p_{1/2}\,\(Z_{cr,4}\simeq 270.5\)$ \footnote{$Z_{cr,i}$ are determined from (\ref{5.9}) with $R(Z)$ defined as in (\ref{6.1}).}. So for $Z=300$ it is only the $s$-channel, where $\d_{tot}(l,k)$ undergoes corresponding resonant jumps by $\pi$. It should be remarked however, that each $Z_{cr,i}$ given above correspond to the case of external Coulomb source with charge $Z=Z_{cr,i}$ and radius $R_i=R(Z_{cr,i})$. Hence, this is not exactly  the case under consideration with $Z=300$ and $R(Z=300)$ fixed, rather it is  a qualitative picture of what happens with 4 lowest $s$-levels in this case. They are absent in the discrete spectrum and show up as positronic resonances, which rough  disposition can be  understood as a result of diving of corresponding levels at $Z_{cr,i}$. But their exact positions on the $k$-axis can be found only via thorough restoration of the form of $ \d_{tot}(0,k) $.

 In particular, the two first low-energy narrow jumps  in Fig.\ref{phase300}(a) correspond to resonances, which  are caused by diving of $2s_{1/2}$ and  $3p_{1/2}$, what happens quite close to $Z=300$. At the same time, the jumps caused by diving of $1s_{1/2}$ and  $2p_{1/2}$ have been already gradually smoothed and almost merged together, hence look like one big $2\,\pi$-jump, which is already significantly shifted to the region of larger $k$. In the other channels with $l \geq 1$ there are no dived levels for such $Z$, and so $\d_{tot}(l,k)$ in these channels look like a smooth decreasing function, which behavior  roughly resembles the one of WKB-phase in Fig.\ref{WKB}.

In  Fig.\ref{phase600} the behavior  of total phases in pertinent channels is presented for $Z=600$.  For this $Z$ diving of discrete levels occurs already  up to $l=3$. The dived levels  naturally group into pairs of different parity. For $l=0$ these are $\{1s_{1/2}\,, 2p_{1/2}\}\,, \dots\,, \{5s_{1/2}\,, 6p_{1/2}\}$\, and the last unpaired $6s_{1/2}$ with corresponding $Z_{cr} \simeq 576.4$. For $l=1$ this is the set $\{2p_{3/2}\,, 3d_{3/2}\}\,, \dots\,, \{5p_{3/2}\,, 6d_{3/2}\}$\,, with the last unpaired $6p_{3/2}\, \(Z_{cr}\simeq 581.2\)$, while for $l=2$ one has $\{3d_{5/2}\,, 4f_{5/2}\}\,, \dots\,,  \{5p_{3/2}\,, 6f_{5/2}\}$\,, with the last $6f_{5/2}\, \(Z_{cr}\simeq 564.5\)$. Diving stops at $l=3$ with the pair $\{4f_{7/2}\,, 5g_{7/2}\}$\,, corresponding to critical charges $\simeq 578.6$ and $\simeq 582.7$. The meaning of $Z_{cr,i}$ in this case is the same as for $Z=300$. The behavior  of total phases in these channels is quite similar and differs only by the number of resonant jumps, hence  in the appearance of curves. For the $s$-channel the total number of dived levels exceeds 11, and so $\d_{tot}(0,k)$ undergoes the corresponding number of jumps by $\pi$,  which at small $k$ practically overlap each other. At the same time, for $l=2$ and $l=3$ all the jumps are quite clearly pronounced, transparent and lie in one-to-one correspondence with the sequence of dived levels. In the other channels with $l \geq 4$ there are no dived levels, so $\d_{tot}(l,k)$ transforms into  a smooth decreasing function  similar to  the  WKB-phase (compare Fig.\ref{phase600}(g) with Fig.\ref{WKB}).

Thus, for any $l$ the total phase  $ \d_{tot}(l,k) $ turns out to be regular everywhere on the whole half-axis $0 \leq k \leq \infty$ and for $k \to \inf$ decreases sufficiently fast to provide the convergence of the phase integral (\ref{3.42}). So the latter could be quite reliably evaluated by means of the standard numerical recipes with a special account for the region of small $k$ caused by  extremely narrow low-energy resonances.

\section{Evaluation of  the discrete spectrum for the potential (\ref{1.5a})}

Quite similar to the case of the total elastic phase $ \d_{tot}(l,k) $, it  suffices to consider only $(u_l\,, q_l)$-pair, since  the  contribution of $(p_l\,, v_l)$-pair   can be again achieved via crossing symmetry of initial DC. For the external potential (\ref{1.5a}) the discrete levels with $ |\e| < 1$ are found via equations in terms of the Tricomi function $\P(b\,,c\,, z)$, which are valid for any $\vk_l$.   By means of denotations
\beq\begin{gathered}
\label{5.1}
\g=\sqrt{1-\e^2} \ , \quad b_l=\vk_l- \e\, Q/\g \ , \quad c_l=1+2\vk_l \ , \\
\P=\P(b_l\,, c_l\,, 2 \g R) \ , \quad \P_+=\P(b_l+1\,, c_l\,, 2 \g R) \
\end{gathered}\eeq
and
\beq\begin{gathered}
\label{5.2}
\x=\sqrt{(\e+V_0)^2-1} \ ,
\end{gathered}\eeq
the  equation  for $(u_l\,, q_l)$-levels  reads
\begin{widetext}
\beq\label{5.3}
(l+1-Q/\g)\,{\P_+ \over \P}={ \sqrt{(1+\e)\,(\e+V_0-1)}\,J_{l+3/2}(\x R)-\sqrt{(1-\e)\,(\e+V_0+1)}\,J_{l+1/2}(\x R) \over \sqrt{(1+\e)\,(\e+V_0-1)}\,J_{l+3/2}(\x R)+\sqrt{(1-\e)\,(\e+V_0+1)}\,J_{l+1/2}(\x R)}
\eeq
However, for imagine $\vk_l=i\, \eta_l$ another form of equations turns out to be more pertinent for concrete calculations, since it deals with the Kummer function instead of the Tricomi one
\beq\label{5.5}
{\mathrm{Im}\[X_l b_l \F_+\] \over \mathrm{Im}\[X_l \F\]}=(l+1+Q/\g)\,{ \sqrt{(1-\e)\,(\e+V_0+1)}\,J_{l+1/2}(\x R)-\sqrt{(1+\e)\,(\e+V_0-1)}\,J_{l+3/2}(\x R) \over \sqrt{(1-\e)\,(\e+V_0+1)}\,J_{l+1/2}(\x R)+\sqrt{(1+\e)\,(\e+V_0-1)}\,J_{l+3/2}(\x R)} \ ,
\eeq
\end{widetext}
where
\beq\begin{gathered}
\label{5.7}
b_l=i\,\eta_l- \e\, Q/\g \ , \quad c_l=1+2\,i\,\eta_l \ ,
\\ \F=\F (b_l\,, c_l\,, 2 \g R) \ , \quad \F_+=\F (b_l+1\,, c_l\,, 2 \g R) \ ,
\end{gathered}\eeq
and
\beq\begin{gathered}
\label{5.8}
X_l=\mathrm{e}^{i\eta_l\ln(2\,\g R)}\,{\G(b_l) \over \G(c_l)} \ ,
\end{gathered}\eeq
with the same $\g$ and $\x$.

Discrete levels on the threshold of the lower continuum with $\e=-1$ cannot be found by means of the confluent hypergeometric functions, since the latters become ill-defined. But since such levels are directly connected with the corresponding critical charges $Z_{cr,i}$ of the external source, there exists another well-known  procedure, which deals specially with this case.  For details see, e.g., Ref.~\cite{Greiner2012}. The result is that such levels appear only for $\vk_l=i\,\eta_l$, while
the critical charges for both parities $(\pm)$ are found in this case from equations
\begin{multline}\label{5.9}
2\, z_1\,K_{2 i \eta_l}\(\sqrt{8\,Q R}\)\,J_{\pm} \ \mp  \\ \mp \ \[\sqrt{2\,Q R}\,\(K_{1+2\, i \eta_l}\(\sqrt{8\,Q R}\) + K_{1-2 i \eta_l}\(\sqrt{8\,Q R}\)\)  \pm \right. \\ \left. \pm \ 2\,k\,K_{2 i \eta_l}\(\sqrt{8\,Q R}\)\]\,J_{\mp}=0 \ ,
\end{multline}
where $K_{\n}(z)$ is the McDonald function, $\eta_l$ is defined in (\ref{4.29}),
and
\beq\begin{gathered}\label{5.10}
 z_1=\sqrt{Q^2-2\, Q R} \ , \\  J_+=J_{l+3/2}(z_1) \ ,  \quad J_-=J_{l+1/2}(z_1) \ .
\end{gathered}\eeq
The meaning of eqs. (\ref{5.9},\ref{5.10}) is twofold. First, for the  given charge $Z$ and $R=R(Z)$ by means of these eqs. the existence of levels with $\e=-1$  can be checked. The answer is positive if and only if the current $Z$ coincides with one of $Z_{cr,i}$, otherwise there are no levels at the lower threshold. Second, by solving these eqs. with respect to $Z$ and $R=R(Z)$ one finds the complete set of critical charges $Z_{cr,i}$.

 The asymptotical behavior of the discrete spectrum for both $(u_l\,, q_l)$- and $(p_l\,, v_l)$-pairs in the   vicinity of the condensation point $\e_{n,l} \to 1$  reproduces the  Coulomb-Schroedinger problem for a point-like source with the same $Z$ including $O\(Q^4\)$-fine-structure correction  and $O\(Q^{2l+4}\)$-correction coming from non-vanishing  size of the Coulomb source
\begin{widetext}
\begin{multline}\label{5.11}
1-\e_{n_r,l} ={Q^2 \over 2\,(n_r+l+1)^2} \ + \ Q^4\, \({1 \over 2 (l+1) (n_r+l+1)^3} - {3 \over 8 (n_r+l+1)^4}\)
 \ - \\  - \
{Q^2 \over (n_r+l+1)^2}\, \( {2\, Q R \over n_r+l+1 }\)^{2l+2}\, {(n_r + 2l+1)! \over n_r!} { (2l+2)\,(2l+3) \over (2l+3)!^2}  \ , \quad n_r \to \inf \ .
\end{multline}
\end{widetext}
For any $l$  in this asymptotics the terms with Bohr levels and fine-structure correction are exactly summable over $n_r$ starting from certain $n_0$ via
\beq\begin{gathered}
\label{5.12}
\sum_{n_r=n_0}^{\inf}{1 \over (n_r+a)^2}=\p^{(1)}\(n_0 + a\) \ , \\ \sum_{n_r=n_0}^{\inf}{1 \over (n_r+a)^3}=-{1 \over 2}\,\p^{(2)}\(n_0 + a\) \ , \\ \sum_{n_r=n_0}^{\inf}{1 \over (n_r+a)^4}={1 \over 6}\,\p^{(3)}\(n_0 + a\) \ ,
\end{gathered}\eeq
where $\p^{(n)}(z)=d^n \p(z)/d z^n\, , \ \p(z)=\G'(z)/\G(z)$.
The correction from non-zero size of the source  in (\ref{5.11}) is also exactly summable over $n_r\geqslant n_0$ in terms of $\p^{(n)}\(n_0 + a\)$, but the general expression for arbitrary $l$ is very cumbersome.

Thus, for any $l$ the global strategy of summation over discrete spectrum in order to find its contribution to the partial $\E_{VP,l}(Z)$ turns out to be the following. First, summation is performed separately for $(u_l\,, q_l)$- and $(p_l\,, v_l)$-pairs. In the first step for each pair one finds  the set of lowest levels  with $n_r < n_0(l)$, where $n_0(l)$ is subject of condition that for $n_r \geqslant n_0(l)$ the levels coincide with the asymptotics (\ref{5.11}) with a given accuracy. Since for any non-vanishing accuracy $n_0(l)$ is finite, summation over this part of the discrete spectrum poses no problems. In the next step, the remaining infinite part of the spectrum is summed up by means of (\ref{5.11},\ref{5.12}). Proceeding this way we find with the given accuracy the sum
\beq
\label{5.15}
S(l)=\sum\limits_{\pm} \sum\limits_{-1 \leqslant \e_{n,l}^{\pm}<1} \(1-\e_{n,l}^{\pm}\) \ ,
\eeq
which is the discrete spectrum contribution to the partial VP-energy (\ref{3.31}). It should be noted however that for small or even moderate $l$ and especially for the case $l+1 < Q$ with $\vk_l=i\eta_l$ accounting for nonzero size of the Coulomb source  in this calculation takes care, since  the more current $Z$ and $R(Z)$, the more $n_0(l)$. In actual calculations up to $Z \simeq 1000$ the value of $n_0(l)$ exceeds thousands or even dozens of thousands, and since  all the levels with $n_r < n_0(l)$ must be find by solving the corresponding equations numerically, it takes some definite time \footnote{There exists another procedure of "quasi-exact"\, summation over discrete spectrum  in such DC problems, elaborated by K.S. and Y.V., which takes account for  nonzero size of the Coulomb source in  essentially non-perturbative way and works much more rapidly. Within this approach one needs to calculate exactly via eqs. (\ref{5.3},\ref{5.5}) only a few number, not more than one hundred for reasonable accuracies lying in the interval $10^{-16}-10^{-20}$, of the lowest levels lying  below $0.99$, afterwards the summation for the rest is performed in a closed form by means of (\ref{5.11},\ref{5.12}). However, this method requires for a thorough analysis of  behavior of confluent hypergeometric functions in non-trivial asymptotical regimes  combined with a number of additional original tricks and so will be presented separately.}.

\section{Renormalized VP-energy  for the Coulomb  source (\ref{1.5a}).}
\begin{figure*}[t!]
\subfigure[]{
		\includegraphics[width=1.3\columnwidth]{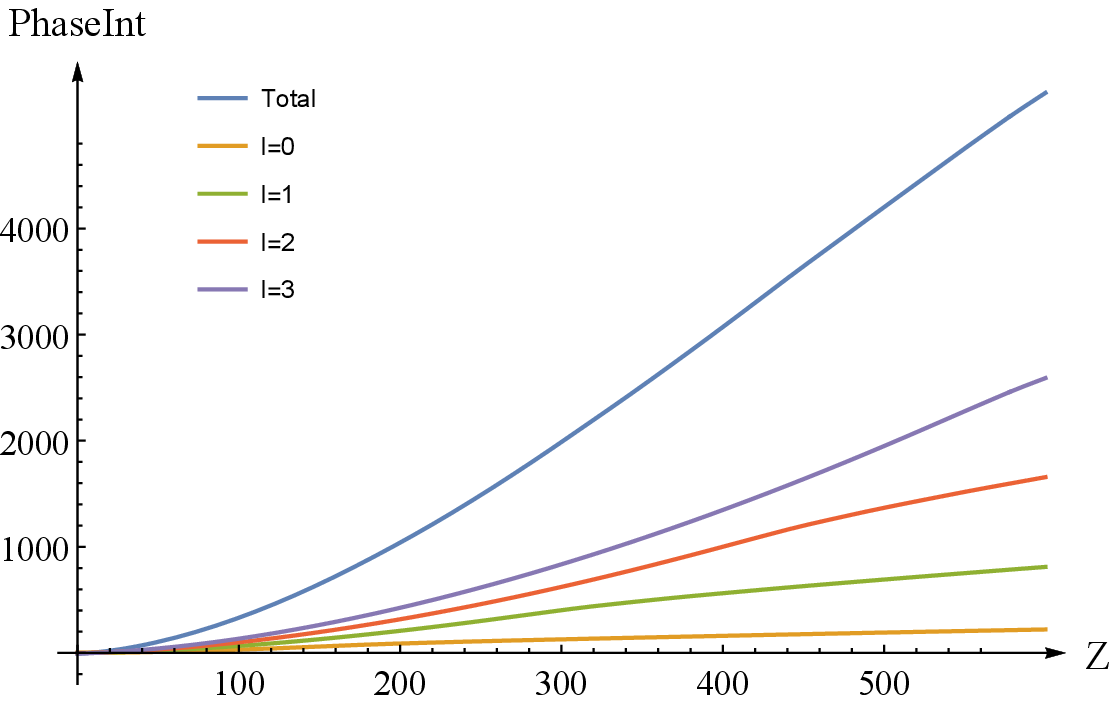}
}
\vfill
\subfigure[]{
		\includegraphics[width=1.3\columnwidth]{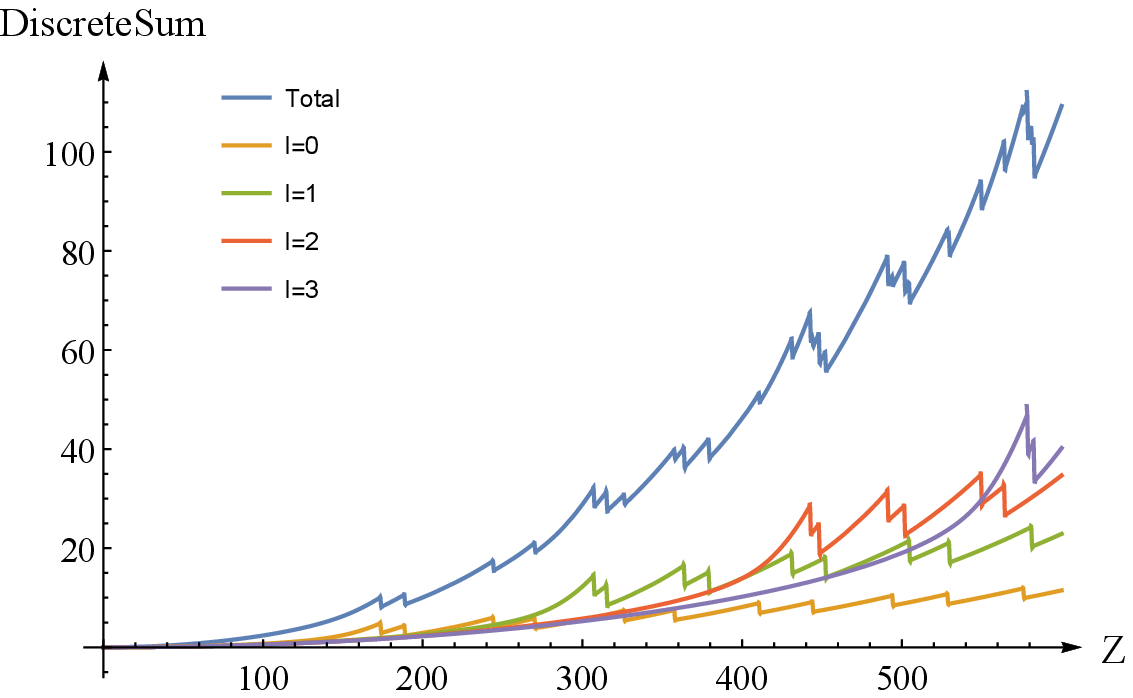}
}
\vfill
\subfigure[]{
		\includegraphics[width=1.3\columnwidth]{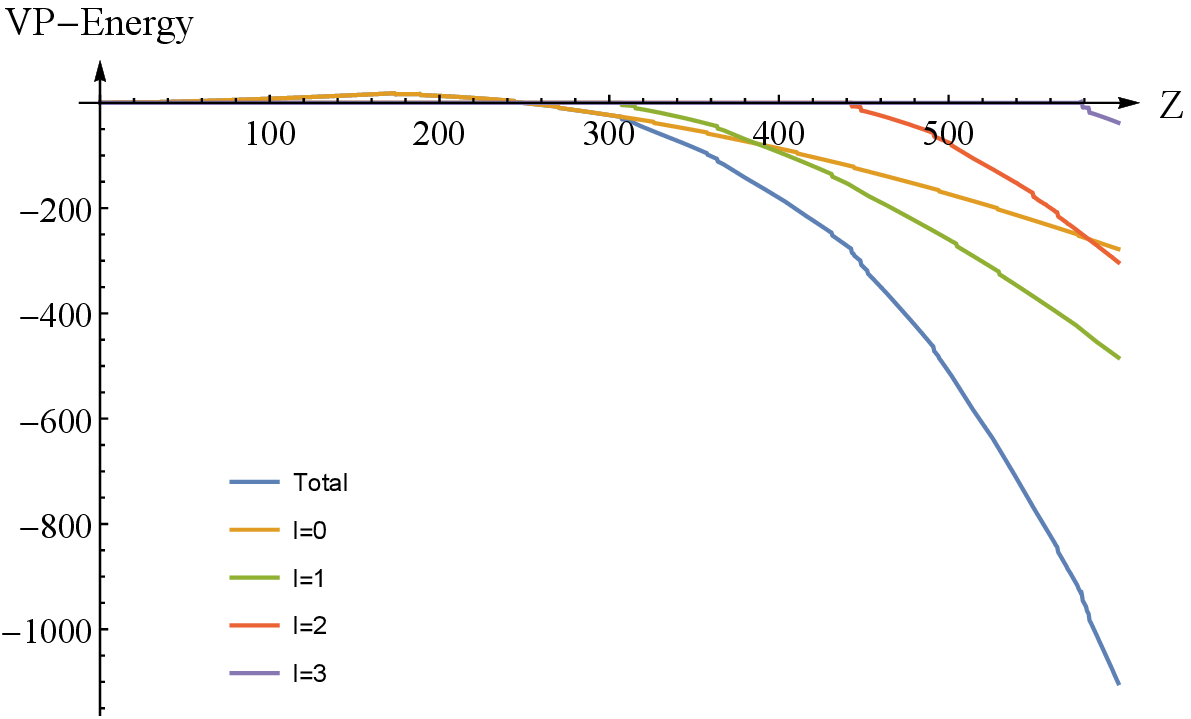}
}
\caption{(Color online) $(l+1)\,I(l)\, , (l+1)\,S(l)\, , \E_{VP,l}^{ren}(Z)$ and  $I_{tot}=\sum_{l=0}^{3}(l+1)\,I(l)\, , S_{tot}=\sum_{l=3}^{4}(l+1)\,S(l)\, , \E_{VP,tot}^{ren}(Z)=\sum_{l=0}^{3}\E_{VP,l}^{ren}(Z) $ on the interval  $0 < Z < 600$ . }
	\label{Int-Sum-VP-600}	
\end{figure*}

For greater clarity  of results  we restrict this presentation to the case of charged sphere (\ref{1.5a}) on the interval $0 < Z < 600$ with the numerical coefficient in eq. (\ref{1.8}) chosen as\footnote{With such a choice for a charged ball with $Z=170$ the lowest $1s_{1/2}$-level lies precisely at $\e_{1s}=-0.99999$. Furthermore, it is quite close to $1.23$, which is the most commonly used coefficient in heavy nuclei physics.}
\beq
\label{6.1}
R(Z)=1.228935\, (2.5\, Z)^{1/3} \ \hbox{fm} \ .
\eeq
On this interval of $Z$  the main contribution comes from the partial channels with $l=0\,, \dots\,,3$, in which a non-zero number of discrete levels has already reached the lower continuum. In Fig.\ref{Int-Sum-VP-600} there are shown the specific features of partial  phase integrals,
partial sums over discrete levels and renormalized partial VP-energies.

 As it follows from  Fig.\ref{Int-Sum-VP-600}(a),  phase integrals increase monotonically with growing  $ Z $ and are always positive. The clearly seen  bending, which is in fact nothing else but the negative jump in the derivative of the curve, for each channel  starts  when the first discrete level from this channel attains the lower continuum.   The origin of this effect  is the sharp jump by  $\pi$ in $\d_{tot}(l,k)$ due to the resonance just born. In each $l$-channel such effect is most pronounced for the first level diving, which takes place at  $Z_{cr,1}\simeq 173.6$ for $l=0$, at  $Z_{cr,5}\simeq 307.4$ for $l=1$, at  $Z_{cr,15}\simeq 442.7$ for $l=2$ and at $Z_{cr,26}\simeq 578.6$ for $l=3$.

  The subsequent levels diving also leads to jumps in the derivative of $I(l)$, but they turn out to be much less pronounced already, since with increasing $Z_{cr}$ there shows up the effect of  ``catalyst poisoning'' --- just below the threshold of the lower continuum for $Z=Z_{cr}+\D Z \ , \ \D Z \ll Z_{cr}$, the  resonance broadening and its rate of further diving into the lower continuum are exponentially slower in agreement  with the well-known result~\cite{Zeldovich1972}, according to which the resonance width just under the threshold behaves like $\sim \exp \( - \sqrt{Z_{cr}/\D Z}\)$. This effect leads to that for each subsequent resonance the region of the phase jump by $\pi$ with increasing  $Z$ grows exponentially slower, and derivative of the phase integral changes in the same way. If not for this effect, then every next level reaching the lower continuum would lead to the same negative jump in the derivative as the first one, and the phase integral curve in the overcritical region would have a continuously  increasing negative curvature with all the ensuing consequences for the rate of decrease of the total vacuum energy $\E_{VP}^{ren}(Z) $. At the same time, before first levels diving the curves of $I(l)$ in all the partial  channels reveal an almost quadratical growth. For the last channel with $l=3$ this growth takes place during almost the whole interval $0 < Z < 600$, since the first level diving in this channel occurs at $Z \simeq 578.6$, which is very close to $Z=600$ and simultaneously  large enough  already to be subject of  Zeldovich-Popov effect discussed just above.

The  behavior of the total bound energy of discrete levels per each partial channel $S(l)$  is shown in Fig.\ref{Int-Sum-VP-600}(b). $S(l)$  are  discontinuous functions with jumps emerging each time, when the charge of the source reaches the subsequent critical value and the corresponding discrete level dives into the lower continuum. At this moment  the bound energy loses exactly two units of $mc^2 $, which in the final answer must be multiplied by the degeneracy factor $l+1$. Due to this factor the jumps in the curves of $S(l)$ are more pronounced with growing $l$. On the intervals between two neighboring  $ Z_{cr} $ the bound energy is always positive and increases monotonically, since there grow the bound energies of all the  discrete levels, while on the intervals between $Z=0$ up to first levels diving their growth is almost quadratic.

The partial VP-energies $ \E_{VP,l}^{ren}(Z) $ are shown in Fig.\ref{Int-Sum-VP-600}(c). Note that the behavior of $s$-channel is different from the others, since in this channel the structure of the renormalization coefficient  $ \z_0 $ differs from those with $l \geqslant 1$ by the   perturbative (Uehling) contribution to VP-energy $\E^{(1)}_{VP}(Z)$.
It is indeed the latter term in the total VP-energy, which is responsible for an almost quadratic growth with $Z$ of VP-energy on the interval $0 < Z < Z_{cr,1}$.
The global change in the behavior of $ \E_{VP,0}^{ren}(Z) $ from the perturbative quadratic growth for $Z \ll Z_{cr,1}$, when the dominant contribution comes from  $\E^{(1)}_{VP}(Z)$, to the regime of decrease into the negative range with increasing $Z$ beyond  $Z_{cr,1}$ is shown  below in Sect.VIII in Figs.\ref{VP240},\ref{VPball240}.
At the same time, in the  channels  with $l\geqslant 1$ the quadratic perturbative contribution  is absent. Therefore upon renormalization  (\ref{3.58}-\ref{3.60}), which removes the quadratic component in $I(l)$ and $S(l)$, just after first level diving $\E_{VP,l}^{ren}(Z) $ reveal with increasing $Z$ a well-pronounced decrease into the negative range.

To specify the dependence of partial VP-energies $\E_{VP,l}^{ren}(Z) $ on $l$ below there are given two tabs with the required information.
\begin{table*}[ht!]
\caption{Dependence of partial VP-energies on $l$ for $Z=300$ with and without multiplicity factor $l+1$.}
\begin{center}
	\begin{tabular}{|c|c|c|c|c|}
		\hline $l$ & 0 & 1 & 2 & 3  \\[2pt]
			\hline  $ \E_{VP,l}^{ren}(Z) $  &  -23.107040 &  -0.172432 & -0.028663 & -0.007796  \\[2pt]
	\hline  $ \E_{VP,l}^{ren}(Z)/(l+1) $  & -23.107040 & -0.086216 & -0.009554 & -0.001949    \\[2pt]		 \hline Number of dived levels & 8 & 0 & 0 & 0   \\ [-2pt]
		with multiplicity factor $2j+1$ &&&&\\
				\hline
	\end{tabular}\label{t300}
\end{center}
\end{table*}
\begin{table*}[ht!]
\caption{Dependence of partial VP-energies on $l$ for $Z=600$ with and without multiplicity factor $l+1$.}
\begin{center}
	\begin{tabular}{|c|c|c|c|c|c|c|}
		\hline $l$ & 0 & 1 & 2 & 3 & 4 & 5 \\[2pt]
			\hline  $  \E_{VP,l}^{ren}(Z) $  & -277.969408 & -483.912426 & -302.902746 & -37.850795 & -0.035715 & -0.014984  \\[2pt]
\hline  $ \E_{VP,l}^{ren}(Z)/(l+1) $  & -277.969408 & -241.956213 & -100.967582 & -9.462698 & -0.007143 & -0.002497   \\[2pt]	
	\hline Number of dived levels & 22 & 36 & 36 & 16 & 0 & 0  \\ [-2pt]
		with multiplicity factor $2j+1$ &&&&&\\
				\hline
	\end{tabular}\label{t600}
\end{center}
\end{table*}
From these tabs it is clearly seen that  the main contribution to the total VP-energy for the given $Z$ is produced indeed  by  those partial channels, where a non-zero number of discrete levels has already reached the lower continuum. At the same time, the number of dived levels and the amount of contribution to VP-energy per partial channel do not correlate precisely. It is only the case  $Z=300$, when all the dived levels belong to $s$-channel, and indeed this channel dominates in the total VP-energy. However, already for $Z=600$ the situation is different and with further growth of $Z$ this discrepancy becomes more and more pronounced. Note also that  vanishing of $\E_{VP,l}^{ren}(Z) $ in the channels with  $l$ large enough to prevent levels diving proceeds even faster than predicted by WKB-analysis performed in Sect.III.

The final answers for the total VP-energy achieved this way are
\beq
\label{6.2}
\E_{VP}^{ren}(300)=- 23.3 \ , \quad \E_{VP}^{ren}(600)=-1102.7 \ .
\eeq
On this interval of $Z$ the decrease of the total VP-energy into the negative range proceeds very fast,  but with further growth of $Z$ the decay rate becomes smaller. In particular,
\beq
\label{6.3}
\E_{VP}^{ren}(1000)=- 8398.6 \ .
\eeq
while the reasonable estimate of asymptotical behavior of $\E_{VP}^{ren}(Z)$ as a function of $Z$, achieved from the interval $1000 < Z < 3000$, reads
\beq
\label{6.4}
\E_{VP}^{ren}(Z) \sim - Z^4/R(Z) \ .
\eeq

\section{Charged ball Coulomb source configuration}

Now let us consider the main differences in  VP-energy  between the models of uniformly charged  sphere and ball. Even more appropriate there should be the spherical layer model with uniform charge distribution in the range $R_2\leqslant r \leqslant R_1$ and the potential
\begin{multline}\label{7.1}
V(r)=-Q\, \[ \tt(R_2 - r) { 3\,\(R_1 + R_2\) \over 2\,\(R_1^2 + R_1 R_2 + R_2^2\) } \ + \right. \\ \left.
+ \   \tt(R_2 \leqslant r \leqslant R_1) {3\, R_1^2/2 - r^2/2 - R_2^3/r \over
    R_1^3 - R_2^3 } \ + \right. \\ \left. + \tt(r - R_1) {1 \over r} \]  \ ,
\end{multline}
which includes both sphere and ball as  limiting cases. However, it turns to be excessive, since the results for the layer lie always in between those for sphere and ball. Moreover, even the conversion factor for converting results from sphere and ball to layer can be estimated quite reliably. Therefore in what follows we restrict to the charged ball configuration (\ref{1.6}).

Since the models of charged sphere and ball differ only by the shape of the potential inside the source, the general approach to evaluation the VP-effects for the ball configuration  including the techniques of phase integral method  and renormalization  via fermionic loop remains unchanged. The main difference lies in the inner solutions of DC problem. In the case of the sphere they can be found in analytical form, but  for the ball such option is absent.
In the last case the most straightforward approach is to find numerically for given $Z$ the inner solutions of DC  via direct solving corresponding DE. But it turns out to be time consuming, especially for the phase integrals (\ref{3.42}), since in the latter case for $Z$ lying in the range $100 \leqslant Z \leqslant 1000$ one needs to calculate numerically $\d_{tot}(l,k)$ for  $l$ up to $l_{max}\simeq 10$ and for $k$ up to  $k_{max} \simeq 10000-100000$. For the sets of $(l, k)$ close to $(l_{max}, k_{max})$ the inner solutions are rapidly oscillating functions akin to  Bessel-type ones, but with more complicated behavior, calculation of which with given accuracy  takes time. However, since the inner potential in (\ref{1.6}) is just a parabola, it is possible to apply another method, based on approximation of the inner potential  via step-like function.  Such approximation works quite well and already for a step-like function consisting of  $100 < N < 1000$ separate segments yields the answer, which coincides with exact numerical solution with  the precision required. The main advantage of such approach is that the existing nowadays soft- and hard-ware is able to solve  the appearing recurrences directly in analytical form and so avoid the problems with numerical solving DE.

The simplest grid for such approximation of $V(r)$ inside the ball, that means for  $0 \leqslant r \leqslant R$, can be chosen as follows
\beq\label{7.2}
R_i=\sqrt{{i \over N}}\, R  \ , \quad V_i=- {3 Q \over 2 R } + { i \over 2 N}{Q \over R} \ ,
\eeq
where $1 \leqslant i \leqslant N$ is the numerator of separate segments with constant values of the potential $V_i$. Note that the grid is uniform in the step between subsequent $V_i$
\beq\label{7.3}
\D V=V_i - V_{i-1}={1 \over 2 N}{Q \over R} \ ,
\eeq
but not uniform in the length of subsequent segments in the radial variable. This is made specially to optimize the approximation of parabolic behavior of $V(r)$  on this interval. Moreover, such grid approximates $V(r)$ from above and coincides with $V(r)$ only at points $R_i$. Here it should be noted that there exists a plethora of  other types of parabola approximation via step-like grids, which are slightly different in the rate of convergence to the exact answer, but for our purposes there isn't any need to dive into such details here.

Within such a grid the inner solutions for DC problem in the ball take the following form. As for the case of the sphere, we start with assembling the total phase $\d_{tot}(l,k)$ from $(u_l\, ,q_l)$- and $(p_l\, ,v_l)$-components using the crossing symmetry to restore the contribution from $(p_l\, ,v_l)$-pair via $(u_l\, ,q_l)$-pair. For the latter in the upper continuum with $\e(k)=\sqrt{k^2+1} \geqslant 1$ upon introducing
\beq\label{7.4}
\x_i(k)=\sqrt{(\e(k) -V_i)^2-1}
\eeq
the corresponding solutions on the i-th radial segment ($i=1\,,\dots , N$) with $R_{i-1} \leqslant r \leqslant R_i $ should be written as
\begin{widetext}
\beq
\label{7.5}
\left\lbrace\bal
& u_l(k,r)=\sqrt{\e(k)-V_i+1}\, \[J_{l+1/2} \(\x_i(k)\, r \) + \s_i(l,k)\,Y_{l+1/2} \(\x_i(k)\, r \) \]/\sqrt{r} \ ,\\
& q_l(k,r)=-\sqrt{\e(k)-V_i-1}\, \[J_{l+3/2} \(\x_i(k)\, r \) + \s_i(l,k)\,Y_{l+3/2} \(\x_i(k)\, r \)\] /\sqrt{r} \ ,
\eal\right.
\eeq
with $Y_{\n}(z)$ being the Neumann function.

The  coefficients $\s_i$ in eqs. (\ref{7.5}) are determined from the following recurrence relations
\beq\label{7.7}
 \s_1(l,k)=0 \ ,
\eeq
\beq\label{7.8}
\s_i(l,k)=- U_1(i) /U_2(i) \ ,  \quad 2 \leqslant i \leqslant N \ ,
\eeq
where
 \beq\begin{gathered}
\label{7.9}
\left\lbrace\bal
& U_1(i)=\sqrt{\e(k)-V_{i-1}+1}\,\sqrt{\e(k)-V_i-1}\, \[J_{l+1/2} \(\x_{i-1}(k)\, R_{i-1} \) + \s_{i-1}(l,k)\,Y_{l+1/2} \(\x_{i-1}(k)\, R_{i-1} \) \]\,J_{l+3/2}\(\x_{i}(k)\, R_{i-1} \) \  - \\
& - \ \sqrt{\e(k)-V_{i-1}-1}\,\sqrt{\e(k)-V_i+1}\,  \[J_{l+1/2} \(\x_{i-1}(k)\, R_{i-1} \) + \s_{i-1}(l,k)\,Y_{l+1/2} \(\x_{i-1}(k)\, R_{i-1} \) \]\,J_{l+1/2}\(\x_{i}(k)\, R_{i-1} \) \ , \\
& U_2(i)=\sqrt{\e(k)-V_{i-1}+1}\,\sqrt{\e(k)-V_i-1}\, \[J_{l+1/2} \(\x_{i-1}(k)\, R_{i-1} \) + \s_{i-1}(l,k)\,Y_{l+1/2} \(\x_{i-1}(k)\, R_{i-1} \) \]\,Y_{l+3/2}\(\x_{i}(k)\, R_{i-1} \) \  - \\
& - \ \sqrt{\e(k)-V_{i-1}-1}\,\sqrt{\e(k)-V_i+1}\, \[J_{l+1/2} \(\x_{i-1}(k)\, R_{i-1} \) + \s_{i-1}(l,k)\,Y_{l+1/2} \(\x_{i-1}(k)\, R_{i-1} \) \]\,Y_{l+1/2}\(\x_{i}(k)\, R_{i-1} \)  \ ,
\eal\right.
\end{gathered}
\eeq
Upon solving the recurrences (\ref{7.7}-\ref{7.9}) one finds the set of coefficients $\s_i$, and so the solutions for $(u_l\, ,q_l)$-pair in each segment $R_{i-1} \leqslant r \leqslant R_i $ of the grid, which are sewn together by continuity at points $R_i$ by means of the coefficients $\s_i$. As a result, for the matching point $R=R(Z)$  between inner and outer solutions of DC  one obtains from the  inside
\beq
\label{7.11}
\left\lbrace\bal
& u_l(k,R)=\sqrt{\e(k)+V_0+1}\, \[J_{l+1/2} \(\x R \) + \s_N(l,k)\,Y_{l+1/2} \(\x R \) \]/\sqrt{R} \ ,\\
& q_l(k,R)=-\sqrt{\e(k)+V_0-1}\, \[J_{l+3/2} \(\x R \) + \s_N(l,k)\,Y_{l+3/2} \(\x R \)\] /\sqrt{R} \ ,
\eal\right.
\eeq
where as in Sect.IV $\x=\x(k)=\sqrt{(\e(k) +V_0)^2-1} $, $V_0=Q/R$.
\end{widetext}
At the same time, the outer solutions  remain the same as  for the sphere. Therefore by means of the expression (\ref{7.11}) the procedure of stitching by continuity between inner and outer solutions of DC  for the charged ball in the upper continuum proceeds very simply by means of the following replacement in the matching coefficients $\l^+_{uq}(l,k)$ and $N_R(l,k)$, defined in  (\ref{4.7}) and (\ref{4.33}) for the case of charged sphere,
\beq
\label{7.13}
\left\lbrace\bal
& J_{l+1/2} \(\x R \) \to J_{l+1/2} \(\x R \) + \s_N(l,k)\,Y_{l+1/2} \(\x R \)  \ ,\\
& J_{l+3/2} \(\x R \) \to J_{l+3/2} \(\x R \) + \s_N(l,k)\,Y_{l+3/2} \(\x R \) \ .
\eal\right.
\eeq
Thereafter, the phase shifts $\d^+_{uq}(l,k)$ for the  ball  are obtained from those of the sphere via replacing the  matching coefficients $\l^+_{uq}(l,k)$ and $N_R(l,k)$ by the new ones, determined through substitutions (\ref{7.13}).

In the lower continuum with $\e(k)=-\sqrt{k^2+1} < -1$ for each segment $R_{i-1} \leqslant r \leqslant R_i $ of the grid the  half-axis $0\leq k \leq \inf$  should be again divided in 3 intervals $0\leq k \leq k_1(i)\, , \ k_1(i) \leq k \leq k_2(i)\, , \ k_2(i) \leq k \leq \inf$, where
\beq\begin{gathered}\label{7.15}
k_1(i)=\sqrt{(1+V_i)^2-1}< k_2(i)=\sqrt{(1-V_i)^2-1} \ , \\  1\leqslant i \leqslant N \ ,
\end{gathered}\eeq
\begin{widetext}
while $k_1(N)\, , k_2(N)$ coincide with $k_1\,, k_2$,  defined earlier  in eq. (\ref{4.19}). Upon introducing
\beq\label{7.16}
\tx_i(k)=\sqrt{1-(\e(k) -V_i)^2}
\eeq
the solutions for the i-th segment $R_{i-1} \leqslant r \leqslant R_i $ in this case are written as follows
 \beq\begin{gathered}
\label{7.17}
u_l(k,r)=\sqrt{\Big||\e(k)|+V_i-1\Big|}\,\left\lbrace\bal
&  J_{l+1/2}\(\x_i(k)\, r\) + \s_{i1}(l,k)\,Y_{l+1/2}\(\x_i(k)\, r\) \ , \quad  0\leq k \leq k_1(i) \ ,  \\
&  I_{l+1/2}\(\tx_i(k)\, r\) + \s_{i2}(l,k))\, K_{l+1/2}\(\tx_i(k)\, r\) \ , \quad  k_1(i) \leq k \leq k_2(i)  \ , \\
&  J_{l+1/2}\(\x_i(k)\, r\) + \s_{i3}(l,k)\,Y_{l+1/2}\(\x_i(k)\, r\) \ , \quad  k_2(i) \leq k \leq \inf \ ,
\eal\right.\\
\end{gathered}\eeq
 \beq\begin{gathered}
\label{7.18}
q_l(k,r)=\sqrt{\Big||\e(k)|+V_i+1\Big|}\,\left\lbrace\bal
& -\( J_{l+3/2}\(\x_i(k)\, r\) + \s_{i1}(l,k)\,Y_{l+3/2}\(\x_i(k)\, r\)\) \ , \quad  0\leq k \leq k_1(i) \ ,  \\
&  I_{l+3/2}\(\tx_i(k)\, r\) - \s_{i2}(l,k)\, K_{l+3/2}\(\tx_i(k)\, r\) \ , \quad  k_1(i) \leq k \leq k_2(i)  \ , \\
&  J_{l+3/2}\(\x_i(k)\, r\) + \s_{i3}(l,k)\,Y_{l+3/2}\(\x_i(k)\, r\) \ , \quad  k_2(i) \leq k \leq \inf \ ,
\eal\right.\\
\end{gathered}\eeq
\end{widetext}
Stitching the solutions (\ref{7.17},\ref{7.18}) separately  by continuity at points $R_i\,, i=1, \dots , N-1$, one obtains the recurrence relations for the coefficients $\s_{ij}(l,k)\, , j=1,2,3$, which are solved with initial conditions
\beq\label{7.21}
 \s_{1j}(l,k)=0 \ , \quad j=1,2,3 \ .
\eeq
Afterwards the phase shifts $\d^-_{uq}(l,k)$ for the  ball  are obtained from those for  the sphere by the same procedure of replacing the  matching coefficients $\l^-_{uq}(l,k)$ and $N_R(l,k)$ by the new ones, determined through the set of substitutions, similar to (\ref{7.13}). The main difference is that this procedure should be now implemented separately for each of 3 intervals $0\leq k \leq k_1(N)\, , \ k_1(N) \leq k \leq k_2(N)\, , \ k_2(N) \leq k \leq \inf$.

Proceeding this way, one obtains for $\d_{tot}(l,k)$ within the step-like approximation (\ref{7.2}) of the potential inside the ball  an analytical expression, which in explicit form looks quite cumbersome, but poses no problems  for evaluation in any point  $(l,k)$. Moreover, it can be easily verified by means of  the WKB-analysis, presented in Sect. III,  that the leading-order asymptotics for $k \to \inf$ of the total phase for the ball is still $O(1/(k R)^3)$. $\d_{tot}(l,k \to 0)$ also remains finite, but now instead of  compact explicit answers (\ref{4.53},\ref{4.55}) for the sphere in the ball case they are given by quite lengthy expressions, which nevertheless allow for an effective numerical evaluation. So the calculation of phase integrals (\ref{3.42}) for the ball configuration within the step-like approximation is implemented for any $l$ without any serious loss of time and accuracy compared to  the sphere.

The discrete spectrum for the charged ball configuration is found from the corresponding eqs. for the sphere (\ref{5.3},\ref{5.5}) with the replacement (\ref{7.13}), where instead of $\e(k)=\sqrt{k^2+1} \geqslant 1$ in all the expressions including the recurrences (\ref{7.7}-\ref{7.9}) one should replace $\e(k) \to \e\, , \ |\e|<1$.
 Proceeding further this way, for the levels with $\e=-1$ and so for the corresponding critical charges one obtains the equations similar to eqs.(\ref{5.9},\ref{5.10}), where instead of $J_{\pm}$ defined in (\ref{5.10}) one should insert now the combinations
 \beq\begin{gathered}
\label{7.22}
 J_{\pm} + \g_N(l)\,Y_{\pm} \ ,
 \end{gathered}\eeq
where
 \beq\begin{gathered}
\label{7.23}
Y_-=Y_{l+1/2}(z_1) \ , \ Y_+=Y_{l+3/2}(z_1) \ ,
 \end{gathered}\eeq
$z_1$  is defined in (\ref{5.10}), while $\g_N(l)$ are found from the recurrences (\ref{7.7}-\ref{7.9}) with the replacement $\e(k) \to -1$.  All the square roots in recurrences remain well-defined, since the levels attain the lower threshold only for $V_0>2$, hence for $|V_i|\geqslant V_0 >2$ for all $1\leqslant i \leqslant N$. As it was already stated in Sect. IV, in  the case under consideration $V_0$  exceeds actually several dozens or even hundreds, so the condition  $V_0>2$ is always satisfied.

Applying such procedure to the search for critical charges in the ball configuration with the same relation (\ref{6.1}) for $R(Z)$, one finds for the grid with $N=1000$ segments the results, which match  those obtained via direct numerical  solution of DE in all the characters given below. In particular, for the lowest $1s$- and  $2p$-levels one finds $ Z_{cr,1} \simeq 170.0048 $ and $ Z_{cr,2} \simeq 183.0756  $ correspondingly. Compared to the case of the sphere with $ Z_{cr,1} \simeq 173.613 $ and $ Z_{cr,2} \simeq 188.5497  $ in the ball case the diving points reveal a small  shift to the left by $ \D Z_{cr,1} \simeq 3.608 $ and $ \D Z_{cr,2} \simeq  5.4741 $. Actually, the last circumstance is the common feature of all the dived levels in the ball case, but with each subsequent diving point it becomes less and less pronounced.

So the step-like approximation (\ref{7.2}) provides quite effective dealing with the discrete spectrum for the ball configuration too. Further steps in order to find the bound states contribution (\ref{5.15}) to the partial VP-energy (\ref{3.31}) for the ball are the same as described in  Sect.V for  the sphere \footnote{The procedure of "quasi-exact"\, summation over discrete spectrum  in such DC problems,  which takes account for  nonzero size of the Coulomb source in  essentially non-perturbative way, can be extended for the ball case too.}. The general behavior of obtained this way $\E^{ren}_{VP}(Z)$ for the ball configuration looks quite similar to that for the sphere with two  main differences. The first one is that all the negative jumps in $\E^{ren}_{VP}(Z)$, caused by discrete levels diving into the lower continuum, are slightly shifted to the left. The second is that the general magnitude of VP-energy in the ball case is approximately $(6/5) \times$VP-energy for the sphere. Note that the same result (\ref{2.22}) holds for the perturbative contributions $\E^{(1)}_{VP}(Z)$ to VP-energy under condition (\ref{2.18}).

\section{Spontaneous positron emission}

\begin{figure*}[ht!]
\subfigure[]{
		\includegraphics[width=\columnwidth]{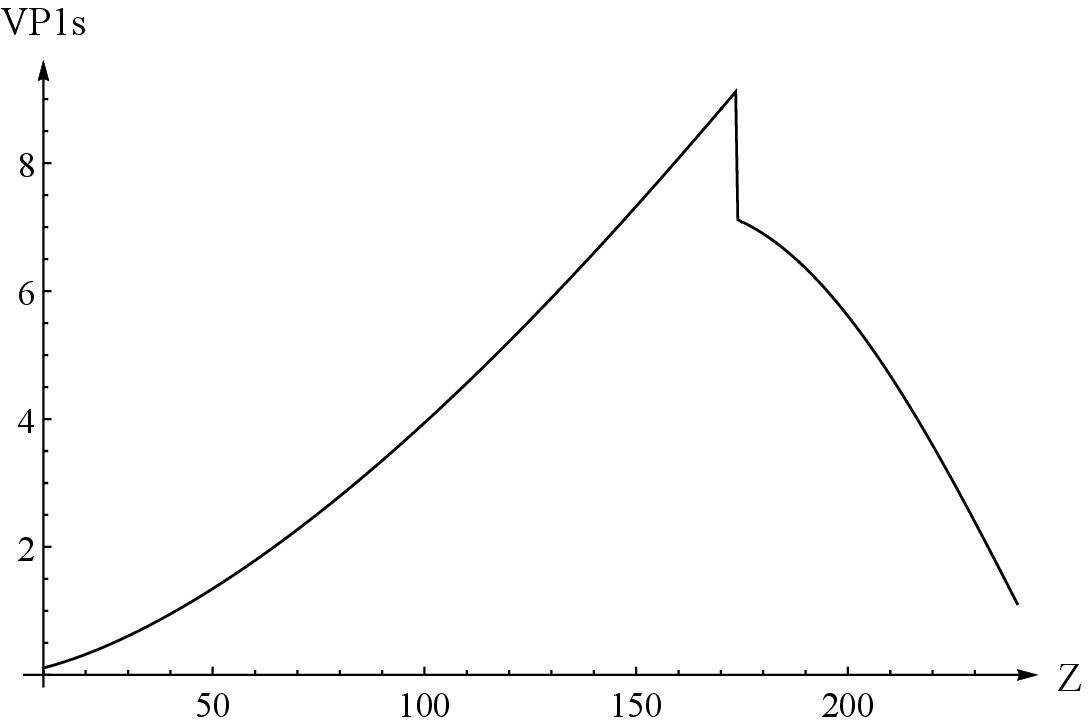}
}
\hfill
\subfigure[]{
		\includegraphics[width=\columnwidth]{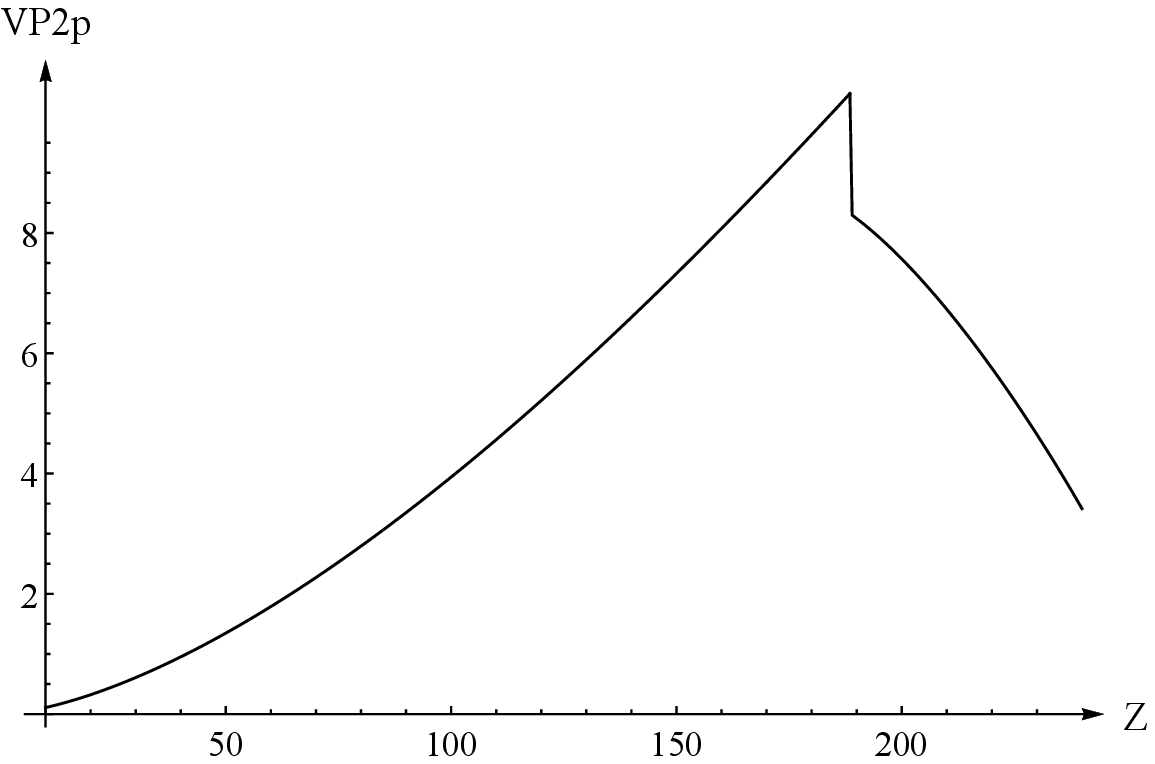}
}
\vfill
\subfigure[]{
		\includegraphics[width=1.2\columnwidth]{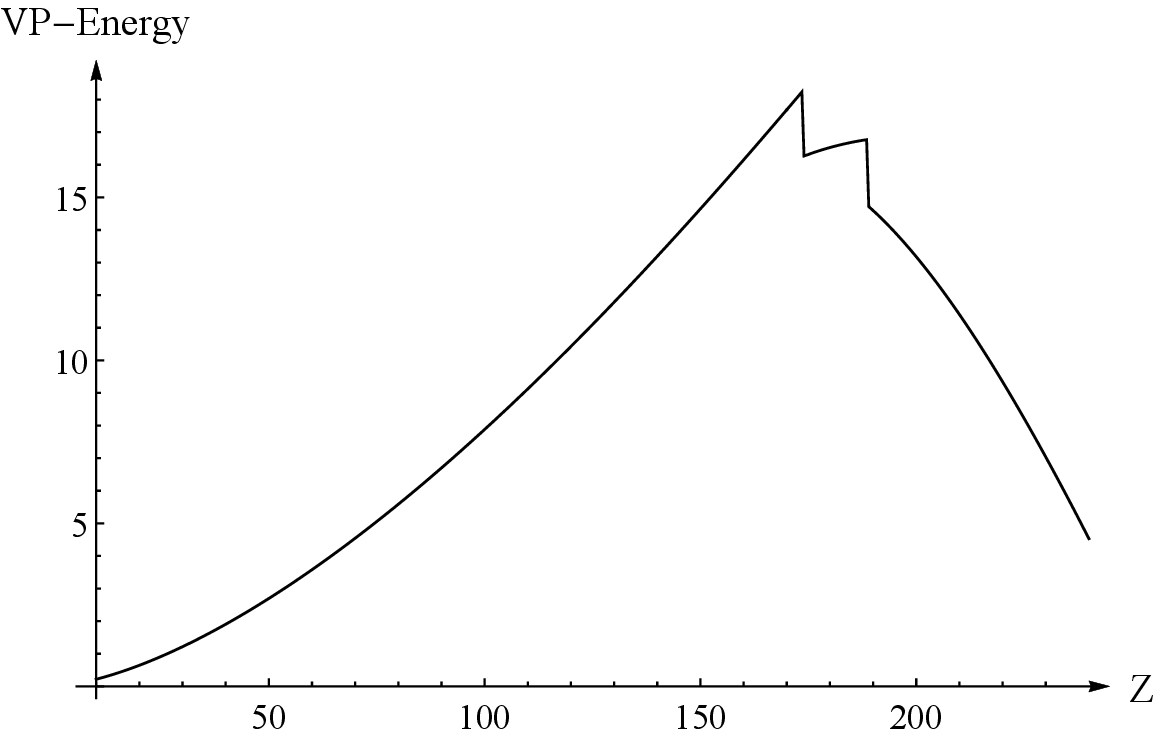}
}
\caption{$ \E_{VP,+}^{ren}(Z)\,, \ \E_{VP,-}^{ren}(Z)\,, \ \E_{VP}^{ren}(Z)  $  on the interval  $10 \leqslant Z \leqslant 240$ (sphere).  }
	\label{VP240}	
\end{figure*}

\begin{figure*}[ht!]
\subfigure[]{
		\includegraphics[width=\columnwidth]{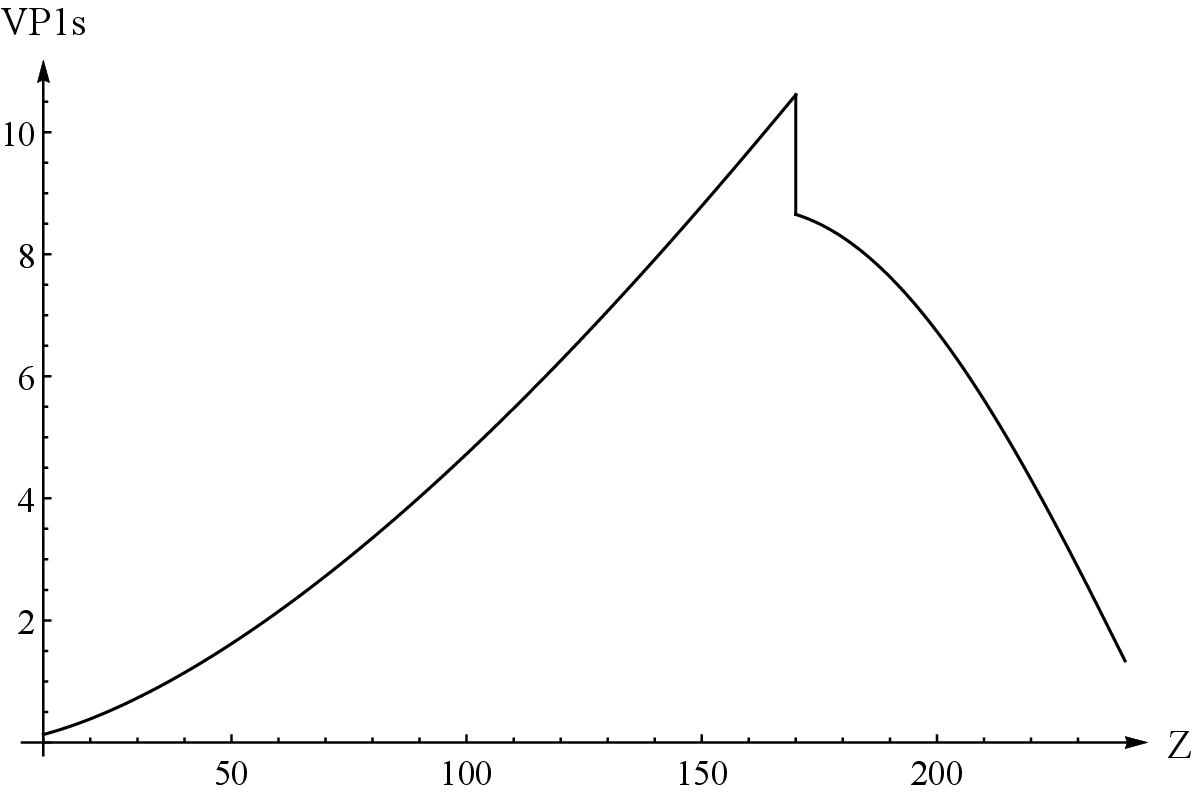}
}
\hfill
\subfigure[]{
		\includegraphics[width=\columnwidth]{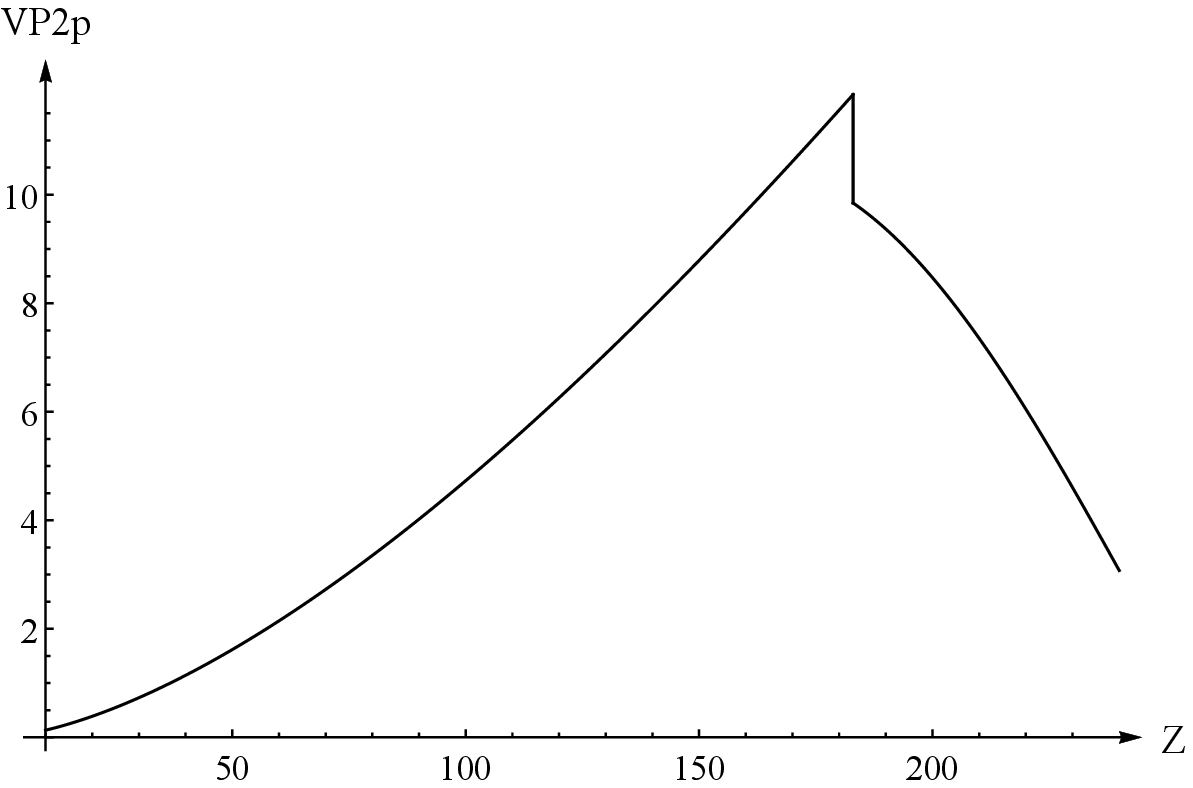}
}
\caption{$ \E_{VP,+}^{ren}(Z)\,, \ \E_{VP,-}^{ren}(Z)\,$  on the interval  $10 \leqslant Z \leqslant 240$ (ball).}
	\label{VPball240}	
\end{figure*}

Now --- having dealt with the general properties of VP-effects  this way --- let us consider more thoroughly  the interval $10 \leqslant Z \leqslant 240$, when only two first levels $1s_{1/2}\,, \ 2p_{1/2}$ with opposite parity $(\pm)$ have already dived into the lower continuum at $Z_{cr,1}$ and $Z_{cr,2}$. This issue is of special interest in view of intimate relation  between spontaneous positron emission and lepton number conservation~\cite{Krasnov2022}. The plots of fixed parity $\E_{VP,\pm}^{ren}(Z)$ for the charged sphere and ball source configurations are presented in Figs.\ref{VP240},\ref{VPball240}, while their similarity has been discussed just before.

General theory (see Refs.~\cite{Greiner1985a,Plunien1986,Greiner2012, Rafelski2016} and citations therein) based on the  framework~\cite{Fano1961}, predicts that  after diving such a level transforms into a metastable state with lifetime $\sim 10^{-19}$ sec, afterwards there occurs the spontaneous positron emission accompanied with vacuum shells formation.  An important point here is  that due to spherical symmetry of the source during this process all the angular quantum numbers and parity of the dived level are preserved by metastable state and further by positrons created. Furthermore, the spontaneous emission of positrons should be caused solely by VP-effects  without any other channels of energy transfer.

So for each parity  the energy balance suggests the following picture of this process. The rest mass of positrons is created just after level diving via negative jumps in the VP-energy at corresponding $Z_{cr,i}$, which are exactly equal to $2 \times mc^2$ in accordance with two possible spin projections. However,  to create a real positron scattering state, which provides  a sufficiently large  probability of being in the immediate vicinity of the Coulomb source, it is not enough due to electrostatic reflection  between positron and Coulomb source. This circumstance has been outlined first in Refs. ~\cite{Reinhardt1981,*Mueller1988,Ackad2008}, and explored quite recently with more details in Refs.~\cite{Popov2018,*Novak2018,*Maltsev2018,
Maltsev2019,*Maltsev2020}.

\begin{figure*}[ht!]
\subfigure[]{
		\includegraphics[width=\columnwidth]{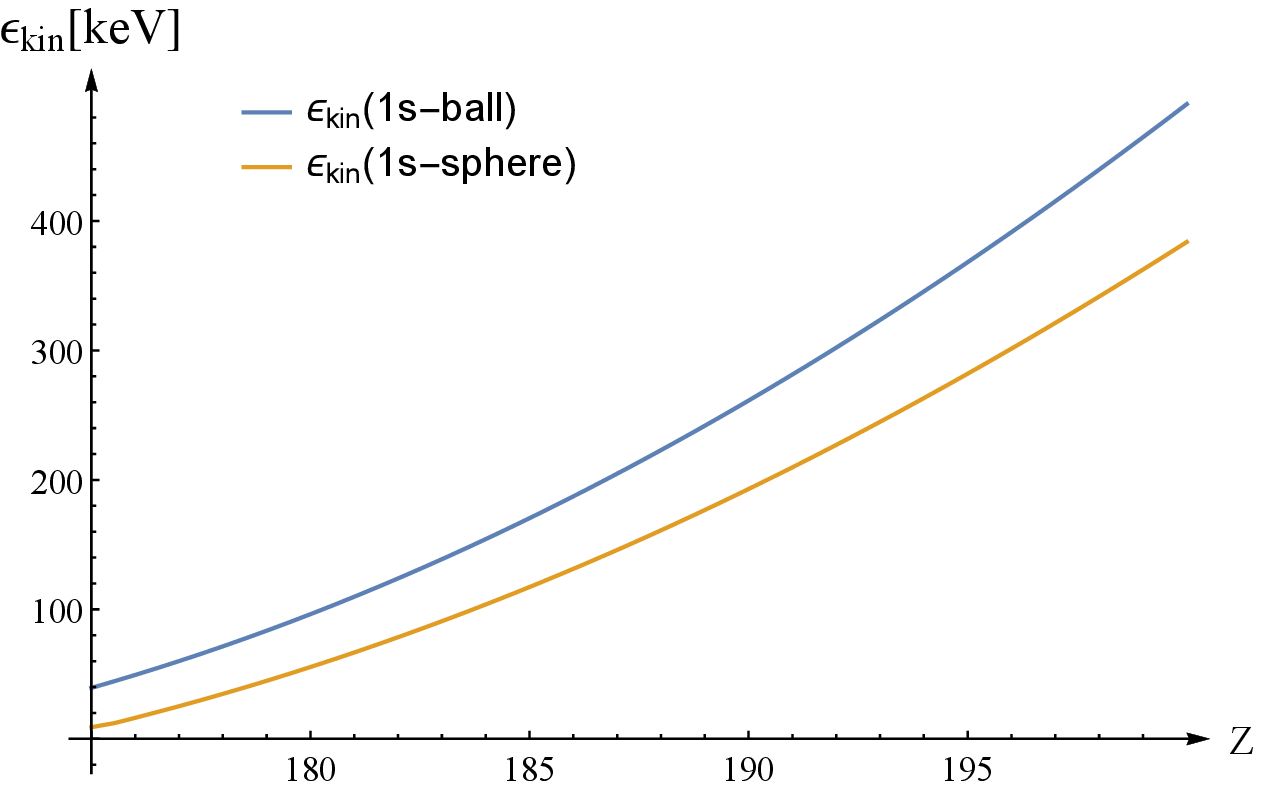}
}
\hfill
\subfigure[]{
		\includegraphics[width=\columnwidth]{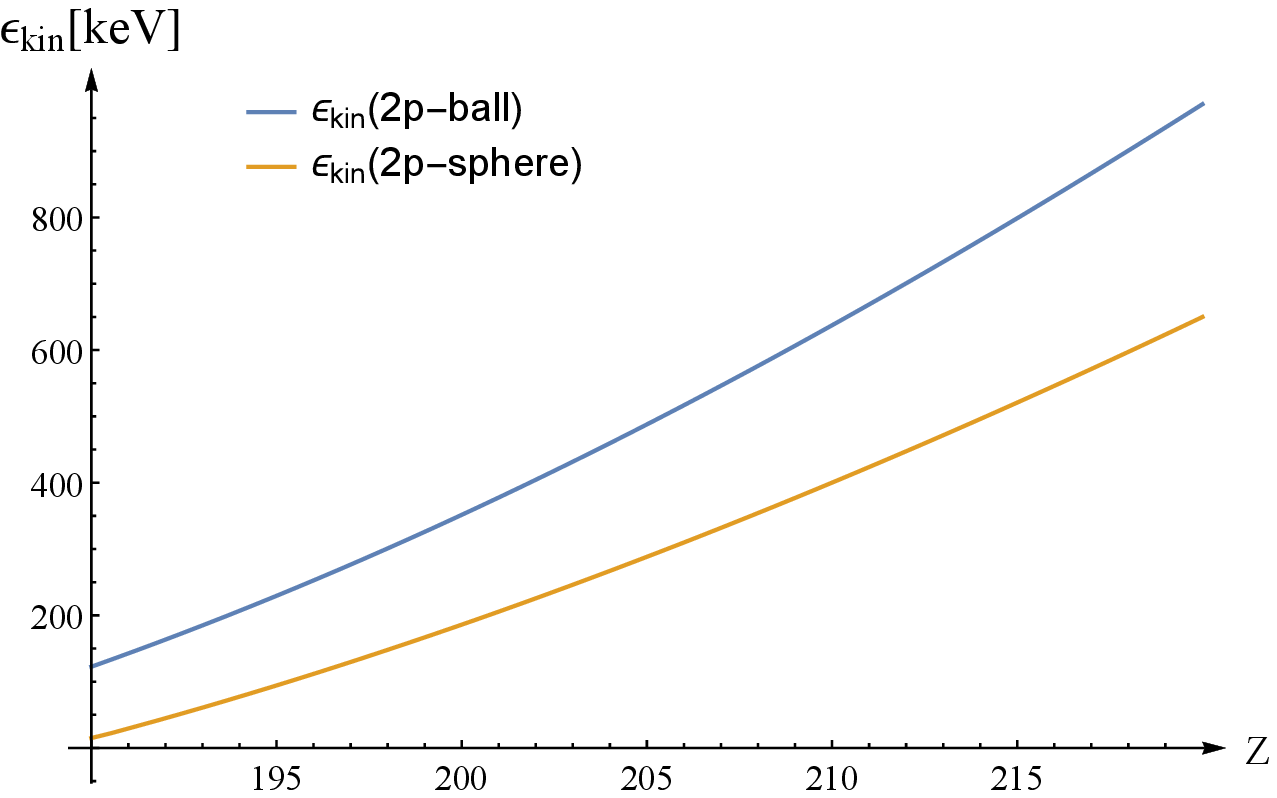}
}
\caption{(Color online)\, $ \e_{kin}(Z)$ with definite parity on the pertinent intervals  of $Z$ for sphere and  ball source configurations and (a): 1s-channel (even); (b): 2p-channel (odd).}
	\label{ekin}	
\end{figure*}

From the analysis presented above there follows that to supply the emerging vacuum positrons with corresponding reflection energy, an additional decrease of  $\E_{VP}^{ren}(Z)$ in each parity channel is required. So we are led to the following energy balance conditions for spontaneous positrons emission after level diving at $Z_{cr,i}$
\beq
\label{6.5}
\E_{VP,\pm}^{ren}(Z_{cr,i}+0)  - \E_{VP,\pm}^{ren}(Z) =2\,\e_{kin}(Z)  \ ,
\eeq
where $\e_{kin}(Z)$ is the positron kinetic energy and it is assumed that for each parity positrons are created in pairs with opposite spin projections. In agreement with Refs.~\cite{Popov2018,*Novak2018,*Maltsev2018,
Maltsev2019,*Maltsev2020}, in the considered range of $Z$ the spontaneous positrons are limited to the energy $0< \e_{kin}(Z) < 800$ (see Fig.\ref{ekin}), while the related natural resonance widths do not exceed a few KeV~\cite{Marsman2011,*Maltsev2020a}\footnote{In the spherically-symmetric case under consideration the widths of resonances can be found directly from the jumps by $\pi$ in $\d_{tot}(l,k)$, considered in Sect.IV.}.

\begin{figure*}[ht!]
\subfigure[]{
		\includegraphics[width=1.5\columnwidth]{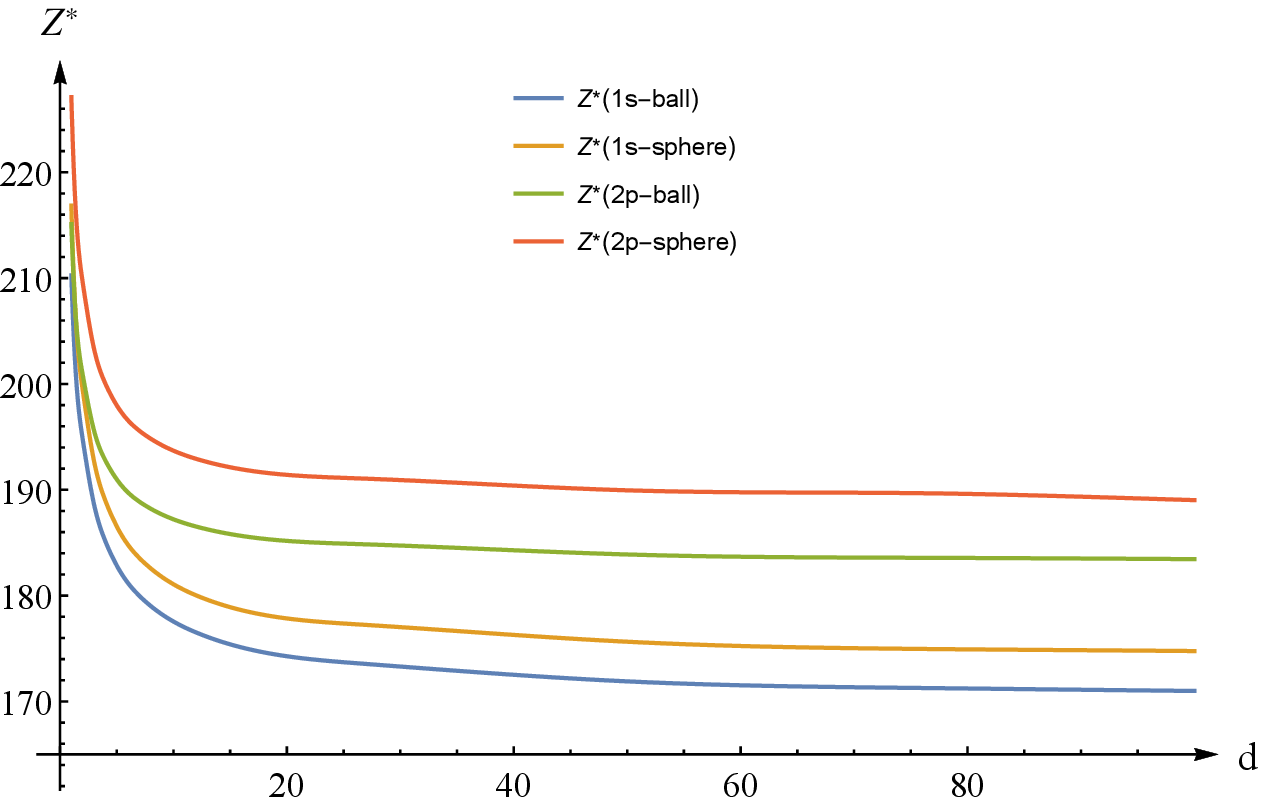}
}
\caption{(Color online)\, $Z^{\ast}(d)$  for  $1\leqslant d \leqslant 100$. }
	\label{Zast-ball-sphere}	
\end{figure*}

 It is also useful to introduce the parameter $d$ by means of relation
\beq
\label{6.7}
\e_{kin}(Z)= Z \a/d \ ,
\eeq
which can be interpreted as the distance from the center of the Coulomb source and the conditional point of the vacuum positron creation\footnote{Due to uncertainty relation there isn't any definite point of the vacuum positron creation in the scattering state with fixed energy. However, treatment of the parameter $d$ via distance between conditional point of the vacuum positron creation and the Coulomb  center turns out to be quite pertinent  and besides, can be reliably justified at least in  the quasiclassical approximation.}.
Upon solving   eqs. (\ref{6.5}) with respect to $Z$, one finds that the vacuum positron emission  is quite sensitive to  $d$, as it follows from Tabs.\ref{tast},\ref{tastball} and Fig.\ref{Zast-ball-sphere}.
\begin{table*}[ht!]
\caption{$Z^{\ast}(d)$ for $2/3 \leqslant d \leqslant 137$ (sphere).}
\begin{center}
	\begin{tabular}{|c|c|c|c|c|c|c|c|c|c|c|c|c|c|}
		\hline $d$ &  2/3 & 1 & 2 & 3 & 4 & 5 & 7 & 10 & 20 & 50 & 80 & 100 & 137 \\[2pt]
			\hline  $ Z^{\ast}(1s) $  &  233.0 & 216.8 & 199.4 & 192.7 & 189.0 & 186.5 & 183.5 & 181.1 & 177.8 & 175.6 & 174.9 & 174.8 & 174.0 \\[2pt]
	\hline $Z^{\ast}(2p)$   & 243.4 & 227.1 & 209.6 & 203.3  & 200.0 & 198.0 & 195.6 & 193.7 & 191.4 & 189.9 & 189.6 & 189.0 & 188.8 \\
				\hline
	\end{tabular}\label{tast}
\end{center}
\end{table*}
\begin{table*}[ht!]
\caption{$Z^{\ast}(d)$ for $2/3 \leqslant d \leqslant 137$ (ball).}
\begin{center}
	\begin{tabular}{|c|c|c|c|c|c|c|c|c|c|c|c|c|c|}
		\hline $d$ &  2/3 & 1 & 2 & 3 & 4 & 5 & 7 & 10 & 20 & 50 & 80 & 100 & 137 \\[2pt]
			\hline  $ Z^{\ast}(1s) $  &  224.0 & 210.3 & 194.8 & 188.6 & 185.1 & 182.8 & 180.0 & 177.6 & 174.3 & 171.9 & 171.2 & 171.0 & 170.7 \\[2pt]
	\hline $Z^{\ast}(2p)$   &  228.4 & 215.1 & 200.9 & 195.6 & 192.8 & 191.0 & 188.9 &  187.2 & 185.2 & 183.9 & 183.6 & 183.4 &  183.3 \\
				\hline
	\end{tabular}\label{tastball}
\end{center}
\end{table*}
The  reasonable choice for  $d$  is approximately one electron Compton length $\l_C$ ($=1$ in the units accepted). The latter is a rough estimate for the average radius of vacuum shells, created simultaneously with the positron emission.  Therefore it provides the most favorable conditions for  positron production, from the point of view of both the charge distribution and creation of a specific  lepton pair.

However, the calculations performed show unambiguously that for such $d$ the emission of vacuum positrons  cannot occur earlier than $Z$ exceeds $Z^{\ast}(1s)=217\,, Z^{\ast}(2p)=227$ (sphere), $Z^{\ast}(1s)=211\, , Z^{\ast}(2p)=215$ (ball) (to compare with  $Z_{cr,1}=173.6\,, Z_{cr,2}=188.5$ (sphere), $Z_{cr,1}=170\,, Z_{cr,2}=183.1$ (ball)). Note also that $Z^{\ast}$ increase very rapidly for $d < \l_C$. In particular, already for  $d=2/3$ one obtains $Z^{\ast}(1s)=233\, , Z^{\ast}(2p)=244$ (sphere), $Z^{\ast}(1s)=224\, , Z^{\ast}(2p)=229$ (ball). Such $Z^{\ast}$ lie far beyond  the interval $170 \leqslant Z \leqslant 192 $, which is nowadays the main region of theoretical and experimental activity in heavy ions collisions aimed at the study of such VP-effects ~\cite{Popov2018,Maltsev2019,FAIR2009,Ter2015,MA2017169}. In the ball source configuration the condition $170 \leqslant Z^{\ast} \leqslant 192 $ is fulfilled  by $1s$-channel  for  $d > 3 \l_C$ and  $d > 5 \l_C$ by $2p$-channel, but although such result can considered as the encouraging one, it should be noted that it is achieved within the model (\ref{1.6}) with a strongly underestimated  size of the Coulomb source in comparison with the real two heavy-ion configuration.

It would be worth to note that if allowed by the lepton number, spontaneous positrons can  by no means be emitted also just beyond the corresponding diving point. In particular,  for $d=137$ (one Bohr radius) one obtains  $Z^{\ast}(1s) \simeq 174.0\, , Z^{\ast}(2p) \simeq 188.8$ (sphere),\,$Z^{\ast}(1s) \simeq 170.7\, , Z^{\ast}(2p) \simeq 183.3$ (ball), what is already quite close to corresponding $Z_{cr,i}$. But in this case they appear in states localized far enough from the Coulomb source with small $\e_{kin} \sim 0.01$ and so cannot be distinguished from those emerging due to nuclear conversion  pairs production.  To the contrary, vacuum positrons created with  $d \simeq \l_C$  possess a number of specific properties~\cite{Popov2018,*Novak2018,*Maltsev2018,
Maltsev2019,*Maltsev2020,Greiner1985a}, which allow for an unambiguous detection, but the charge of the Coulomb source should be taken in this case not less than  $Z^{\ast} \simeq 210$. The negative result of early investigations at GSI~\cite{Mueller1994} can be  at least partially explained by the last circumstance. Moreover, estimating the self-energy contribution to the radiative part of QED effects due to virtual photon exchange near the lower continuum  shows~\cite{Roenko2018} that it is just a perturbative correction to essentially nonlinear VP-effects caused by  fermionic loop and so yields only a small increase of  $Z_{cr,i}$ and of other VP-quantities under question.

\section{Concluding remarks}

To conclude it should be  mentioned first  that as in the case of the one-dimensional ``hydrogen atom'' ~\cite{Davydov2017,Sveshnikov2017,Voronina2017}, as well as in two-dimensional  ``planar graphene-based hetero-structures'' ~\cite{Davydov2018a,Davydov2018b,Sveshnikov2019a,*Sveshnikov2019b,Voronina2019a,*Voronina2019b,Grashin2020a, *Grashin2020b}, the  evaluation of VP-energy by means of renormalization via fermionic loop is implemented via relations  (\ref{3.58} -\ref{3.60}) similar to, but actually without any references to VP-density and vacuum shells formation.  So the renormalization via fermionic loop turns out to be a universal tool, that removes the divergence of the theory both in the purely perturbative and in the essentially non-perturbative regimes of vacuum polarization by the external EM-field. Moreover, such approach can be easily extended to the study of VP-effects in more complicated problems, e.g., with two Coulomb centers or including axial magnetic field,  when the spherical symmetry is lost and so there remains only  $j_z$ as a conserved quantum number.

The substantial decrease of $\E_{VP}^{ren}(Z)$ in the overcritical region $\sim - Z^4/R$ can be reliably justified via the properties of  partial terms in the series (\ref{3.58}). Each   $\E^{ren}_{VP,l}(Z)$ in (\ref{3.58}) has the structure, which by omitting the  degeneracy factor $2(l+1)$ is quite similar to $\E_{VP}^{ren}(Z)$ in 1+1 D ~\cite{Davydov2017,Sveshnikov2017,Voronina2017}. The direct consequence of the latter  is that in the overcritical region the negative contribution from the renormalization term $\z_l Z^2$  turns out to be the dominant one in $\E^{ren}_{VP,l}(Z)$, since in this region  the growth rate of the non-renormalized partial VP-energies (\ref{3.31}), as in 1+1 and 2+1 D, is estimated as $\sim Z^\n \ , \ 1<\n<2$. Such estimate is clearly seen in plots shown in Fig.\ref{Int-Sum-VP-600}, where both the phase integral and the sum over discrete spectrum reveal an almost quadratic growth before first level diving, but afterwards their curves undergo either bending or negative jumps, which noticeably change the slope of the curves. So in essence the decrease of $\E_{VP}^{ren}$ in the overcritical region is governed  by the non-perturbative changes in the VP-density for $Z > Z_{cr,1}$ due to discrete levels diving into the lower continuum.  Furthermore,  the total number of partial channels, in which the levels have already sunk into the lower continuum,  grows approximately linearly with increasing $Z$. At the same time, indeed these channels yield the main contribution to  VP-energy (see Tabs.\ref{t300},\ref{t600}).   So the  rate of decrease of the total VP-energy into negative range acquires an additional factor    $\sim Z^2$, which in turn leads to the final answer   $\E^{ren}_{VP}(Z) \sim - Z^4/R$  in the overcritical region.

Lepton number also poses serious questions for both theory and experiment dealing with Coulomb super-criticality, since the emitted positrons must carry away the lepton number equal to $(-1)\times$their total number. Hence, the   corresponding amount of positive lepton number should be shared by VP-density, concentrated in vacuum shells. In this case instead of integer lepton number of real particles  there appears the lepton number VP-density. Otherwise either the lepton number conservation  in such processes must be broken, or the positron emission  prohibited. So any reliable answer concerning the spontaneous positron emission --- either positive or negative --- is important for our understanding of the nature of this number, since so far leptons show up as point-like particles with no indications on existence of any kind intrinsic structure. Therefore the reasonable conditions, under which the vacuum positron emission can be unambiguously detected on the nuclear conversion pairs background, should play an exceptional role  in slow  ions collisions, aimed at the search of such events~\cite{Krasnov2022}.

\section{Acknowledgments}

The authors are very indebted to Dr. O.V.Pavlovsky, P.A.Grashin  and A.A.Krasnov from MSU Department of Physics and to A.S.Davydov from Kurchatov Center for interest and helpful discussions.  This work has been supported in part by the RF Ministry of Sc. $\&$ Ed.  Scientific Research Program, projects No. 01-2014-63889, A16-116021760047-5, and by RFBR grant No. 14-02-01261. The research is carried out using the equipment of the shared research facilities of HPC computing resources at Lomonosov Moscow State University.

\section{Appendix. $I(l)$
for $l \gg Q$
}

For these purposes we replace first $ \d_{tot}(l,k) $   by the total WKB-phase  (\ref{3.36}). In the limit  (\ref{3.33a}) the difference between $ \d_{tot}(l,k) $ and $ \d^\mathrm{WKB}_{tot}(l,k) $ shows up only in  oscillations of the exact phase for large  $ k $, caused by diffraction on a sphere of the radius  $R$, but they are smoothed upon integration over $dk$, and so this difference can be freely ignored.

Proceeding further, we rewrite $I(l)$ as
\beq
\label{3.42a}
I(l)={1 \over \pi}\, \int\limits_1^{\inf} \!  d\e \ \d^\mathrm{WKB}_{tot}(l,\e)
 \eeq
and introduce  in (\ref{3.42a}) an intermediate UV-cutoff $\L$. The latter is necessary to provide the possibility of exchange the sequence of integrations. As a result, upon introducing the turning point
\beq
\label{3.43}
r_0={ (l+1)^2 - Q^2 \over 2\,Q} \gg R \ ,
\eeq
and the subsidiary function
\beq
\label{3.44}
W(l,r)=\sqrt{1+ (l+1)^2/r^2} \ ,
\eeq
the expression (\ref{3.42a})  can  be represented as
\begin{widetext}
\begin{multline}
\label{3.45}
I(l)={2 \over \pi}\, \lim_{\L \to \inf}\, \Bigg\{ \[
\int\limits_0^{r_0} \! dr  \int\limits_{V(r)+ W(l,r) \geqslant 1}^{\L} \! d\e \ + \ \int\limits_{r_0}^{\inf} \! dr  \int\limits_{1}^{\L} \! d\e \]\, \sqrt{\(\e - V(r)\)^2-W^2(l,r)} \ + \\ + \
\int\limits_0^{\inf} \! dr  \int\limits_{-V(r)+W(l,r) \geqslant 1}^{\L} \! d\e \ \sqrt{\(\e+ V(r)\)^2-W^2(l,r)} \ - \\ - \ 2\, \int\limits_{0}^{\inf} \! dr  \int\limits_{1}^{\L} \! d\e \ \sqrt{\e^2-W^2(l,r)} \Bigg\} \ ,
\end{multline}
where the first term originates from $\d_+(l,k)$, the second one comes from $\d_-(l,k)$, while the last one from $\d_0(l,k)$. Thereafter we replace  $\e - V(r) \to t\, W(l,r)$ in the first term, $\e + V(r) \to t\, W(l,r)$ in the second, and $\e \to t\, W(l,r)$ in the last one, what gives
\begin{multline}
\label{3.46}
I(l)={2 \over \pi}\, \lim_{\L \to \inf}\, \Bigg\{ \[
\int\limits_0^{r_0} \! dr \ W^2(l,r)\,  \int\limits_{1}^{{\L-V(r) \over W(l,r)}} \! dt \ + \ \int\limits_{r_0}^{\inf} \! dr \ W^2(l,r)\,  \int\limits_{{1-V(r) \over W(l,r)}}^{{\L-V(r) \over W(l,r)}} \! dt \]\, \sqrt{t^2-1} \ + \\ + \
\int\limits_0^{\inf} \! dr \ W^2(l,r)\, \int\limits_{1}^{{\L+V(r) \over W(l,r)}} \! dt \ \sqrt{t^2-1} \ - \\ - \ 2\, \int\limits_{0}^{\inf} \! dr \  W^2(l,r)\, \int\limits_{1}^{\L / W(l,r)} \! dt \ \sqrt{t^2-1} \Bigg\} \ .
\end{multline}
Introducing further the subsidiary function
\beq
\label{3.47}
Y(x)=\1/2\, \[x \sqrt{x^2-1} - \ln \(x+\sqrt{x^2-1}\)\]   \ ,
\eeq
we recast the expression (\ref{3.46}) in the form
 \begin{multline}
\label{3.48}
I(l)={2 \over \pi}\, \lim_{\L \to \inf}\, \Bigg\{
\int\limits_0^{\inf} \! dr \  W^2(l,r)\, \[ Y\({\L + V(r)\over W(l,r)}\) + Y\({\L - V(r)\over W(l,r)}\)  \ -  \ 2\, Y\({\L\over W(l,r)}\) \] \ - \\ - \ \int\limits_{r_0}^{\inf} \! dr \ W^2(l,r)\,  Y\({1 - V(r)\over W(l,r)}\) \Bigg\} \ .
\end{multline}
\end{widetext}
Since by construction there holds $|V(r)| \ll \L$ for all $0 \leq r \leq \inf$, we can freely expand the square bracket in (\ref{3.48})  in the power series with the expansion  parameter $V(r)/W(l,r)$ in the vicinity of the point $\L$. Thereafter by noticing that for $\L \to \inf$ in this expansion there survives only the term with second derivative $Y''(\L)=\L/\sqrt{\L^2 -1} \to 1$, one obtains
\beq
\label{3.49}
I(l)=I_0 + \D I(l)   \ ,
\eeq
where
\beq
\label{3.50}
I_0={2 \over \pi}\, \int\limits_0^{\inf} \! dr \ V^2(r)   \ ,
\eeq
and
\beq
\label{3.51}
\D I(l)= -{2 \over \pi}\,\int\limits_{r_0}^{\inf}\! dr \ W^2(l,r)\, Y\({1 - V(r)\over W(l,r)}\)   \ .
\eeq
The most important for the present analysis feature of $I_0$ is that for any extended Coulomb-like source the integral in the r.h.s. of (\ref{3.50}) converges  and that it is $O(Q^2)$-quantity. The concrete value of $I_0$ depends on the profile of the Coulomb source chosen (sphere, ball, or spherical layer), but there is no need to dive into these details here.

The remaining integral (\ref{3.51}) can be easily calculated analytically and so the leading-order-WKB answer for $I(l)$ subject to  condition (\ref{3.33a}) reads
 \begin{multline}
\label{3.52}
I(l)=I_0 - 2 \(l+1-\vk_l\) = \\ ={2 \over \pi}\, \int\limits_0^{\inf} \! dr \ V^2(r) - {  Q^2 \over l+1 } - { Q^4 \over 4 (l+1)^3} + O\({Q^6 \over (l+1)^5}\) \ .
 \end{multline}
The next-to-leading orders of the WKB-approximation for the total phase lead to corrections, which should be proportional to $Q^4/(l+1)^3$~\cite{Lazur2005, *Zon2012}. Hence, there follows from (\ref{3.52}) that with growing $l$ the partial phase integral should behave as follows
\beq\label{3.53}
 I(l) \to {2 \over \pi}\,\int\limits_0^{\infty} \! dr \ V^2(r)-{ Q^2 \over l+1} + O\(Q^4 \over (l+1)^3\) \ , \quad l \to \inf \ .
\eeq

 \bibliography{VP3DC}

\begin{thebibliography}{53}%
\makeatletter
\providecommand \@ifxundefined [1]{%
 \@ifx{#1\undefined}
}%
\providecommand \@ifnum [1]{%
 \ifnum #1\expandafter \@firstoftwo
 \else \expandafter \@secondoftwo
 \fi
}%
\providecommand \@ifx [1]{%
 \ifx #1\expandafter \@firstoftwo
 \else \expandafter \@secondoftwo
 \fi
}%
\providecommand \natexlab [1]{#1}%
\providecommand \enquote  [1]{``#1''}%
\providecommand \bibnamefont  [1]{#1}%
\providecommand \bibfnamefont [1]{#1}%
\providecommand \citenamefont [1]{#1}%
\providecommand \href@noop [0]{\@secondoftwo}%
\providecommand \href [0]{\begingroup \@sanitize@url \@href}%
\providecommand \@href[1]{\@@startlink{#1}\@@href}%
\providecommand \@@href[1]{\endgroup#1\@@endlink}%
\providecommand \@sanitize@url [0]{\catcode `\\12\catcode `\$12\catcode
  `\&12\catcode `\#12\catcode `\^12\catcode `\_12\catcode `\%12\relax}%
\providecommand \@@startlink[1]{}%
\providecommand \@@endlink[0]{}%
\providecommand \url  [0]{\begingroup\@sanitize@url \@url }%
\providecommand \@url [1]{\endgroup\@href {#1}{\urlprefix }}%
\providecommand \urlprefix  [0]{URL }%
\providecommand \Eprint [0]{\href }%
\providecommand \doibase [0]{http://dx.doi.org/}%
\providecommand \selectlanguage [0]{\@gobble}%
\providecommand \bibinfo  [0]{\@secondoftwo}%
\providecommand \bibfield  [0]{\@secondoftwo}%
\providecommand \translation [1]{[#1]}%
\providecommand \BibitemOpen [0]{}%
\providecommand \bibitemStop [0]{}%
\providecommand \bibitemNoStop [0]{.\EOS\space}%
\providecommand \EOS [0]{\spacefactor3000\relax}%
\providecommand \BibitemShut  [1]{\csname bibitem#1\endcsname}%
\let\auto@bib@innerbib\@empty
%</preamble>
\bibitem [{\citenamefont {Rafelski}\ \emph {et~al.}(2017)\citenamefont
  {Rafelski}, \citenamefont {Kirsch}, \citenamefont {M\"uller}, \citenamefont
  {Reinhardt},\ and\ \citenamefont {Greiner}}]{Rafelski2016}%
  \BibitemOpen
  \bibfield  {author} {\bibinfo {author} {\bibfnamefont {J.}~\bibnamefont
  {Rafelski}}, \bibinfo {author} {\bibfnamefont {J.}~\bibnamefont {Kirsch}},
  \bibinfo {author} {\bibfnamefont {B.}~\bibnamefont {M\"uller}}, \bibinfo
  {author} {\bibfnamefont {J.}~\bibnamefont {Reinhardt}}, \ and\ \bibinfo
  {author} {\bibfnamefont {W.}~\bibnamefont {Greiner}},\ }\enquote {\bibinfo
  {title} {Probing {QED} {Vacuum} with {Heavy} {Ions}},}\ in\ \href {\doibase
  10.1007/978-3-319-44165-8_17} {\emph {\bibinfo {booktitle} {New {Horizons} in
  {Fundamental} {Physics}}}},\ \bibinfo {series and number} {{FIAS}
  {Interdisciplinary} {Science} {Series}}\ (\bibinfo  {publisher} {Springer},\
  \bibinfo {year} {2017})\ pp.\ \bibinfo {pages} {211--251}\BibitemShut
  {NoStop}%
\bibitem [{\citenamefont {Kuleshov}\ \emph
  {et~al.}(2015{\natexlab{a}})\citenamefont {Kuleshov}, \citenamefont {Mur},
  \citenamefont {Narozhny}, \citenamefont {Fedotov},\ and\ \citenamefont
  {Lozovik}}]{Kuleshov2015a}%
  \BibitemOpen
  \bibfield  {author} {\bibinfo {author} {\bibfnamefont {V.~M.}\ \bibnamefont
  {Kuleshov}}, \bibinfo {author} {\bibfnamefont {V.~D.}\ \bibnamefont {Mur}},
  \bibinfo {author} {\bibfnamefont {N.~B.}\ \bibnamefont {Narozhny}}, \bibinfo
  {author} {\bibfnamefont {A.~M.}\ \bibnamefont {Fedotov}}, \ and\ \bibinfo
  {author} {\bibfnamefont {Y.~E.}\ \bibnamefont {Lozovik}},\ }\href {\doibase
  10.1134/s0021364015040098} {\bibfield  {journal} {\bibinfo  {journal} {Jetp
  Lett.}\ }\textbf {\bibinfo {volume} {101}},\ \bibinfo {pages} {264} (\bibinfo
  {year} {2015}{\natexlab{a}})}\BibitemShut {NoStop}%
\bibitem [{\citenamefont {Kuleshov}\ \emph
  {et~al.}(2015{\natexlab{b}})\citenamefont {Kuleshov}, \citenamefont {Mur},
  \citenamefont {Narozhny}, \citenamefont {Fedotov}, \citenamefont {Lozovik},\
  and\ \citenamefont {Popov}}]{Kuleshov2015b}%
  \BibitemOpen
  \bibfield  {author} {\bibinfo {author} {\bibfnamefont {V.~M.}\ \bibnamefont
  {Kuleshov}}, \bibinfo {author} {\bibfnamefont {V.~D.}\ \bibnamefont {Mur}},
  \bibinfo {author} {\bibfnamefont {N.~B.}\ \bibnamefont {Narozhny}}, \bibinfo
  {author} {\bibfnamefont {A.~M.}\ \bibnamefont {Fedotov}}, \bibinfo {author}
  {\bibfnamefont {Y.~E.}\ \bibnamefont {Lozovik}}, \ and\ \bibinfo {author}
  {\bibfnamefont {V.~S.}\ \bibnamefont {Popov}},\ }\href {\doibase
  10.3367/ufne.0185.201508d.0845} {\bibfield  {journal} {\bibinfo  {journal}
  {Physics-Uspekhi}\ }\textbf {\bibinfo {volume} {58}},\ \bibinfo {pages} {785}
  (\bibinfo {year} {2015}{\natexlab{b}})}\BibitemShut {NoStop}%
\bibitem [{\citenamefont {Godunov}\ \emph {et~al.}(2017)\citenamefont
  {Godunov}, \citenamefont {Machet},\ and\ \citenamefont
  {Vysotsky}}]{Godunov2017}%
  \BibitemOpen
  \bibfield  {author} {\bibinfo {author} {\bibfnamefont {S.~I.}\ \bibnamefont
  {Godunov}}, \bibinfo {author} {\bibfnamefont {B.}~\bibnamefont {Machet}}, \
  and\ \bibinfo {author} {\bibfnamefont {M.~I.}\ \bibnamefont {Vysotsky}},\
  }\href {\doibase 10.1140/epjc/s10052-017-5325-4} {\bibfield  {journal}
  {\bibinfo  {journal} {Eur. Phys. J. C}\ }\textbf {\bibinfo {volume} {77}},\
  \bibinfo {pages} {77:782} (\bibinfo {year} {2017})}\BibitemShut {NoStop}%
\bibitem [{\citenamefont {Davydov}\ \emph {et~al.}(2017)\citenamefont
  {Davydov}, \citenamefont {Sveshnikov},\ and\ \citenamefont
  {Voronina}}]{Davydov2017}%
  \BibitemOpen
  \bibfield  {author} {\bibinfo {author} {\bibfnamefont {A.}~\bibnamefont
  {Davydov}}, \bibinfo {author} {\bibfnamefont {K.}~\bibnamefont {Sveshnikov}},
  \ and\ \bibinfo {author} {\bibfnamefont {Y.}~\bibnamefont {Voronina}},\
  }\href {\doibase 10.1142/S0217751X17500543} {\bibfield  {journal} {\bibinfo
  {journal} {Int. J. Mod. Phys. A}\ }\textbf {\bibinfo {volume} {32}},\
  \bibinfo {pages} {1750054} (\bibinfo {year} {2017})}\BibitemShut {NoStop}%
\bibitem [{\citenamefont {Voronina}\ \emph
  {et~al.}(2017{\natexlab{a}})\citenamefont {Voronina}, \citenamefont
  {Davydov},\ and\ \citenamefont {Sveshnikov}}]{Sveshnikov2017}%
  \BibitemOpen
  \bibfield  {author} {\bibinfo {author} {\bibfnamefont {Y.}~\bibnamefont
  {Voronina}}, \bibinfo {author} {\bibfnamefont {A.}~\bibnamefont {Davydov}}, \
  and\ \bibinfo {author} {\bibfnamefont {K.}~\bibnamefont {Sveshnikov}},\
  }\href {\doibase 10.1134/S004057791711006X} {\bibfield  {journal} {\bibinfo
  {journal} {Theor. Math. Phys.}\ }\textbf {\bibinfo {volume} {193}},\ \bibinfo
  {pages} {1647} (\bibinfo {year} {2017}{\natexlab{a}})}\BibitemShut {NoStop}%
\bibitem [{\citenamefont {Voronina}\ \emph
  {et~al.}(2017{\natexlab{b}})\citenamefont {Voronina}, \citenamefont
  {Davydov},\ and\ \citenamefont {Sveshnikov}}]{Voronina2017}%
  \BibitemOpen
  \bibfield  {author} {\bibinfo {author} {\bibfnamefont {Y.}~\bibnamefont
  {Voronina}}, \bibinfo {author} {\bibfnamefont {A.}~\bibnamefont {Davydov}}, \
  and\ \bibinfo {author} {\bibfnamefont {K.}~\bibnamefont {Sveshnikov}},\
  }\href {\doibase 10.1134/S1547477117050144} {\bibfield  {journal} {\bibinfo
  {journal} {Phys. Part. Nucl. Lett.}\ }\textbf {\bibinfo {volume} {14}},\
  \bibinfo {pages} {698 } (\bibinfo {year} {2017}{\natexlab{b}})}\BibitemShut
  {NoStop}%
\bibitem [{\citenamefont {Popov}\ \emph {et~al.}(2018)\citenamefont {Popov},
  \citenamefont {Bondarev}, \citenamefont {Kozhedub}, \citenamefont {Maltsev},
  \citenamefont {Shabaev}, \citenamefont {Tupitsyn}, \citenamefont {Ma},
  \citenamefont {Plunien},\ and\ \citenamefont {St{\"o}hlker}}]{Popov2018}%
  \BibitemOpen
  \bibfield  {author} {\bibinfo {author} {\bibfnamefont {R.}~\bibnamefont
  {Popov}}, \bibinfo {author} {\bibfnamefont {A.}~\bibnamefont {Bondarev}},
  \bibinfo {author} {\bibfnamefont {Y.}~\bibnamefont {Kozhedub}}, \bibinfo
  {author} {\bibfnamefont {I.}~\bibnamefont {Maltsev}}, \bibinfo {author}
  {\bibfnamefont {V.}~\bibnamefont {Shabaev}}, \bibinfo {author} {\bibfnamefont
  {I.}~\bibnamefont {Tupitsyn}}, \bibinfo {author} {\bibfnamefont
  {X.}~\bibnamefont {Ma}}, \bibinfo {author} {\bibfnamefont {G.}~\bibnamefont
  {Plunien}}, \ and\ \bibinfo {author} {\bibfnamefont {T.}~\bibnamefont
  {St{\"o}hlker}},\ }\href {\doibase 10.1140/epjd/e2018-90056-4} {\bibfield
  {journal} {\bibinfo  {journal} {Eur. Phys. J. D}\ }\textbf {\bibinfo {volume}
  {72}},\ \bibinfo {pages} {115} (\bibinfo {year} {2018})}\BibitemShut
  {NoStop}%
\bibitem [{\citenamefont {Novak}\ \emph {et~al.}(2018)\citenamefont {Novak},
  \citenamefont {Kholodov}, \citenamefont {Surzhykov}, \citenamefont
  {Artemyev},\ and\ \citenamefont {St{\"o}hlker}}]{Novak2018}%
  \BibitemOpen
  \bibfield  {author} {\bibinfo {author} {\bibfnamefont {O.}~\bibnamefont
  {Novak}}, \bibinfo {author} {\bibfnamefont {R.}~\bibnamefont {Kholodov}},
  \bibinfo {author} {\bibfnamefont {A.}~\bibnamefont {Surzhykov}}, \bibinfo
  {author} {\bibfnamefont {A.~N.}\ \bibnamefont {Artemyev}}, \ and\ \bibinfo
  {author} {\bibfnamefont {T.}~\bibnamefont {St{\"o}hlker}},\ }\href {\doibase
  https://doi.org/10.1103/PhysRevA.97.032518} {\bibfield  {journal} {\bibinfo
  {journal} {Phys. Rev. A}\ }\textbf {\bibinfo {volume} {97}},\ \bibinfo
  {pages} {032518} (\bibinfo {year} {2018})}\BibitemShut {NoStop}%
\bibitem [{\citenamefont {Maltsev}\ \emph {et~al.}(2018)\citenamefont
  {Maltsev}, \citenamefont {Shabaev}, \citenamefont {Popov}, \citenamefont
  {Kozhedub}, \citenamefont {Plunien}, \citenamefont {Ma},\ and\ \citenamefont
  {St{\"o}hlker}}]{Maltsev2018}%
  \BibitemOpen
  \bibfield  {author} {\bibinfo {author} {\bibfnamefont {I.~A.}\ \bibnamefont
  {Maltsev}}, \bibinfo {author} {\bibfnamefont {V.~M.}\ \bibnamefont
  {Shabaev}}, \bibinfo {author} {\bibfnamefont {R.~V.}\ \bibnamefont {Popov}},
  \bibinfo {author} {\bibfnamefont {Y.~S.}\ \bibnamefont {Kozhedub}}, \bibinfo
  {author} {\bibfnamefont {G.}~\bibnamefont {Plunien}}, \bibinfo {author}
  {\bibfnamefont {X.}~\bibnamefont {Ma}}, \ and\ \bibinfo {author}
  {\bibfnamefont {T.}~\bibnamefont {St{\"o}hlker}},\ }\href {\doibase
  10.1103/PhysRevA.98.062709} {\bibfield  {journal} {\bibinfo  {journal} {Phys.
  Rev. A}\ }\textbf {\bibinfo {volume} {98}},\ \bibinfo {pages} {062709}
  (\bibinfo {year} {2018})}\BibitemShut {NoStop}%
\bibitem [{\citenamefont {Roenko}\ and\ \citenamefont
  {Sveshnikov}(2018)}]{Roenko2018}%
  \BibitemOpen
  \bibfield  {author} {\bibinfo {author} {\bibfnamefont {A.}~\bibnamefont
  {Roenko}}\ and\ \bibinfo {author} {\bibfnamefont {K.}~\bibnamefont
  {Sveshnikov}},\ }\href {\doibase https://doi.org/10.1103/PhysRevA.97.012113}
  {\bibfield  {journal} {\bibinfo  {journal} {Phys. Rev. A}\ }\textbf {\bibinfo
  {volume} {97}},\ \bibinfo {pages} {012113} (\bibinfo {year}
  {2018})}\BibitemShut {NoStop}%
\bibitem [{\citenamefont {Maltsev}\ \emph {et~al.}(2019)\citenamefont
  {Maltsev}, \citenamefont {Shabaev}, \citenamefont {Popov}, \citenamefont
  {Kozhedub}, \citenamefont {Plunien}, \citenamefont {Ma}, \citenamefont
  {St\"ohlker},\ and\ \citenamefont {Tumakov}}]{Maltsev2019}%
  \BibitemOpen
  \bibfield  {author} {\bibinfo {author} {\bibfnamefont {I.~A.}\ \bibnamefont
  {Maltsev}}, \bibinfo {author} {\bibfnamefont {V.~M.}\ \bibnamefont
  {Shabaev}}, \bibinfo {author} {\bibfnamefont {R.~V.}\ \bibnamefont {Popov}},
  \bibinfo {author} {\bibfnamefont {Y.~S.}\ \bibnamefont {Kozhedub}}, \bibinfo
  {author} {\bibfnamefont {G.}~\bibnamefont {Plunien}}, \bibinfo {author}
  {\bibfnamefont {X.}~\bibnamefont {Ma}}, \bibinfo {author} {\bibfnamefont
  {T.}~\bibnamefont {St\"ohlker}}, \ and\ \bibinfo {author} {\bibfnamefont
  {D.~A.}\ \bibnamefont {Tumakov}},\ }\href {\doibase
  10.1103/PhysRevLett.123.113401} {\bibfield  {journal} {\bibinfo  {journal}
  {Phys. Rev. Lett.}\ }\textbf {\bibinfo {volume} {123}},\ \bibinfo {pages}
  {113401} (\bibinfo {year} {2019})}\BibitemShut {NoStop}%
\bibitem [{\citenamefont {Popov}\ \emph {et~al.}(2020)\citenamefont {Popov},
  \citenamefont {Shabaev}, \citenamefont {Telnov}, \citenamefont {Tupitsyn},
  \citenamefont {Maltsev}, \citenamefont {Kozhedub}, \citenamefont {Bondarev},
  \citenamefont {Kozin}, \citenamefont {Ma}, \citenamefont {Plunien},
  \citenamefont {St\"ohlker}, \citenamefont {Tumakov},\ and\ \citenamefont
  {Zaytsev}}]{Maltsev2020}%
  \BibitemOpen
  \bibfield  {author} {\bibinfo {author} {\bibfnamefont {R.~V.}\ \bibnamefont
  {Popov}}, \bibinfo {author} {\bibfnamefont {V.~M.}\ \bibnamefont {Shabaev}},
  \bibinfo {author} {\bibfnamefont {D.~A.}\ \bibnamefont {Telnov}}, \bibinfo
  {author} {\bibfnamefont {I.~I.}\ \bibnamefont {Tupitsyn}}, \bibinfo {author}
  {\bibfnamefont {I.~A.}\ \bibnamefont {Maltsev}}, \bibinfo {author}
  {\bibfnamefont {Y.~S.}\ \bibnamefont {Kozhedub}}, \bibinfo {author}
  {\bibfnamefont {A.~I.}\ \bibnamefont {Bondarev}}, \bibinfo {author}
  {\bibfnamefont {N.~V.}\ \bibnamefont {Kozin}}, \bibinfo {author}
  {\bibfnamefont {X.}~\bibnamefont {Ma}}, \bibinfo {author} {\bibfnamefont
  {G.}~\bibnamefont {Plunien}}, \bibinfo {author} {\bibfnamefont
  {T.}~\bibnamefont {St\"ohlker}}, \bibinfo {author} {\bibfnamefont {D.~A.}\
  \bibnamefont {Tumakov}}, \ and\ \bibinfo {author} {\bibfnamefont {V.~A.}\
  \bibnamefont {Zaytsev}},\ }\href {\doibase 10.1103/PhysRevD.102.076005}
  {\bibfield  {journal} {\bibinfo  {journal} {Phys. Rev. D}\ }\textbf {\bibinfo
  {volume} {102}},\ \bibinfo {pages} {076005} (\bibinfo {year}
  {2020})}\BibitemShut {NoStop}%
\bibitem [{\citenamefont {Greiner}\ \emph {et~al.}(1985)\citenamefont
  {Greiner}, \citenamefont {M\"uller},\ and\ \citenamefont
  {Rafelski}}]{Greiner1985a}%
  \BibitemOpen
  \bibfield  {author} {\bibinfo {author} {\bibfnamefont {W.}~\bibnamefont
  {Greiner}}, \bibinfo {author} {\bibfnamefont {B.}~\bibnamefont {M\"uller}}, \
  and\ \bibinfo {author} {\bibfnamefont {J.}~\bibnamefont {Rafelski}},\ }\href
  {http://link.springer.com/book/10.1007/978-3-642-82272-8} {\emph {\bibinfo
  {title} {Quantum Electrodynamics of Strong Fields}}},\ \bibinfo {edition}
  {2nd}\ ed.\ (\bibinfo  {publisher} {Springer},\ \bibinfo {address} {Berlin},\
  \bibinfo {year} {1985})\BibitemShut {NoStop}%
\bibitem [{\citenamefont {Plunien}\ \emph {et~al.}(1986)\citenamefont
  {Plunien}, \citenamefont {M\"uller},\ and\ \citenamefont
  {Greiner}}]{Plunien1986}%
  \BibitemOpen
  \bibfield  {author} {\bibinfo {author} {\bibfnamefont {G.}~\bibnamefont
  {Plunien}}, \bibinfo {author} {\bibfnamefont {B.}~\bibnamefont {M\"uller}}, \
  and\ \bibinfo {author} {\bibfnamefont {W.}~\bibnamefont {Greiner}},\ }\href
  {\doibase 10.1016/0370-1573(86)90020-7} {\bibfield  {journal} {\bibinfo
  {journal} {Phys. Rep.}\ }\textbf {\bibinfo {volume} {134}},\ \bibinfo {pages}
  {87 } (\bibinfo {year} {1986})}\BibitemShut {NoStop}%
\bibitem [{\citenamefont {Greiner}\ and\ \citenamefont
  {Reinhardt}(2009)}]{Greiner2012}%
  \BibitemOpen
  \bibfield  {author} {\bibinfo {author} {\bibfnamefont {W.}~\bibnamefont
  {Greiner}}\ and\ \bibinfo {author} {\bibfnamefont {J.}~\bibnamefont
  {Reinhardt}},\ }\href {\doibase /10.1007/978-3-540-87561-1} {\emph {\bibinfo
  {title} {Quantum Electrodynamics}}},\ \bibinfo {edition} {4th}\ ed.\
  (\bibinfo  {publisher} {Springer-Verlag Berlin Heidelberg},\ \bibinfo {year}
  {2009})\BibitemShut {NoStop}%
\bibitem [{\citenamefont {Ruffini}\ \emph {et~al.}(2010)\citenamefont
  {Ruffini}, \citenamefont {Vereshchagin},\ and\ \citenamefont
  {Xue}}]{Ruffini2010}%
  \BibitemOpen
  \bibfield  {author} {\bibinfo {author} {\bibfnamefont {R.}~\bibnamefont
  {Ruffini}}, \bibinfo {author} {\bibfnamefont {G.}~\bibnamefont
  {Vereshchagin}}, \ and\ \bibinfo {author} {\bibfnamefont {S.-S.}\
  \bibnamefont {Xue}},\ }\href {\doibase 10.1016/j.physrep.2009.10.004}
  {\bibfield  {journal} {\bibinfo  {journal} {Phys. Rep.}\ }\textbf {\bibinfo
  {volume} {487}},\ \bibinfo {pages} {1 } (\bibinfo {year} {2010})}\BibitemShut
  {NoStop}%
\bibitem [{\citenamefont {Gumberidze}\ \emph {et~al.}(2009)\citenamefont
  {Gumberidze}, \citenamefont {St{\"o}hlker}, \citenamefont {Beyer},
  \citenamefont {Bosch}, \citenamefont {Bräuning-Demian}, \citenamefont
  {Hagmann}, \citenamefont {Kozhuharov}, \citenamefont {K\"uhl}, \citenamefont
  {Mann}, \citenamefont {Indelicato}, \citenamefont {Quint}, \citenamefont
  {Schuch},\ and\ \citenamefont {Warczak}}]{FAIR2009}%
  \BibitemOpen
  \bibfield  {author} {\bibinfo {author} {\bibfnamefont {A.}~\bibnamefont
  {Gumberidze}}, \bibinfo {author} {\bibfnamefont {T.}~\bibnamefont
  {St{\"o}hlker}}, \bibinfo {author} {\bibfnamefont {H.~F.}\ \bibnamefont
  {Beyer}}, \bibinfo {author} {\bibfnamefont {F.}~\bibnamefont {Bosch}},
  \bibinfo {author} {\bibfnamefont {A.}~\bibnamefont {Bräuning-Demian}},
  \bibinfo {author} {\bibfnamefont {S.}~\bibnamefont {Hagmann}}, \bibinfo
  {author} {\bibfnamefont {C.}~\bibnamefont {Kozhuharov}}, \bibinfo {author}
  {\bibfnamefont {T.}~\bibnamefont {K\"uhl}}, \bibinfo {author} {\bibfnamefont
  {R.}~\bibnamefont {Mann}}, \bibinfo {author} {\bibfnamefont {P.}~\bibnamefont
  {Indelicato}}, \bibinfo {author} {\bibfnamefont {W.}~\bibnamefont {Quint}},
  \bibinfo {author} {\bibfnamefont {R.}~\bibnamefont {Schuch}}, \ and\ \bibinfo
  {author} {\bibfnamefont {A.}~\bibnamefont {Warczak}},\ }\href {\doibase
  https://doi.org/10.1016/j.nimb.2008.10.079} {\bibfield  {journal} {\bibinfo
  {journal} {Nucl. Instr. {\&} Meth. in Phys. Research B}\ }\textbf {\bibinfo
  {volume} {267}},\ \bibinfo {pages} {248} (\bibinfo {year}
  {2009})}\BibitemShut {NoStop}%
\bibitem [{\citenamefont {Ter-Akopian}\ \emph {et~al.}(2015)\citenamefont
  {Ter-Akopian}, \citenamefont {Greiner}, \citenamefont {Meshkov},
  \citenamefont {Oganessian}, \citenamefont {Reinhardt},\ and\ \citenamefont
  {Trubnikov}}]{Ter2015}%
  \BibitemOpen
  \bibfield  {author} {\bibinfo {author} {\bibfnamefont {G.~M.}\ \bibnamefont
  {Ter-Akopian}}, \bibinfo {author} {\bibfnamefont {W.}~\bibnamefont
  {Greiner}}, \bibinfo {author} {\bibfnamefont {I.}~\bibnamefont {Meshkov}},
  \bibinfo {author} {\bibfnamefont {Y.}~\bibnamefont {Oganessian}}, \bibinfo
  {author} {\bibfnamefont {J.}~\bibnamefont {Reinhardt}}, \ and\ \bibinfo
  {author} {\bibfnamefont {G.}~\bibnamefont {Trubnikov}},\ }\href {\doibase
  10.1142/S0218301315500160} {\bibfield  {journal} {\bibinfo  {journal} {Int.
  J. Mod. Phys. E}\ }\textbf {\bibinfo {volume} {24}},\ \bibinfo {pages}
  {1550016} (\bibinfo {year} {2015})}\BibitemShut {NoStop}%
\bibitem [{\citenamefont {Ma}\ \emph {et~al.}(2017)\citenamefont {Ma},
  \citenamefont {Wen}, \citenamefont {Zhang}, \citenamefont {Yu}, \citenamefont
  {Cheng}, \citenamefont {Yang}, \citenamefont {Huang}, \citenamefont {Wang},
  \citenamefont {Zhu}, \citenamefont {Cai}, \citenamefont {Zhao}, \citenamefont
  {Mao}, \citenamefont {Yang}, \citenamefont {Zhou}, \citenamefont {Xu},
  \citenamefont {Yuan}, \citenamefont {Xia}, \citenamefont {Zhao},
  \citenamefont {Xiao},\ and\ \citenamefont {Zhan}}]{MA2017169}%
  \BibitemOpen
  \bibfield  {author} {\bibinfo {author} {\bibfnamefont {X.}~\bibnamefont
  {Ma}}, \bibinfo {author} {\bibfnamefont {W.}~\bibnamefont {Wen}}, \bibinfo
  {author} {\bibfnamefont {S.}~\bibnamefont {Zhang}}, \bibinfo {author}
  {\bibfnamefont {D.}~\bibnamefont {Yu}}, \bibinfo {author} {\bibfnamefont
  {R.}~\bibnamefont {Cheng}}, \bibinfo {author} {\bibfnamefont
  {J.}~\bibnamefont {Yang}}, \bibinfo {author} {\bibfnamefont {Z.}~\bibnamefont
  {Huang}}, \bibinfo {author} {\bibfnamefont {H.}~\bibnamefont {Wang}},
  \bibinfo {author} {\bibfnamefont {X.}~\bibnamefont {Zhu}}, \bibinfo {author}
  {\bibfnamefont {X.}~\bibnamefont {Cai}}, \bibinfo {author} {\bibfnamefont
  {Y.}~\bibnamefont {Zhao}}, \bibinfo {author} {\bibfnamefont {L.}~\bibnamefont
  {Mao}}, \bibinfo {author} {\bibfnamefont {J.}~\bibnamefont {Yang}}, \bibinfo
  {author} {\bibfnamefont {X.}~\bibnamefont {Zhou}}, \bibinfo {author}
  {\bibfnamefont {H.}~\bibnamefont {Xu}}, \bibinfo {author} {\bibfnamefont
  {Y.}~\bibnamefont {Yuan}}, \bibinfo {author} {\bibfnamefont {J.}~\bibnamefont
  {Xia}}, \bibinfo {author} {\bibfnamefont {H.}~\bibnamefont {Zhao}}, \bibinfo
  {author} {\bibfnamefont {G.}~\bibnamefont {Xiao}}, \ and\ \bibinfo {author}
  {\bibfnamefont {W.}~\bibnamefont {Zhan}},\ }\href {\doibase
  https://doi.org/10.1016/j.nimb.2017.03.129} {\bibfield  {journal} {\bibinfo
  {journal} {Nucl. Instr. {\&} Meth. in Phys. Research B}\ }\textbf {\bibinfo
  {volume} {408}},\ \bibinfo {pages} {169} (\bibinfo {year}
  {2017})}\BibitemShut {NoStop}%
\bibitem [{\citenamefont {Krasnov}\ and\ \citenamefont
  {Sveshnikov}(2022)}]{Krasnov2022}%
  \BibitemOpen
  \bibfield  {author} {\bibinfo {author} {\bibfnamefont {A.}~\bibnamefont
  {Krasnov}}\ and\ \bibinfo {author} {\bibfnamefont {K.}~\bibnamefont
  {Sveshnikov}},\ }\href@noop {} {\bibfield  {journal} {\bibinfo  {journal} {In
  preparation}\ } (\bibinfo {year} {2022})}\BibitemShut {NoStop}%
\bibitem [{\citenamefont {Wichmann}\ and\ \citenamefont
  {Kroll}(1956)}]{Wichmann1956}%
  \BibitemOpen
  \bibfield  {author} {\bibinfo {author} {\bibfnamefont {E.~H.}\ \bibnamefont
  {Wichmann}}\ and\ \bibinfo {author} {\bibfnamefont {N.~M.}\ \bibnamefont
  {Kroll}},\ }\href {\doibase 10.1103/PhysRev.101.843} {\bibfield  {journal}
  {\bibinfo  {journal} {Phys. Rev.}\ }\textbf {\bibinfo {volume} {101}},\
  \bibinfo {pages} {843} (\bibinfo {year} {1956})}\BibitemShut {NoStop}%
\bibitem [{\citenamefont {Gyulassy}(1975)}]{Gyulassy1975}%
  \BibitemOpen
  \bibfield  {author} {\bibinfo {author} {\bibfnamefont {M.}~\bibnamefont
  {Gyulassy}},\ }\href {\doibase 10.1016/0375-9474(75)90554-0} {\bibfield
  {journal} {\bibinfo  {journal} {Nucl. Phys. A}\ }\textbf {\bibinfo {volume}
  {244}},\ \bibinfo {pages} {497 } (\bibinfo {year} {1975})}\BibitemShut
  {NoStop}%
\bibitem [{\citenamefont {Brown}\ \emph
  {et~al.}(1975{\natexlab{a}})\citenamefont {Brown}, \citenamefont {Cahn},\
  and\ \citenamefont {McLerran}}]{McLerran1975a}%
  \BibitemOpen
  \bibfield  {author} {\bibinfo {author} {\bibfnamefont {L.}~\bibnamefont
  {Brown}}, \bibinfo {author} {\bibfnamefont {R.}~\bibnamefont {Cahn}}, \ and\
  \bibinfo {author} {\bibfnamefont {L.}~\bibnamefont {McLerran}},\ }\href
  {\doibase 10.1103/PhysRevD.12.581} {\bibfield  {journal} {\bibinfo  {journal}
  {Phys. Rev. D}\ }\textbf {\bibinfo {volume} {12}},\ \bibinfo {pages} {581}
  (\bibinfo {year} {1975}{\natexlab{a}})}\BibitemShut {NoStop}%
\bibitem [{\citenamefont {Brown}\ \emph
  {et~al.}(1975{\natexlab{b}})\citenamefont {Brown}, \citenamefont {Cahn},\
  and\ \citenamefont {McLerran}}]{McLerran1975b}%
  \BibitemOpen
  \bibfield  {author} {\bibinfo {author} {\bibfnamefont {L.}~\bibnamefont
  {Brown}}, \bibinfo {author} {\bibfnamefont {R.}~\bibnamefont {Cahn}}, \ and\
  \bibinfo {author} {\bibfnamefont {L.}~\bibnamefont {McLerran}},\ }\href
  {\doibase 10.1103/PhysRevD.12.596} {\bibfield  {journal} {\bibinfo  {journal}
  {Phys. Rev. D}\ }\textbf {\bibinfo {volume} {12}},\ \bibinfo {pages} {596}
  (\bibinfo {year} {1975}{\natexlab{b}})}\BibitemShut {NoStop}%
\bibitem [{\citenamefont {Brown}\ \emph
  {et~al.}(1975{\natexlab{c}})\citenamefont {Brown}, \citenamefont {Cahn},\
  and\ \citenamefont {McLerran}}]{McLerran1975c}%
  \BibitemOpen
  \bibfield  {author} {\bibinfo {author} {\bibfnamefont {L.}~\bibnamefont
  {Brown}}, \bibinfo {author} {\bibfnamefont {R.}~\bibnamefont {Cahn}}, \ and\
  \bibinfo {author} {\bibfnamefont {L.}~\bibnamefont {McLerran}},\ }\href
  {\doibase 10.1103/PhysRevD.12.609} {\bibfield  {journal} {\bibinfo  {journal}
  {Phys. Rev. D}\ }\textbf {\bibinfo {volume} {12}},\ \bibinfo {pages} {609}
  (\bibinfo {year} {1975}{\natexlab{c}})}\BibitemShut {NoStop}%
\bibitem [{\citenamefont {Itzykson}\ and\ \citenamefont
  {Zuber}(1980)}]{Itzykson1980}%
  \BibitemOpen
  \bibfield  {author} {\bibinfo {author} {\bibfnamefont {C.}~\bibnamefont
  {Itzykson}}\ and\ \bibinfo {author} {\bibfnamefont {J.-B.}\ \bibnamefont
  {Zuber}},\ }\href@noop {} {\emph {\bibinfo {title} {Quantum Field Theory}}}\
  (\bibinfo  {publisher} {McGraw-Hill},\ \bibinfo {year} {1980})\BibitemShut
  {NoStop}%
\bibitem [{\citenamefont {{Rajaraman}}(1982)}]{Rajaraman1982}%
  \BibitemOpen
  \bibfield  {author} {\bibinfo {author} {\bibfnamefont {R.}~\bibnamefont
  {{Rajaraman}}},\ }\href
  {https://inis.iaea.org/search/search.aspx?orig_q=RN:15036991} {\emph
  {\bibinfo {title} {Solitons and Instantons}}},\ \bibinfo {edition} {1st}\
  ed.\ (\bibinfo  {publisher} {North-Holland Publishing Company},\ \bibinfo
  {year} {1982})\BibitemShut {NoStop}%
\bibitem [{\citenamefont {Sveshnikov}(1991)}]{Sveshnikov1991}%
  \BibitemOpen
  \bibfield  {author} {\bibinfo {author} {\bibfnamefont {K.}~\bibnamefont
  {Sveshnikov}},\ }\href {\doibase 10.1016/0370-2693(91)90244-K} {\bibfield
  {journal} {\bibinfo  {journal} {Phys. Lett. B}\ }\textbf {\bibinfo {volume}
  {255}},\ \bibinfo {pages} {255} (\bibinfo {year} {1991})}\BibitemShut
  {NoStop}%
\bibitem [{\citenamefont {Sundberg}\ and\ \citenamefont
  {Jaffe}(2004)}]{Jaffe2004}%
  \BibitemOpen
  \bibfield  {author} {\bibinfo {author} {\bibfnamefont {P.}~\bibnamefont
  {Sundberg}}\ and\ \bibinfo {author} {\bibfnamefont {R.~L.}\ \bibnamefont
  {Jaffe}},\ }\href {\doibase 10.1016/j.aop.2003.08.015} {\bibfield  {journal}
  {\bibinfo  {journal} {Ann. Phys.}\ }\textbf {\bibinfo {volume} {309}},\
  \bibinfo {pages} {442} (\bibinfo {year} {2004})}\BibitemShut {NoStop}%
\bibitem [{\citenamefont {Davydov}\ \emph
  {et~al.}(2018{\natexlab{a}})\citenamefont {Davydov}, \citenamefont
  {Sveshnikov},\ and\ \citenamefont {Voronina}}]{Davydov2018b}%
  \BibitemOpen
  \bibfield  {author} {\bibinfo {author} {\bibfnamefont {A.}~\bibnamefont
  {Davydov}}, \bibinfo {author} {\bibfnamefont {K.}~\bibnamefont {Sveshnikov}},
  \ and\ \bibinfo {author} {\bibfnamefont {Y.}~\bibnamefont {Voronina}},\
  }\href {\doibase 10.1142/S0217751X18500057} {\bibfield  {journal} {\bibinfo
  {journal} {Int. J. Mod. Phys. A}\ }\textbf {\bibinfo {volume} {33}},\
  \bibinfo {pages} {1850005} (\bibinfo {year}
  {2018}{\natexlab{a}})}\BibitemShut {NoStop}%
\bibitem [{\citenamefont {Lazur}\ \emph {et~al.}(2005)\citenamefont {Lazur},
  \citenamefont {Reity},\ and\ \citenamefont {Rubish}}]{Lazur2005}%
  \BibitemOpen
  \bibfield  {author} {\bibinfo {author} {\bibfnamefont {V.}~\bibnamefont
  {Lazur}}, \bibinfo {author} {\bibfnamefont {O.}~\bibnamefont {Reity}}, \ and\
  \bibinfo {author} {\bibfnamefont {V.}~\bibnamefont {Rubish}},\ }\href
  {\doibase https://doi.org/10.1007/s11232-005-0090-1} {\bibfield  {journal}
  {\bibinfo  {journal} {Theor. Math. Phys.}\ }\textbf {\bibinfo {volume}
  {143}},\ \bibinfo {pages} {559 } (\bibinfo {year} {2005})}\BibitemShut
  {NoStop}%
\bibitem [{\citenamefont {Zon}\ and\ \citenamefont {Kornev}(2012)}]{Zon2012}%
  \BibitemOpen
  \bibfield  {author} {\bibinfo {author} {\bibfnamefont {B.}~\bibnamefont
  {Zon}}\ and\ \bibinfo {author} {\bibfnamefont {A.}~\bibnamefont {Kornev}},\
  }\href {\doibase https://doi.org/10.1007/s11232-012-0046-1} {\bibfield
  {journal} {\bibinfo  {journal} {Theor. Math. Phys.}\ }\textbf {\bibinfo
  {volume} {171}},\ \bibinfo {pages} {478 } (\bibinfo {year}
  {2012})}\BibitemShut {NoStop}%
\bibitem [{\citenamefont {Voronina}\ \emph
  {et~al.}(2019{\natexlab{a}})\citenamefont {Voronina}, \citenamefont
  {Komissarov},\ and\ \citenamefont {Sveshnikov}}]{Voronina2019c}%
  \BibitemOpen
  \bibfield  {author} {\bibinfo {author} {\bibfnamefont {Y.}~\bibnamefont
  {Voronina}}, \bibinfo {author} {\bibfnamefont {I.}~\bibnamefont
  {Komissarov}}, \ and\ \bibinfo {author} {\bibfnamefont {K.}~\bibnamefont
  {Sveshnikov}},\ }\href {\doibase https://doi.org/10.1016/j.aop.2019.02.014}
  {\bibfield  {journal} {\bibinfo  {journal} {Ann. Phys.}\ }\textbf {\bibinfo
  {volume} {404}},\ \bibinfo {pages} {132 } (\bibinfo {year}
  {2019}{\natexlab{a}})}\BibitemShut {NoStop}%
\bibitem [{\citenamefont {Voronina}\ \emph
  {et~al.}(2019{\natexlab{b}})\citenamefont {Voronina}, \citenamefont
  {Komissarov},\ and\ \citenamefont {Sveshnikov}}]{Voronina2019d}%
  \BibitemOpen
  \bibfield  {author} {\bibinfo {author} {\bibfnamefont {Y.}~\bibnamefont
  {Voronina}}, \bibinfo {author} {\bibfnamefont {I.}~\bibnamefont
  {Komissarov}}, \ and\ \bibinfo {author} {\bibfnamefont {K.}~\bibnamefont
  {Sveshnikov}},\ }\href {\doibase 10.1103/PhysRevA.99.062504} {\bibfield
  {journal} {\bibinfo  {journal} {Phys. Rev. A}\ }\textbf {\bibinfo {volume}
  {99}},\ \bibinfo {pages} {062504} (\bibinfo {year}
  {2019}{\natexlab{b}})}\BibitemShut {NoStop}%
\bibitem [{\citenamefont {Berestetskii}\ \emph {et~al.}(2012)\citenamefont
  {Berestetskii}, \citenamefont {Pitaevskii},\ and\ \citenamefont
  {Lifshitz}}]{landau2012qed}%
  \BibitemOpen
  \bibfield  {author} {\bibinfo {author} {\bibfnamefont {V.}~\bibnamefont
  {Berestetskii}}, \bibinfo {author} {\bibfnamefont {L.}~\bibnamefont
  {Pitaevskii}}, \ and\ \bibinfo {author} {\bibfnamefont {E.}~\bibnamefont
  {Lifshitz}},\ }\href@noop {} {\emph {\bibinfo {title} {Quantum
  Electrodynamics}}},\ Vol.~\bibinfo {volume} {4}\ (\bibinfo  {publisher}
  {Elsevier},\ \bibinfo {year} {2012})\BibitemShut {NoStop}%
\bibitem [{\citenamefont {Davydov}\ \emph
  {et~al.}(2018{\natexlab{b}})\citenamefont {Davydov}, \citenamefont
  {Sveshnikov},\ and\ \citenamefont {Voronina}}]{Davydov2018a}%
  \BibitemOpen
  \bibfield  {author} {\bibinfo {author} {\bibfnamefont {A.}~\bibnamefont
  {Davydov}}, \bibinfo {author} {\bibfnamefont {K.}~\bibnamefont {Sveshnikov}},
  \ and\ \bibinfo {author} {\bibfnamefont {Y.}~\bibnamefont {Voronina}},\
  }\href {\doibase 10.1142/S0217751X18500045} {\bibfield  {journal} {\bibinfo
  {journal} {Int. J. Mod. Phys. A}\ }\textbf {\bibinfo {volume} {33}},\
  \bibinfo {pages} {1850004} (\bibinfo {year}
  {2018}{\natexlab{b}})}\BibitemShut {NoStop}%
\bibitem [{\citenamefont {Sveshnikov}\ \emph
  {et~al.}(2019{\natexlab{a}})\citenamefont {Sveshnikov}, \citenamefont
  {Voronina}, \citenamefont {Davydov},\ and\ \citenamefont
  {Grashin}}]{Sveshnikov2019a}%
  \BibitemOpen
  \bibfield  {author} {\bibinfo {author} {\bibfnamefont {K.}~\bibnamefont
  {Sveshnikov}}, \bibinfo {author} {\bibfnamefont {Y.}~\bibnamefont
  {Voronina}}, \bibinfo {author} {\bibfnamefont {A.}~\bibnamefont {Davydov}}, \
  and\ \bibinfo {author} {\bibfnamefont {P.}~\bibnamefont {Grashin}},\ }\href
  {\doibase doi.org/10.1134/S0040577919030024} {\bibfield  {journal} {\bibinfo
  {journal} {Theor. Math. Phys.}\ }\textbf {\bibinfo {volume} {198}},\ \bibinfo
  {pages} {331} (\bibinfo {year} {2019}{\natexlab{a}})}\BibitemShut {NoStop}%
\bibitem [{\citenamefont {Sveshnikov}\ \emph
  {et~al.}(2019{\natexlab{b}})\citenamefont {Sveshnikov}, \citenamefont
  {Voronina}, \citenamefont {Davydov},\ and\ \citenamefont
  {Grashin}}]{Sveshnikov2019b}%
  \BibitemOpen
  \bibfield  {author} {\bibinfo {author} {\bibfnamefont {K.}~\bibnamefont
  {Sveshnikov}}, \bibinfo {author} {\bibfnamefont {Y.}~\bibnamefont
  {Voronina}}, \bibinfo {author} {\bibfnamefont {A.}~\bibnamefont {Davydov}}, \
  and\ \bibinfo {author} {\bibfnamefont {P.}~\bibnamefont {Grashin}},\ }\href
  {\doibase doi.org/10.1134/S0040577919040056} {\bibfield  {journal} {\bibinfo
  {journal} {Theor. Math. Phys.}\ }\textbf {\bibinfo {volume} {199}},\ \bibinfo
  {pages} {533} (\bibinfo {year} {2019}{\natexlab{b}})}\BibitemShut {NoStop}%
\bibitem [{\citenamefont {Mohr}\ \emph {et~al.}(1998)\citenamefont {Mohr},
  \citenamefont {Plunien},\ and\ \citenamefont {Soff}}]{Mohr1998}%
  \BibitemOpen
  \bibfield  {author} {\bibinfo {author} {\bibfnamefont {P.~J.}\ \bibnamefont
  {Mohr}}, \bibinfo {author} {\bibfnamefont {G.}~\bibnamefont {Plunien}}, \
  and\ \bibinfo {author} {\bibfnamefont {G.}~\bibnamefont {Soff}},\ }\href
  {\doibase 10.1016/S0370-1573(97)00046-X} {\bibfield  {journal} {\bibinfo
  {journal} {Phys. Rep.}\ }\textbf {\bibinfo {volume} {293}},\ \bibinfo {pages}
  {227 } (\bibinfo {year} {1998})}\BibitemShut {NoStop}%
\bibitem [{\citenamefont {Bateman}\ and\ \citenamefont
  {Erdelyi}(1953)}]{Bateman1953}%
  \BibitemOpen
  \bibfield  {author} {\bibinfo {author} {\bibfnamefont {H.}~\bibnamefont
  {Bateman}}\ and\ \bibinfo {author} {\bibfnamefont {A.}~\bibnamefont
  {Erdelyi}},\ }\href@noop {} {\emph {\bibinfo {title} {Higher Transcendental
  Functions}}},\ Vol.\ \bibinfo {volume} {1-2}\ (\bibinfo  {publisher} {Mc
  Graw-Hill, New York},\ \bibinfo {year} {1953})\BibitemShut {NoStop}%
\bibitem [{\citenamefont {Zeldovich}\ and\ \citenamefont
  {Popov}(1972)}]{Zeldovich1972}%
  \BibitemOpen
  \bibfield  {author} {\bibinfo {author} {\bibfnamefont {Y.~B.}\ \bibnamefont
  {Zeldovich}}\ and\ \bibinfo {author} {\bibfnamefont {V.~S.}\ \bibnamefont
  {Popov}},\ }\href {http://stacks.iop.org/0038-5670/14/i=6/a=R01} {\bibfield
  {journal} {\bibinfo  {journal} {Soviet Physics Uspekhi}\ }\textbf {\bibinfo
  {volume} {14}},\ \bibinfo {pages} {673} (\bibinfo {year} {1972})}\BibitemShut
  {NoStop}%
\bibitem [{\citenamefont {Fano}(1961)}]{Fano1961}%
  \BibitemOpen
  \bibfield  {author} {\bibinfo {author} {\bibfnamefont {U.}~\bibnamefont
  {Fano}},\ }\href {\doibase 10.1103/PhysRev.124.1866} {\bibfield  {journal}
  {\bibinfo  {journal} {Phys. Rev.}\ }\textbf {\bibinfo {volume} {124}},\
  \bibinfo {pages} {1866} (\bibinfo {year} {1961})}\BibitemShut {NoStop}%
\bibitem [{\citenamefont {Reinhardt}\ \emph {et~al.}(1981)\citenamefont
  {Reinhardt}, \citenamefont {M\"uller},\ and\ \citenamefont
  {Greiner}}]{Reinhardt1981}%
  \BibitemOpen
  \bibfield  {author} {\bibinfo {author} {\bibfnamefont {J.}~\bibnamefont
  {Reinhardt}}, \bibinfo {author} {\bibfnamefont {B.}~\bibnamefont {M\"uller}},
  \ and\ \bibinfo {author} {\bibfnamefont {W.}~\bibnamefont {Greiner}},\ }\href
  {\doibase 10.1103/PhysRevA.24.103} {\bibfield  {journal} {\bibinfo  {journal}
  {Phys. Rev. A}\ }\textbf {\bibinfo {volume} {24}},\ \bibinfo {pages} {103}
  (\bibinfo {year} {1981})}\BibitemShut {NoStop}%
\bibitem [{\citenamefont {M{\"u}ller}\ \emph {et~al.}(1988)\citenamefont
  {M{\"u}ller}, \citenamefont {de~Reus}, \citenamefont {Reinhardt},
  \citenamefont {M{\"u}ller}, \citenamefont {Greiner},\ and\ \citenamefont
  {Soff}}]{Mueller1988}%
  \BibitemOpen
  \bibfield  {author} {\bibinfo {author} {\bibfnamefont {U.}~\bibnamefont
  {M{\"u}ller}}, \bibinfo {author} {\bibfnamefont {T.}~\bibnamefont {de~Reus}},
  \bibinfo {author} {\bibfnamefont {J.}~\bibnamefont {Reinhardt}}, \bibinfo
  {author} {\bibfnamefont {B.}~\bibnamefont {M{\"u}ller}}, \bibinfo {author}
  {\bibfnamefont {W.}~\bibnamefont {Greiner}}, \ and\ \bibinfo {author}
  {\bibfnamefont {G.}~\bibnamefont {Soff}},\ }\href {\doibase
  10.1103/PhysRevA.37.1449} {\bibfield  {journal} {\bibinfo  {journal} {Phys.
  Rev. A}\ }\textbf {\bibinfo {volume} {37}},\ \bibinfo {pages} {1449}
  (\bibinfo {year} {1988})}\BibitemShut {NoStop}%
\bibitem [{\citenamefont {Ackad}\ and\ \citenamefont
  {Horbatsch}(2008)}]{Ackad2008}%
  \BibitemOpen
  \bibfield  {author} {\bibinfo {author} {\bibfnamefont {E.}~\bibnamefont
  {Ackad}}\ and\ \bibinfo {author} {\bibfnamefont {M.}~\bibnamefont
  {Horbatsch}},\ }\href {\doibase 10.1103/PhysRevA.78.062711} {\bibfield
  {journal} {\bibinfo  {journal} {Phys. Rev. A}\ }\textbf {\bibinfo {volume}
  {78}},\ \bibinfo {pages} {062711} (\bibinfo {year} {2008})}\BibitemShut
  {NoStop}%
\bibitem [{\citenamefont {Marsman}\ and\ \citenamefont
  {Horbatsch}(2011)}]{Marsman2011}%
  \BibitemOpen
  \bibfield  {author} {\bibinfo {author} {\bibfnamefont {A.}~\bibnamefont
  {Marsman}}\ and\ \bibinfo {author} {\bibfnamefont {M.}~\bibnamefont
  {Horbatsch}},\ }\href {\doibase 10.1103/PhysRevA.84.032517} {\bibfield
  {journal} {\bibinfo  {journal} {Phys. Rev. A}\ }\textbf {\bibinfo {volume}
  {84}},\ \bibinfo {pages} {032517} (\bibinfo {year} {2011})}\BibitemShut
  {NoStop}%
\bibitem [{\citenamefont {Maltsev}\ \emph {et~al.}(2020)\citenamefont
  {Maltsev}, \citenamefont {Shabaev}, \citenamefont {Zaytsev},\ and\
  \citenamefont {et~al.}}]{Maltsev2020a}%
  \BibitemOpen
  \bibfield  {author} {\bibinfo {author} {\bibfnamefont {I.}~\bibnamefont
  {Maltsev}}, \bibinfo {author} {\bibfnamefont {V.}~\bibnamefont {Shabaev}},
  \bibinfo {author} {\bibfnamefont {V.}~\bibnamefont {Zaytsev}}, \ and\
  \bibinfo {author} {\bibnamefont {et~al.}},\ }\href {\doibase
  10.1134/S0030400X2008024X} {\bibfield  {journal} {\bibinfo  {journal} {Opt.
  Spectrosc.}\ }\textbf {\bibinfo {volume} {128}},\ \bibinfo {pages} {1100}
  (\bibinfo {year} {2020})}\BibitemShut {NoStop}%
\bibitem [{\citenamefont {M{\"u}ller-Nehler}\ and\ \citenamefont
  {Soff}(1994)}]{Mueller1994}%
  \BibitemOpen
  \bibfield  {author} {\bibinfo {author} {\bibfnamefont {U.}~\bibnamefont
  {M{\"u}ller-Nehler}}\ and\ \bibinfo {author} {\bibfnamefont {G.}~\bibnamefont
  {Soff}},\ }\href {\doibase 10.1016/0370-1573(94)90068-X} {\bibfield
  {journal} {\bibinfo  {journal} {Phys.Rep.}\ }\textbf {\bibinfo {volume}
  {246}},\ \bibinfo {pages} {101} (\bibinfo {year} {1994})}\BibitemShut
  {NoStop}%
\bibitem [{\citenamefont {Voronina}\ \emph
  {et~al.}(2019{\natexlab{c}})\citenamefont {Voronina}, \citenamefont
  {Sveshnikov}, \citenamefont {Grashin},\ and\ \citenamefont
  {Davydov}}]{Voronina2019a}%
  \BibitemOpen
  \bibfield  {author} {\bibinfo {author} {\bibfnamefont {Y.}~\bibnamefont
  {Voronina}}, \bibinfo {author} {\bibfnamefont {K.}~\bibnamefont
  {Sveshnikov}}, \bibinfo {author} {\bibfnamefont {P.}~\bibnamefont {Grashin}},
  \ and\ \bibinfo {author} {\bibfnamefont {A.}~\bibnamefont {Davydov}},\ }\href
  {\doibase https://doi.org/10.1016/j.physe.2018.08.013} {\bibfield  {journal}
  {\bibinfo  {journal} {Physica E}\ }\textbf {\bibinfo {volume} {106}},\
  \bibinfo {pages} {298 } (\bibinfo {year} {2019}{\natexlab{c}})}\BibitemShut
  {NoStop}%
\bibitem [{\citenamefont {Voronina}\ \emph
  {et~al.}(2019{\natexlab{d}})\citenamefont {Voronina}, \citenamefont
  {Sveshnikov}, \citenamefont {Grashin},\ and\ \citenamefont
  {Davydov}}]{Voronina2019b}%
  \BibitemOpen
  \bibfield  {author} {\bibinfo {author} {\bibfnamefont {Y.}~\bibnamefont
  {Voronina}}, \bibinfo {author} {\bibfnamefont {K.}~\bibnamefont
  {Sveshnikov}}, \bibinfo {author} {\bibfnamefont {P.}~\bibnamefont {Grashin}},
  \ and\ \bibinfo {author} {\bibfnamefont {A.}~\bibnamefont {Davydov}},\ }\href
  {\doibase https://doi.org/10.1016/j.physe.2018.09.026} {\bibfield  {journal}
  {\bibinfo  {journal} {Physica E}\ }\textbf {\bibinfo {volume} {109}},\
  \bibinfo {pages} {209 } (\bibinfo {year} {2019}{\natexlab{d}})}\BibitemShut
  {NoStop}%
\bibitem [{\citenamefont {Grashin}\ and\ \citenamefont
  {Sveshnikov}(2020{\natexlab{a}})}]{Grashin2020a}%
  \BibitemOpen
  \bibfield  {author} {\bibinfo {author} {\bibfnamefont {P.}~\bibnamefont
  {Grashin}}\ and\ \bibinfo {author} {\bibfnamefont {K.}~\bibnamefont
  {Sveshnikov}},\ }\href {\doibase 10.1016/j.aop.2020.168094} {\bibfield
  {journal} {\bibinfo  {journal} {Ann. of Phys.}\ }\textbf {\bibinfo {volume}
  {532}},\ \bibinfo {pages} {168094} (\bibinfo {year}
  {2020}{\natexlab{a}})}\BibitemShut {NoStop}%
\bibitem [{\citenamefont {Grashin}\ and\ \citenamefont
  {Sveshnikov}(2020{\natexlab{b}})}]{Grashin2020b}%
  \BibitemOpen
  \bibfield  {author} {\bibinfo {author} {\bibfnamefont {P.}~\bibnamefont
  {Grashin}}\ and\ \bibinfo {author} {\bibfnamefont {K.}~\bibnamefont
  {Sveshnikov}},\ }\href {\doibase 10.1002/andp.201900351} {\bibfield
  {journal} {\bibinfo  {journal} {Ann. der Physik}\ }\textbf {\bibinfo {volume}
  {415}},\ \bibinfo {pages} {351} (\bibinfo {year}
  {2020}{\natexlab{b}})}\BibitemShut {NoStop}%
\end{thebibliography}%

\end{document}